\documentclass[10pt,journal,compsoc]{IEEEtran}

\ifCLASSOPTIONcompsoc
  \usepackage[nocompress]{cite}
\else
  \usepackage{cite}
\fi

\usepackage{algorithm}
\usepackage[noend]{algpseudocode}

\usepackage{amsmath}
\usepackage{amssymb}
\usepackage{enumitem}
\usepackage{bookmark}
\usepackage{makecell}
\usepackage{xcolor}
\usepackage{balance}
\newcommand{\chengk}[1]{#1}

\usepackage{bm}

\usepackage{subcaption}
\usepackage{booktabs}
\usepackage{diagbox}
\usepackage{colortbl}
\usepackage[many]{tcolorbox}
\usepackage{multirow}

\usepackage{xspace}

\newcommand{\attackname}{\textsc{AS2T}\xspace}
\newcommand{\attacknamenospace}{\textsc{AS2T}}

\DeclareMathOperator\arctanh{arctanh}

\usepackage{diagbox}

\begin{document}

\title{\attackname: Arbitrary Source-To-Target Adversarial Attack on Speaker Recognition Systems}

\author{Guangke Chen, Zhe Zhao, Fu Song, Sen Chen, Lingling Fan, and Yang Liu~\IEEEmembership{Senior~Member,~IEEE}
\IEEEcompsocitemizethanks{
\IEEEcompsocthanksitem Guangke Chen is with ShanghaiTech University, Shanghai, China;
Shanghai Institute of Microsystem and Information Technology, Chinese Academy of Sciences, Shanghai, China;
and University of Chinese Academy of Sciences, Beijing, China.
\IEEEcompsocthanksitem Zhe Zhao is with ShanghaiTech University, Shanghai, China.
\IEEEcompsocthanksitem Fu Song (corresponding author) is with ShanghaiTech University, Shanghai, China. Email: songfu@shanghaitech.edu.cn
\IEEEcompsocthanksitem Sen Chen is with Tianjin University, Tianjin, China.
\IEEEcompsocthanksitem Lingling Fan is with Nankai University, Tianjin, China.
\IEEEcompsocthanksitem Yang Liu is with Nanyang Technological University, Singapore.
}
}


\IEEEtitleabstractindextext{%
\begin{abstract}

Recent work has illuminated the vulnerability of speaker recognition systems (SRSs) against adversarial attacks,
raising significant security concerns in deploying SRSs. 
However, they considered only a few settings (e.g., some combinations of source and target speakers), 
leaving many interesting and important settings in real-world attack scenarios alone.
In this work, we present \attackname, the first attack in this domain which covers all the settings, thus 
allows the adversary to craft adversarial voices using arbitrary source and target speakers for any of three main recognition tasks.
Since none of the existing loss functions can be applied to all the settings,
we explore many candidate loss functions for each setting including the existing and newly designed ones.
We thoroughly evaluate their efficacy 
and find that some existing loss functions are suboptimal. 
Then, to improve the robustness of \attackname towards practical over-the-air attack,
we study the possible distortions occurred in over-the-air transmission,
utilize different transformation functions with different parameters 
to model those distortions, and incorporate them into the generation of adversarial voices.
Our simulated over-the-air evaluation validates the effectiveness of our solution
in producing robust adversarial voices which remain effective
under various hardware devices and various acoustic environments
with different reverberation, ambient noises, and noise levels.
Finally, we leverage \attackname to perform thus far the largest-scale evaluation  
to understand transferability among 14 diverse SRSs. 
The transferability analysis provides many interesting and useful insights
which challenge several findings and conclusion drawn in previous works in the image domain.
Our study also sheds light on future directions of adversarial attacks in the speaker recognition domain.
\end{abstract}

}

\maketitle

\IEEEdisplaynontitleabstractindextext

\IEEEpeerreviewmaketitle

\IEEEraisesectionheading{\section{Introduction}\label{sec:introduction}}

\IEEEPARstart{S}{peaker} recognition (SR) is an automatic procedure of
verifying or identifying individual speakers 
by extracting and interpreting their unique acoustic characteristics~\cite{Homayoon11}.
There are three main SR tasks: close-set identification (CSI),
open-set identification (OSI), and speaker verification (SV).
CSI recognizes unknown speakers from a group of enrolled speakers $G$.
OSI is similar to CSI except that it may regard a speaker as an imposter (i.e., an unenrolled speaker).
SV is similar to OSI except that only one speaker can be enrolled. 
SR has been adopted by open-source platforms (e.g., SpeechBrain~\cite{speech-brain})
and commercial products (e.g., Microsoft Azure~\cite{microsoft-azure-vpr}),
and used in security-critical scenarios such as electrical appliances access control in smart home~\cite{RenSYS16} and
remote voice authentication in financial transaction~\cite{TD-Bank}.

Machine learning including deep learning is the dominant
approach to implement the  state-of-the-art SR systems (SRSs)~\cite{reynolds2000speaker,Sremath2016areview}.
Unsurprisingly, SRSs have been shown to be fragile
to adversarial attacks in some settings~\cite{abs-1801-03339,li2020adversarial,jati2021adversarial,zhang2021attack,
LiZJXZWM020, xie2020enabling,WangGX20,shamsabadi2021foolhd,chen2019real,du2020sirenattack,DBLP:journals/corr/abs-1711-03280}, namely,
adding a tiny perturbation to a voice uttered by one source speaker is misclassified 
by the SRS, but still correctly recognized as the source speaker by ordinary users.

We will denote by $s\to {\tt untar}$ and $s\to t$ ($s\neq t$) the untargeted and targeted attacks, respectively.
When $s\in G$ (resp. $s\notin G$), the source speaker is one of the enrolled speakers (resp. an unenrolled speaker).
When $t\in G$ (resp. $t={\tt imposter}$), the adversarial voice is recognized as the enrolled speaker $t$
(resp. rejected as an imposter) by the SRS.


\begin{table}[t]\setlength\tabcolsep{10pt}
\centering
  \caption{Existing attacks}\vspace{-2mm}
  \label{tab:existingattack}
  \resizebox{0.45\textwidth}{!}{ \begin{tabular}{|c|c|c|}
    \toprule
  Attack & Task &  Reference   \\ \midrule
  $s \to {\tt untar}$ for $s\in G$ & CSI &    \cite{jati2021adversarial,shamsabadi2021foolhd,DBLP:journals/corr/abs-1711-03280,du2020sirenattack} \\ \midrule
   $s\to t$ for $s,t\in G$ & CSI  &  \cite{shamsabadi2021foolhd,du2020sirenattack,chen2019real} \\ \midrule
  $s \to t$ for $s\not\in G$ and $t\in G$ & SV &   \cite{abs-1801-03339,li2020adversarial,zhang2021attack,LiZJXZWM020,xie2020enabling,WangGX20,chen2019real}  \\ \midrule
  $s\to t$ for $s\notin G$ and $t\in G$ & OSI &   \cite{chen2019real}   \\
    \bottomrule
  \end{tabular}}\vspace{-4mm}
  \end{table}
Prior works make remarkable progress in revealing the serve security implications of adversarial attacks
on both open-source and commercial SRSs. 
However, they considered only a few settings as shown in \tablename~\ref{tab:existingattack},
leaving many interesting and important settings alone, e.g.,
$s \to {\tt imposter}$ for $s \in G$ on the OSI and SV tasks and
$s \to {\tt untar}$ for $s\notin G$ on the OSI and CSI tasks
(cf. \tablename~\ref{tab:s-t-setting} for all the 11 settings).
%
Yet these settings are interesting and important in real-world attack scenarios.
For example, voice-controllable smart home offers for family members hands-free control of smart products,
e.g., thermostat and light.
Some ordinary products are controlled by all family members,
while others are controlled exclusively by some members, such as parents.
If an adversary attempts to acquire the permission of exclusively controlled products,
he needs to perform a targeted attack $s \to t\in G$ for $s\notin G$ on the OSI task (considered in \cite{chen2019real})
where the target speaker $t$ owns the permission.
In contrast, if he intends to obtain the privilege to control ordinary products,
he can simply perform an untargeted attack $s \to {\tt untar}$ for $s\notin G$,
which is less difficult than targeted attack.
Additionally, while the attack $s \to t\in G$ for $s\notin G$ on the SV task considered in previous works
can be exploited to unlock the victim's smartphone, log into the victim's applications, etc.,
the attack $s \to {\tt imposter}$ for $s \in G$ on the SV task can cause Denial-of-Service to the victim by disabling his authentication.

In this work, we aim to tackle the key limitation of prior works by proposing an attack,
named {\bf A}rbitrary {\bf S}ource-{\bf T}o-{\bf T}arget adversarial attack (\attackname),
covering all the above settings. 
Given a source speaker, either enrolled or unenrolled, a benign voice from the source speaker, and a target speaker that is either ${\tt untar}$ (for untargeted attack), or an enrolled or unenrolled speaker (for targeted attack), \attackname produces an adversarial voice with which the adversary can achieve his goal in the intended attack scenario.
The design and implementation of \attackname is motivated by the following research questions:
(RQ1) How to construct adversarial voices given arbitrary source and/or target speakers,
considering that the adversary may own different levels of knowledge about SRSs?
(RQ2) How to launch practical over-the-air attacks
where distortions from both environment and hardware
may disrupt adversarial voices?
(RQ3) How about the attack capability of \attackname under totally black-box setting
in which the adversary does not own any information about the model and cannot frequently query it?

To address RQ1, we formulate the crafting of adversarial voices as a constrained optimization problem
and design appropriate source-/target-oriented loss functions
for each setting, i.e., a combination of source and target speakers and a recognition task.
%
Note that none of existing loss functions in the literature can be applied to all the considered settings. 
For example, the commonly adopted Cross Entropy Loss, defined as the negative
logarithm of the predicted probability of the source class (untargeted attack)
or target class (targeted attack), cannot be used for the attack $s \to {\tt untar}$ with $s\notin G$ on both the OSI and CSI tasks,
as no enrolled speaker of the SR model corresponds to the speaker $s$. 
Therefore, we explore various candidate loss functions including existing and newly designed ones,
and conduct a thorough evaluation to compare their effectiveness
and efficiency for each setting. We find that our loss functions outperform some loss functions adopted by prior works.
According to our results, \attackname is designed to adaptively choose the optimal loss function for each attack setting.
Furthermore, the adversary can freely integrate our loss functions
into an optimization approach according to his prior knowledge about the SR model under attack,
ranging from white-box access to model structure and parameters
to black-box access to model's outputs (decision or score).
Our experimental results with three representative white-box and one state-of-the-art black-box approaches
(after necessary modifications) demonstrate the generic capability of \attackname
to different optimization approaches.

To address RQ2,
we first investigate the major sources of distortions occurred in the over-the-air transmission
which may destruct the effectiveness of adversarial voices,
i.e., reverberation, ambient noise, and equipment distortion.
Based on their nature and properties,
we utilize different proper transformation functions
to model those distortions and incorporate them into \attackname, 
where the loss is computed on the transformed voice,
thus it is expected that the crafted adversarial voices can survive from these distortions.
To guarantee the attacker can use the adversarial voices to launch attack in various environments
and even target multiple victims' devices simultaneously,
we specify a variety of parameters for the transformation functions,
e.g., different Signal-to-Noise ratio for the function simulating the effect of ambient noise.
We empirically confirm that our solution to RQ2 significantly enhance
the robustness of adversarial voices against these distortions under various attack conditions,
thus improve the practicability of \attackname.

\chengk{Under RQ3, the adversary cannot directly and frequently query the target model and thus
has to leverage transfer attacks, i.e., crafting adversarial voices on a white-box source model and transfer them to the totally black-box target model~\cite{chen2019real,jati2021adversarial,li2020adversarial}.}
\chengk{Transfer attack is an obstacle to securely deploying machine learning models due to its ability to perform simple and practical black-box attacks~\cite{Space-Adver,Delving-Transfer}.}
To thoroughly evaluate the capability of \attackname \chengk{under this challenging scenario and study the factors that may influence the success rate of transfer attacks},
we perform a large-scale transferability analysis among 14 SR models, covering
five model architectures, five training datasets,
four input types,
and two scoring backends.
We study the impact of both model-specific factors (architecture, training dataset, and input type)
and attack-specific factors (number of iterations, step\_size, and perturbation budget)
on the transferability success rate of \attackname.
Our analysis provides lots of useful insights and findings \chengk{for better launching transfer attacks}.
For instance, model-specific factors are dominant factors over attack-specific ones;
adversarial examples tend to, but not necessarily, transfer well to the target models
with the same architecture as the source model; the transferability between two models may be asymmetric,
which challenges the decision boundary similarity based explanation of transferable adversarial
examples in the image domain~\cite{Space-Adver, Delving-Transfer},
since similarity is symmetric;
iterative attack does not necessarily produce less transferable adversarial voices than single-step attack,
contradicting the finding in the image domain~\cite{kurakin2016adversarial}.

In summary, we make the following main contributions.
\begin{itemize}[leftmargin=*]
    \item We propose \attackname, an arbitrary source-to-target adversarial attack on SRSs.
    It features source-/target-oriented and novel loss functions
    and enables the adversary to use arbitrary source and target speakers to craft adversarial voices
    to achieve attack goals in adversary-chosen attack scenarios.
    \item We successfully construct robust adversarial voices which remain effective when being played over-the-air
    in various scenes. Our solution is modeling and  incorporating the possible distortions
    into the generation of adversarial voices using different parameterized transformation functions.
    \item To the best of our knowledge, we perform the largest-scale transferability analysis in the speaker recognition domain.
    We discover many valuable insights which can guide the future works in this domain.
\end{itemize}

\begin{figure}[t]
    \includegraphics[width=0.48\textwidth]{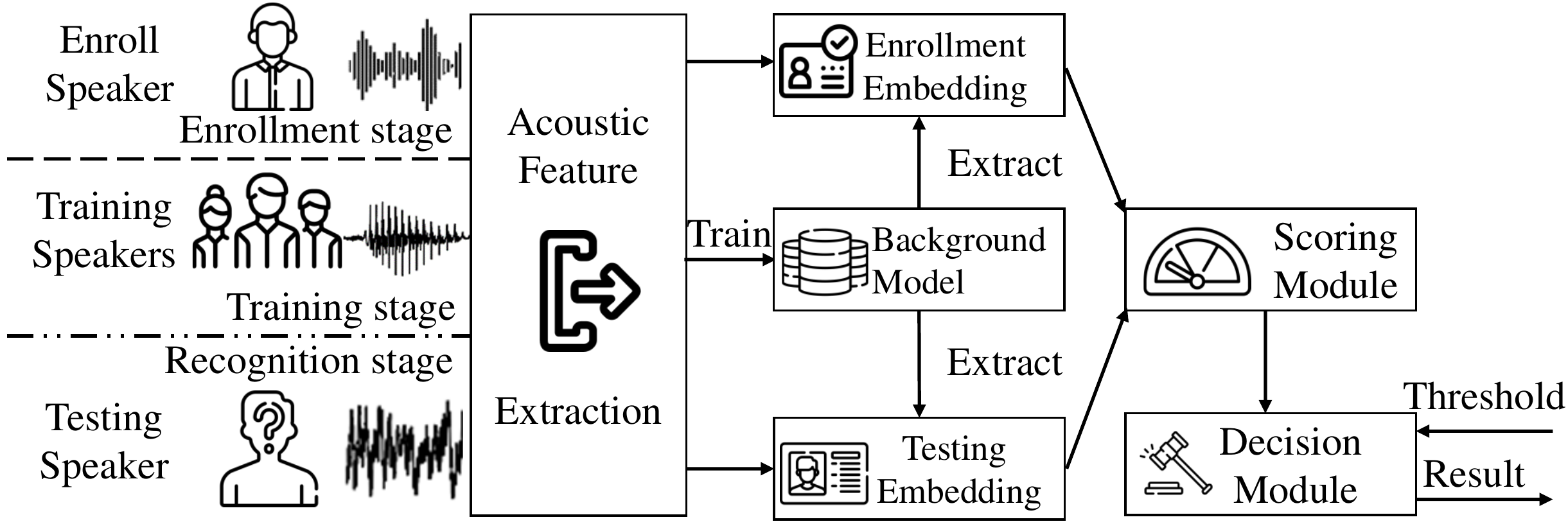}\vspace{-1mm}
    \caption{A generic architecture of embedding-based SRSs}
    \label{fig:typical-SRSs}\vspace{-3mm}
\end{figure}

\begin{table*}
    \centering\setlength\tabcolsep{7pt}
    \caption{Different goals an adversary intends to achieve}\vspace{-2mm}
    \resizebox{.95\textwidth}{!}{%
      \begin{tabular}{|c|c|c|c|c|c|c|}
      \hline
      \multicolumn{2}{|c|}{\textbf{Goal}} & \textbf{Description} & \textbf{Setting} & \makecell{{\bf Source}\\{\bf Speaker}} & \makecell{{\bf Target}\\{\bf Speaker}}  & {\bf ID} \\
      \hline
      \multirow{10}{*}{\textbf{Malicious}} & \multirow{3}{*}{\makecell{{\bf Unauthorized} \\ {\bf Access}}} & \multirow{3}{*}{Obtain privilege} & \multirow{2}{*}{Exclusive privilege owned by $G_1\subset G$} & $s\in G\setminus G1$ & $t\in G_1$ & {S1-1} \\
  \cline{5-7}          &       &       &       & $s\notin G$ & $t\in G_1$ & {S1-2} \\
  \cline{4-7}          &       &       & Low-level privilege shared by $G$ & $s\notin G$ & Untargeted & {S1-3}\\
  \cline{2-7}          & \multirow{3}{*}{\textbf{Denial-of-Service}} & \multirow{3}{*}{\makecell{Hinder accessing services}} & Common services shared by $G$  & $s\in G$ & $t\notin G$ & S2-1 \\
  \cline{4-7}          &       &       & \multirow{2}{*}{Personalized services} & $s\in G$ & $t\notin G$ & S2-2 \\
  \cline{5-7}          &       &       &       & $s\in G$ & Untargeted & S2.3 \\
  \cline{2-7}          & \multirow{2}{*}{\makecell{{\bf Anonymous}\\ {\bf Access}}} & \multirow{2}{*}{\makecell{Hides identity when accessing illegal service}} & ---   & $s\in G$ & Untargeted & S3-1 \\
  \cline{4-7}          &       &       & Cause reputation degrade to a specific speaker & $s\in G$ & $t\in G\setminus\{s\}$ & S3-2 \\
  \cline{2-7}          & \multirow{2}{*}{\textbf{Evasion}} & \multirow{2}{*}{\makecell{Anomalous subject evades detection}} & Single anomalous subject & $s\in G$ & Untargeted & S4-1 \\
  \cline{4-7}          &       &       & Multiple anomalous subjects $G_2\subset G$ & $s\in G$ & $t\in G\setminus G_2$ & S4-2 \\
      \hline
      \textbf{Beneficial} & \textbf{Privacy Protection} & \makecell{Protect privacy against  excessive surveillance} & ---   & $s\in G$ & Untargeted & S5-1 \\
      \hline
      \end{tabular}%
    }
    \label{tab:goal} \vspace{-3mm}
  \end{table*}%

\section{Speaker Recognition Systems}\label{sec:SRSs}
%
%
\subsection{Overview of Speaker Recognition}
Modern and cutting-edge SRSs represent characteristics of speakers as fixed-dimensional vectors,
i.e., speaker embedding~\cite{wang2020simulation}.
\figurename~\ref{fig:typical-SRSs} depicts a generic architecture of embedding-based SRSs,
consisting of training, enrollment, and recognition stages.
In the training stage, a huge number of voices from thousands of training speakers are utilized to train a background model,
which learns a mapping from voices to embeddings in the vector space.
Classic background model utilizes Gaussian Mixture Model (GMM)~\cite{ReynoldsR95,reynolds2000speaker},
which produces identity-vector (ivector) embeddings~\cite{DehakDKBOD09}. 
Recent promising background model
utilizes deep neural networks,
which produce deep embeddings, e.g., dvector~\cite{variani2014deep} or xvector~\cite{snyder2018x}.
In the enrollment stage, the background model maps each enrolling speaker's voice to an \emph{enrollment embedding},
regarded as a unique voice-identity of the enrolling speaker.
In the recognition stage, given a voice of a testing speaker, its \emph{testing embedding} is retrieved from the background model
for scoring. The scoring module computes the similarity between the enrollment
and testing embeddings based on which the result is produced by the decision module.
There are two widely-used scoring approaches: Probabilistic Linear Discriminant Analysis (PLDA)~\cite{nandwana2019analysis}
and COSine Similarity (COSS)~\cite{dehak2010cosine}.
The former one works well in most situations, but needs to be trained using the embeddings of training voices~\cite{wang2020simulation},
while the latter one is a reasonable substitution of PLDA without training.

The acoustic feature extraction module converts raw speech signals to acoustic features
carrying characteristics of the raw signals.
Common feature extraction algorithms include speech spectrogram~\cite{hannun2014deep} 
fBank~\cite{FilterBanks}\cite{li2017deep}, MFCC~\cite{muda2010voice}, and PLP~\cite{hermansky1990perceptual}.
Note that some end-to-end neural network-based SRSs, e.g., SincNet~\cite{SincNet}, 
extract features by the hidden neurons of neural networks instead of an explicit acoustic feature extraction module.

\subsection{Speaker Recognition Tasks}\label{sec:SR-task}
There are three main speaker recognition tasks: 
open-set identification (OSI),
close-set identification (CSI), and speaker verification (SV).
OSI allows multiple speakers
to be enrolled during the enrollment stage,
forming a speaker group $G$.
Given a voice $x$ of a testing speaker, it determines whether $x$ is uttered by one of the enrolled speakers
or none of them, according to the scores of all the enrolled speakers and a preset (score) threshold $\theta$.
Formally, suppose the speaker group $G$ has $n$ speakers $\{1,2,\cdots,n\}$,
the decision module outputs $D(x)$: 
\begin{center}
    $D(x)=\left\{
    \begin{array}{ll}
    \arg\max\limits_{i\in G} \ [S(x)]_i, &\mbox{if } \max\limits_{i\in G} \ [S(x)]_i\geq \theta; \\
    {\tt imposter}, &\mbox{otherwise}.
    \end{array}\right.$
    \end{center}
where $[S(x)]_i$ for $i\in G$ denotes the score of the voice $x$ that is likely uttered by the speaker $i$.
Intuitively, the system classifies the input voice $x$ as the speaker $i$
if and only if the score $[S(x)]_i$ of the speaker $i$ is the largest one among all the enrolled speakers,
and no less than the threshold $\theta$.
If all the scores are less than $\theta$, the system directly rejects the voice, namely,
it is not uttered by any of the enrolled speakers.

CSI and SV accomplish similar tasks as OSI, but with some exceptions.
A CSI system never rejects any input voices, i.e., an input will always be classified as one of the enrolled speakers.
Whereas an SV system can have exactly \emph{one} enrolled speaker
and checks if an input voice is uttered by the enrolled speaker.

\begin{table*}
\begin{minipage}{0.39\textwidth}
 \centering\setlength\tabcolsep{3pt}
    \caption{Settings of source/target speakers}    \vspace{-2mm}
    \label{tab:s-t-setting}
    \resizebox{0.94\textwidth}{!}{%
      \begin{tabular}{|c|c|c|c|c|}
      \toprule
    \textbf{Task} & \textbf{ID} & \textbf{Source Speaker} & \textbf{Target} & \begin{tabular}[c]{@{}c@{}} {\bf Goals} \end{tabular} \\
      \midrule
      \multirow{5}[10]{*}{\textbf{OSI}} & \textbf{C1} & enrolled speaker $s$ & enrolled speaker $t\neq s$ & S1-1, S3-2, S4-2 \\
         & \textbf{C2} & unenrolled speaker &  enrolled speaker  & S1-2 \\
         & \textbf{C3} & enrolled speaker  & imposter  & S2-1, S2-2 \\
        & \textbf{C4} & enrolled speaker  & untargeted & \begin{tabular}[c]{@{}c@{}} S2-3, S3-1, S4-1, S5-1 \end{tabular} \\
       & \textbf{C5} & unenrolled speaker  & untargeted & S1-3 \\
      \midrule
      \multirow{3}[8]{*}{\textbf{CSI}} & \textbf{C6} &  enrolled speaker $s$ & enrolled speaker $t\neq s$ & S1-1, S3-2, S4-2 \\
        & \textbf{C7} &  unenrolled speaker &  enrolled speaker &  S1-2 \\
        & \textbf{C8} &  enrolled speaker  & untargeted & \begin{tabular}[c]{@{}c@{}}S2-3, S3-1, S4-1, S5-1 \end{tabular} \\
      \midrule
\multirow{2}[4]{*}{\textbf{SV}} & \textbf{\chengk{C9}} & enrolled speaker & imposter & \begin{tabular}[c]{@{}c@{}}S2-1, S2-2, S2-3 \\ S3-1, S4-1, S5-1 \end{tabular} \\
         & \textbf{\chengk{C10}} &  unenrolled speaker & enrolled speaker & S1-2, S1-3 \\  
      \bottomrule
      \end{tabular}%
    }
\end{minipage}
\begin{minipage}{0.6\textwidth}
    \centering
    \includegraphics[width=0.9\textwidth]{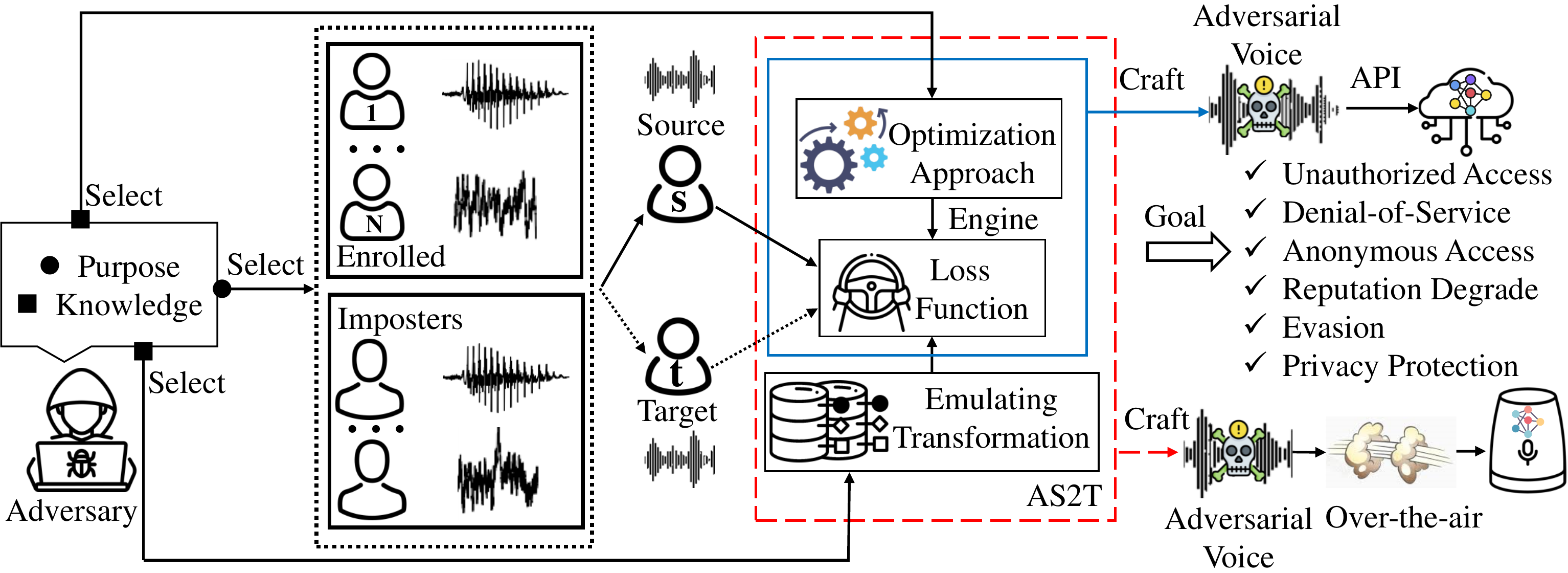}\vspace{-3mm}
   \renewcommand{\tablename}{Fig.}
    \caption{The overview of our attack \attackname}
    \label{fig:attack-overview}
\end{minipage}\vspace{-3mm}
\end{table*}

\section{Our Attack: \attackname}
In this section, we first highlight the motivation and threat model of our work and then present in detail our attack.

\subsection{Motivation and Threat Model}
We assume that the adversary intends to craft an adversarial example from a voice uttered by a source speaker,
so that it is classified as one of the enrolled speakers (untargeted attack)
or the target speaker or imposter (targeted attack)  by the SRS under attack (effectiveness),
but is still recognized as the source speaker by ordinary users  (stealthiness).
\tablename~\ref{tab:goal} summarizes possible goals of the adversary,
including malicious goals, e.g., unauthorized access, Denial-of-Service (Dos), anonymous access, and evasion,
and beneficial goals, e.g., privacy protection.


We emphasize that different goals often require different source and/or target speakers. 
Even for the same goal, the source and target speakers also differ with the settings.
Consider the unauthorized access goal.
When the privilege the adversary intends to obtain is exclusive to some specific enrolled speakers,
e.g., unlocking the victim's smartphone, logging into the victim's applications,
and conducting illegal financial transactions on behalf of the victim,
the adversary should target an enrolled speaker who owns this privilege. 
In contrast, when the privilege is shared by all the enrolled speakers,
e.g., all family members can remotely control some electrical appliances in the smart home,
the adversary can simply perform untargeted attack. 

Therefore, the adversarial attack with arbitrary source and target speakers is important and worthy to explore
considering various attack scenarios in the physical world.
As shown in \tablename~\ref{tab:s-t-setting}, there are 11 combinations of source and target speakers on the three tasks,
differing from the attacks in the image domain. 
Indeed, image recognition is a close-set multi-class problem and prior works on adversarial image attacks correspond to C6 and C8 in \tablename~\ref{tab:s-t-setting}.

In this work, we propose an attack, named Arbitrary Source-To-Target adversarial attack (\attackname),
\chengk{covering all the possible combinations of source speaker and target speaker on all the three SRS tasks, as listed in \tablename~\ref{tab:s-t-setting}.
The source speaker could be an enrolled or unenrolled speaker and the target could be untargeted (untargeted attack), an enrolled speaker or imposter (targeted attack),
leading to $2\times 3$ combinations per task. Note that imposter represents all the unenrolled speakers. 
After excluding the meaningless combination unenrolled $\to$ imposter, there are 5 settings (C1-C5) for OSI task.
Since CSI task never rejects, both enrolled $\to$ imposter and unenrolled $\to$ untargeted are excluded, leading to 3 settings (C6-C8) for CSI task.
Note that unenrolled $\to$ untargeted is excluded since any input voice uttered by any unenrolled speaker will naturally be classified into one of enrolled speakers by CSI task.
Since SV task is a binary classification, its setting is either enrolled $\to$ imposter (C9) or unenrolled $\to$ enrolled (C10).}
The overview of \attackname is depicted in \figurename~\ref{fig:attack-overview}.
\attackname formulates the construction of adversarial voices as an optimization problem (cf. Section~\ref{sec:problem-formulation})
and features source-/target-oriented loss functions (cf. Section~\ref{sec:loss}).
%
%
%
%
%
In addition, an adversary may own different levels of knowledge about the  model under attack,
from white-box access to model structure and parameters,
to black-box access to model outputs (decisions or scores).
The attack should be generic and can work under different levels of knowledge.
We address this in the optimization approach (cf. Section~\ref{sec:optimization}).
Furthermore, the crafted adversarial voices should also remain effective when being played over-the-air
to launch practical attack in the physical world.
The potential distortions occurred in this physical process are likely to destruct the effectiveness of adversarial examples.
We address this challenge by modeling and incorporating 
the possible distortions into \attackname (cf. Section~\ref{sec:practical-attack}). 

\subsection{Problem Formulation}\label{sec:problem-formulation}
The problem of finding an adversarial voice from a voice $x$ uttered by a source speaker $s$,
is formalized as the following constrained optimization problem:
\begin{align*}
    \text{argmin}_{x'} & \ d(x',x) \quad \text{s.t. }   D(x')=t   \mbox{ and }  x'\in [-1,1]
\end{align*}
where $d(x',x)$ is a distance metric quantifying the similarity between $x'$ and $x$ (stealthiness), 
and $t$ is a target speaker ($t$ is automatically selected in an untargeted attack).
%

The above minimization problem is difficult to solve due to the highly non-linear constraint $D(x')=t$.
Therefore, we re-formulate it and turn to solve the following problem:
\begin{align*}
    \text{argmin}_{x'}\quad \mathcal{L}(x', t) + \lambda \times d(x',x)
\end{align*}
where $\mathcal{L}$ is the loss function indicating the effectiveness of the attack
and the hyper-parameter $\lambda$ is a trade-off between the effectiveness and stealthiness of the attack.
In our attack, we utilize $L_p$ norm as the distance metric, i.e., $d(x',x)=(\sum_{i}|x'_i-x_i|^p)^{\frac{1}{p}}$,
which has been widely adopted in previous works, e.g.,~\cite{carlini2017towards,chen2019real,SEC4SR}.
There are many choices of the loss function $\mathcal{L}$.
We will explore various candidate loss functions in Section~\ref{sec:loss} 
and empirically evaluate them in Section~\ref{sec:evaluation}.

\subsection{Loss Function Design}\label{sec:loss}
As aforementioned, none of existing loss functions can be applied to all the settings listed in \tablename~\ref{tab:s-t-setting}.
In this subsection, we explore possible loss functions for each setting.

\subsubsection{Loss functions for OSI task}
We denote by $G=\{1,2,\cdots,n\}$ the group of enrolled speakers. 
Then, the decision space of the OSI task is $\mathcal{D}=G\cup\{{\tt imposter}\}$.

\smallskip\noindent {\bf C1 and C2: Targeted attack ($\bm{t\in G}$).}
Given a benign voice $x$ uttered by a source speaker $s$ that is either an enrolled (i.e., $s\in G$, C1) or unenrolled (i.e., $s\not\in G$, C2) speaker,
we define the following four loss functions for targeted attack on the OSI task
with a target speaker $t\in G$:
\begin{align*}
    \mathcal{L}_{\text{CE}}(x,t) &\triangleq -\log[\sigma(S(x))]_t \qquad  \mathcal{L}_{1}(x,t) \triangleq -[S(x)]_t \\
    \mathcal{L}_{\text{M}}(x,t) &\triangleq {\tt max}_{i\in G,i\neq t}[S(x)]_i - [S(x)]_t \\
    \mathcal{L}_{2}(x,t) &\triangleq {\tt max}\{\theta, {\tt max}_{i\in G,i\neq t}[S(x)]_i\} - [S(x)]_t \nonumber
\end{align*}
where $\sigma$ denotes the softmax function, $\theta$ is a preset (score) threshold, 
and $\mathcal{L}_{\text{CE}}$ and $\mathcal{L}_{\text{M}}$ are the Cross Entropy Loss and the Margin Loss, respectively.
$\mathcal{L}_\text{CE}$ and $\mathcal{L}_\text{M}$ are widely adopted
to craft adversarial examples in the image domain, e.g.,~\cite{goodfellow2014explaining,carlini2017towards}.
Unlike $\mathcal{L}_\text{M}$ which aims to simultaneously increase the score of the target speaker $t$
and reduce the scores of the other enrolled speakers,
$\mathcal{L}_1$ is designed to increase the score of the target speaker $t$ only,
by which we can check the effectiveness of the term
${\tt max}_{i\in G,i\neq t}[S(x)]_i$ in $\mathcal{L}_\text{M}$.
$\mathcal{L}_2$ is designed such that $\mathcal{L}_2(x,t)\leq 0 \Leftrightarrow D(x)=t$.
When $\mathcal{L}_2$ is minimized,
the score $[S(x)]_t$ of the target speaker $t$ is maximized to exceed the threshold $\theta$ and the scores of all the other enrolled speakers,
indicating a successful attack.

\smallskip\noindent {\bf C3: Targeted attack ($\bm{t\not\in G}$).}
Given a benign voice $x$ uttered by a source speaker $s$ that is an enrolled speaker (i.e., $s\in G$),
the following three loss functions are introduced to make its adversarial counterpart being rejected by the OSI task:
\begin{align*}
    \mathcal{L}_{\text{CE}}^{s}(x,t) &\triangleq \log[\sigma(S(x))]_s \qquad    \mathcal{L}_{1}^{s}(x,t) \triangleq [S(x)]_s \\
    \mathcal{L}_{3}(x,t) &\triangleq {\tt max}_{i\in G}[S(x)]_i-\theta
\end{align*}
Note that the parameter $t$ in the above loss functions is not necessary and is used to make
notions consistent.
$\mathcal{L}_{\text{CE}}^{s}$ (resp. $\mathcal{L}_{1}^{s}$) is the negation of $\mathcal{L}_{\text{CE}}$ (resp. $\mathcal{L}_{1}$) in which the target speaker $t$ is replaced by the source speaker $s$,
intended to minimize the score of the source speaker $s$ (Note that $\mathcal{L}_{\text{CE}}$ and $\mathcal{L}_{1}$ are designed to maximize the score of the target speaker $t$).
$\mathcal{L}_{3}$ is designed such that $\mathcal{L}_{3}(x,t)\leq 0\Leftrightarrow D(x)={\tt imposter}$.
Minimizing $\mathcal{L}_{3}$ makes the scores of all the enrolled speakers be less than the threshold $\theta$,
thus the adversarial voice is rejected, indicating a successful attack.

\smallskip\noindent{\bf C4: Untargeted attack ($\bm{s\in G}$).}
Given a benign voice $x$ uttered by a source speaker $s$ such that $s\in G$,
the untargeted attack may craft an adversarial voice such that:
1) it is rejected by the OSI task.
This case is equivelent to ``Targeted attack (${t\not\in G}$)",
hence is omitted here;
or 2) it is recognized as another enrolled speaker $t\in G$.
We define the following five loss functions:
\begin{align*}
    \mathcal{L}_{\text{CE}}^{s}(x,t) &\triangleq \log[\sigma(S(x))]_s \qquad    \mathcal{L}_{1}^{s}(x,t) \triangleq [S(x)]_s \\
    \mathcal{L}_{\text{M}}^{s}(x,t) &\triangleq [S(x)]_s-{\tt max}_{i\in G,i\neq s}[S(x)]_i \\
    \mathcal{L}_{2}^{s}(x,t) &\triangleq {\tt max}\{\theta, [S(x)]_s\}-{\tt max}_{i\in G,i\neq s}[S(x)]_i \\
    \mathcal{L}_{4}^{s}(x,t) &\triangleq -{\tt max}_{i\in G,i\neq s}[S(x)]_i
\end{align*}
$\mathcal{L}_{\text{M}}^s$ is the negation of $\mathcal{L}_{\text{M}}$ in which the target speaker $t$ is replaced by the source speaker $s$,
intended to reduce the score of the source speaker $s$ while increase the scores of other enrolled speakers
(Remark that $\mathcal{L}_{\text{M}}$ is designed to increase the score of the target speaker $t$ while reduce the scores of all the other enrolled speakers).
$\mathcal{L}_{1}^{s}$ and $\mathcal{L}_{4}^{s}$ are the two terms of $\mathcal{L}_{\text{M}}^s$
used to check their effectiveness in $\mathcal{L}_{\text{M}}^s$. 
$\mathcal{L}_{2}^{s}$ is designed such that $\mathcal{L}_{2}^{s}(x,t)\leq 0 \Leftrightarrow D(x)\in G\setminus \{s\}$.
When minimizing $\mathcal{L}_{2}^{s}$, we intend to find an enrolled speaker $s'$ ($s'\neq s$) whose score is the largest one and exceeds the threshold $\theta$,
hence an adversarial voice is recognized as the speaker $s'$, indicating a successful untargeted attack.

\smallskip\noindent {\bf C5: Untargeted attack ($\bm{s\not\in G}$).}
Given a voice $x$ uttered by a source speaker $s$ such that $s$ is not an enrolled speaker (i.e., $s\not\in G$), the adversary attempts to craft an adversarial voice
such that it is accepted as an arbitrary enrolled speaker $t\in G$.
The loss function to achieve this goal is defined as: $\mathcal{L}_{3}^-(x,t)\triangleq-\mathcal{L}_{3}(x,t) = \theta-{\tt max}_{i\in G}[S(x)]_i$.

\subsubsection{Loss functions for CSI task}
\smallskip\noindent{\bf C6 and C7: Targeted attack.}
The CSI task recognizes any input voice as one of the enrolled speakers, i.e.,
the decision space is $G$.
Therefore, when launching targeted attack against SRSs performing the CSI task,
an adversary can choose any enrolled speaker as the target speaker $t\in G$.
The loss function can be derived from the ones defined for targeted attack with $t\in G$ on the OSI task (i.e., C1 and C2)
by ignoring the threshold $\theta$, i.e, the loss functions
$\mathcal{L}_\text{CE}$, $\mathcal{L}_\text{M}$, and $\mathcal{L}_1$, no matter how the source speaker $s$
is chosen, either one of the enrolled speakers (i.e., $s\in G$) or an unenrolled one (i.e., $s\not\in G$).


\smallskip\noindent{\bf \chengk{C8}: Untargeted attack.}
The loss functions can be derived from the ones defined for untargeted attack
on the OSI task with $s\in G$ (i.e., C4) by ignoring the threshold $\theta$, namely, the loss functions
$\mathcal{L}_{\text{CE}}^{s}$, $\mathcal{L}_{\text{M}}^{s}$, $\mathcal{L}_{1}^{s}$, and $\mathcal{L}_{4}^{s}$.




\subsubsection{Loss functions for SV task}
The SV task involves only one enrolled speaker
and determines if an input voice is uttered by the enrolled speaker or not.
Hence, an adversary may potentially aim to achieve two opposite goals: 
1) a voice uttered by the enrolled speaker is rejected as an imposter;
2) a voice uttered by an unenrolled speaker is recognized as the enrolled speaker.

\smallskip\noindent{\bf \chengk{C9}: Enrolled speaker $\to$ imposter.}
Two loss functions which can be used to achieve this goal are formulated as:
\begin{align*}
    \mathcal{L}_{\text{BCE}}(x,t) &\triangleq -\log (1-\varphi(S(x))) \qquad    \mathcal{L}_{3B}(x,t) \triangleq  S(x) - \theta
\end{align*}
where $\varphi$ denotes the sigmoid function and $\mathcal{L}_{\text{BCE}}$ is the binary Cross Entropy Loss function.
Note that $\mathcal{L}_{3B}$ is adapted from $\mathcal{L}_{3}$ by assuming the speaker group $G$ is singleton.
$\mathcal{L}_{\text{BCE}}$ (resp. $\mathcal{L}_{3B}$) is the special case of $\mathcal{L}_{\text{CE}}$ (resp. $\mathcal{L}_{3}$) for the binary classification task SV.

\smallskip\noindent{\bf \chengk{C10}: Unenrolled speaker $\to$ enrolled speaker.}
Two loss functions for this goal can be derived from the above loss functions with minor modifications:
\begin{align*}
    \mathcal{L}'_{\text{BCE}}(x,t) &\triangleq -\log (\varphi(S(x)))\qquad   \mathcal{L}^{-}_{3B}(x,t) &\triangleq \theta - S(x)
\end{align*}
Remark that $\mathcal{L}^{-}_{3B}(x,t)$ is the special case of $\mathcal{L}^{-}_{3}(x,t)$ for the binary classification task SV.


\subsection{Optimization Approachs}\label{sec:optimization}
After formulating the generation of an adversarial example as an optimization problem with various source-/target-oriented loss functions,
it remains to solve the optimization problem. In general,
one can freely leverage any existing approaches to solve the optimization problem and craft adversarial examples.
In this work, to increase the attack capability of \attackname and make it applicable to different levels of knowledge about the victim model,
we consider the following solving approaches:
FGSM~\cite{goodfellow2014explaining}, PGD~\cite{madry2017towards}, and CW~\cite{carlini2017towards} that were developed in the image domain,
and FAKEBOB~\cite{chen2019real} that was tailored for SRSs.
FGSM, PGD, and FAKEBOB use the standard gradient descent to solve the optimization problem,
while CW utilizes Adam optimizer~\cite{kingma2014adam}.
FGSM, PGD, and CW are representative white-box approaches, while FAKEBOB is the state-of-the-art black-box one.
Below we integrate them into  \attackname with necessary modifications.
Since the ways that CW copes with the box-constraint and stealthiness differ from
that of FGSM, PGD, and FAKEBOB, we first discuss how to adapt CW and then the others.

\smallskip
\noindent{\bf Adapting CW}.
CW deals with the box constraint $x'\in [-1,1]$ by introducing a new free variable $z$ 
and optimizing over this new variable $z$ instead of  $x'$.
Inspired by this trick, we define the variable $z=\arctanh(x')$.
Since $-1\leq x'\leq 1$, we have $-\infty\leq z\leq \infty$, thus
the box-constraint is removed when we optimize over $z$.
Note that our new free variable $z$ is different from the one used in CW,
due to the range difference between voices and images, i.e., $[-1,1]$ vs. $[0,1]$.

To achieve the stealthiness, CW minimizes the adversarial perturbation $\delta=x'-x$
by finding an optimal trade-off hyper-parameter $\lambda$ using a binary search.
Intuitively, a large $\lambda$ will instruct the optimization to focus more on reducing the distance $d(x',x)$,
leading to better stealthiness, but meanwhile pay less attention to the loss $\mathcal{L}$,
undermining the effectiveness of the attack.
We follow this practice and adopt $L_2$ norm~\cite{carlini2017towards} as the distance metric, i.e., $d(x',x)\triangleq\sqrt{\sum_{i}(x'_i-x_i)^2}$.

Since the objective function to be minimized is the combination of two terms,
to focus only on reducing the perturbation after the attack succeeds
(i.e., the loss $\mathcal{L}$ is small enough),
CW adds a clamping operation to the loss function, i.e., ${\tt max}(\mathcal{L} + \kappa, 0)$
where $\kappa$ controls the strength of the adversarial examples.
Due to the clamping operation, only part of our defined loss functions are suitable for CW approach,
including $\mathcal{L}_2$, $\mathcal{L}_3$, $\mathcal{L}_2^s$, $\mathcal{L}_3^-$, $\mathcal{L}_\text{M}$ (on CSI),
$\mathcal{L}_\text{M}^s$ (on CSI), $\mathcal{L}_{3B}$, and $\mathcal{L}_{3B}^-$.
Other loss functions are excluded since they do not satisfy $\mathcal{L}(x,t)\leq 0 \Rightarrow D(x')=t$ (targeted attack)
or $\mathcal{L}(x,t)\leq 0 \Rightarrow D(x')\neq s$ (untargeted attack).

\smallskip
\noindent{\bf Adapting FGSM, PGD, and FAKEBOB}.
Unlike the CW approach (i.e., $\lambda\neq 0$),
FGSM, PGD, and FAKEBOB set $\lambda=0$ and cope with the stealthiness and box-constraint
by clipping the intermediate example at each iteration of gradient descent.
Following this practice, we define the clipping function as
\[\emph{clip}_{x,\varepsilon}(x')\triangleq{\tt min}(x+\varepsilon,1,{\tt max}(x',x-\varepsilon,-1))\]
where $\varepsilon$ is the upper bound of the perturbation magnitude measured in $L_\infty$ norm.
Note that our clipping function is different from the one exploited by FGSM and PGD,
due to the range difference between images and voices.

FGSM is an one-iteration approach with the perturbation budget $\varepsilon$ as the step\_size,
while PGD is an iterative version of FGSM with a small step size $\alpha$.
FAKEBOB is similar to PGD except that it is a black-box approach and estimates gradients via Natural Evolution Strategy~\cite{IlyasEAL18}.
Furthermore, 
FAKEBOB 
proposes the first algorithm to estimate the score threshold $\theta$ for the SV and OSI tasks.

Although FGSM, PGD, and CW were initially proposed as attack algorithms
using loss functions $\mathcal{L}_\text{CE}$
and  $\mathcal{L}_\text{M}$, we highlight that
\attackname only utilizes them as optimization approaches
since neither $\mathcal{L}_\text{CE}$ nor $\mathcal{L}_\text{M}$
fits to the purpose of \attackname: adversarial attack with arbitrary source/target speakers.
We desgin suitable loss functions and integrate them into FGSM, PGD, and CW.
This is the same for FAKEBOB,
since our attack \attackname covers more cases of source and target speakers,
i.e., C1, C3-C5, C7-C8, and C9.
Even for the cases covered by FAKEBOB,
\attackname provides more choices of loss functions.
The common loss functions between our work and FAKEBOB are only $\mathcal{L}_2$, $\mathcal{L}_3^-$,
$\mathcal{L}_\text{M}$ (for C6), $\mathcal{L}_\text{M}^s$ (for C8), and $\mathcal{L}_{3B}^-$.
Furthermore, we empirically show that one of the loss functions adopted by FAKEBOB
achieves inferior performance than ours (cf. Section~\ref{sec:evaluation}).

\subsection{Robust Over-the-Air  \attackname}\label{sec:practical-attack}

In practice, adversarial voices can be fed to SR models via different ways,
including API and air channel.
API attack
directly feeds adversarial voices in the form of audio file to models
via exposed API interfaces,
thus it will not introduce any disruption to adversarial voices.
In contrast, over-the-air attack, where adversarial voices are played/recorded by loudspeakers/microphones
and transmitted over the air,
is a lossy channel.
The distortion occurred in the transmission could largely destruct the effectiveness of adversarial examples.
Thus, an over-the-air attack is generally more practical and realistic yet more challenging
than the API attack.

In this subsection, we first investigate the sources of distortions that may destruct the effectiveness of adversarial voices played over-the-air
and then present our solution to enhance our attack \attackname towards robust over-the-air attack.


\subsubsection{Sources of Distortions}
\begin{figure}[t]
\centering
    \includegraphics[width=0.38\textwidth]{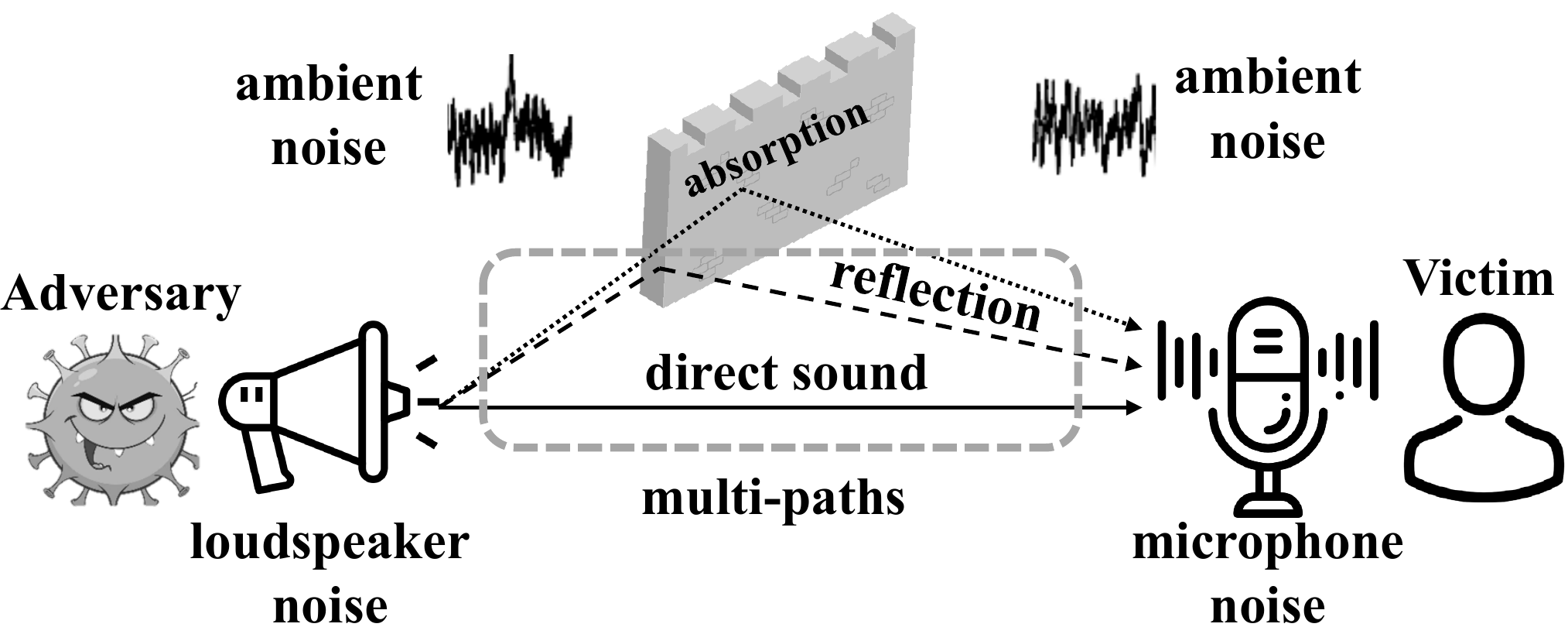}\vspace{-1mm}
    \caption{The acoustic model of over-the-air attack}
    \label{fig:over-the-air-model}\vspace{-3mm}
\end{figure}

\figurename~\ref{fig:over-the-air-model} depicts the acoustic model for over-the-air attack.
When an adversarial voice is played by a loudspeaker, it is transmitted over the air, and finally recorded by the victim's microphone.
There are three sources of distortions:
equipment distortion, ambient noise, and reverberation.

\smallskip\noindent {\bf Equipment distortion.}
Equipment distortion is introduced by loudspeakers and microphones due to their frequency-selectivity feature.
Specifically, their frequency response is non-uniform across the frequency band
with amplification in some frequency ranges and attenuation in others,
which may distort the adversarial voices and undermine the attack.
In addition, different loudspeakers and microphones may exhibit different frequency responses,
thus incur different equipment distortions.
We can model the equipment distortion on adversarial voices in the time domain
via loudspeakers and microphones' impulse response $h$.
The adversarial voice disrupted by equipment distortion is formulated as $x^{adv}\otimes h$ where $\otimes$ denotes convolution operation.

\smallskip\noindent {\bf Ambient noise.}
In practice, ambient noise includes ambient human voice, background music, traffic noise, and so on, depending on the specific attack scenario (e.g., living room, office, airport, and mall), but is inevitable in playback and recording environment.
%
These noises, denoted by $n$, influence the effectiveness of adversarial voices by additively changing their magnitude,
i.e., from $x^{adv}$ to $x^{adv}+n$.
However, the influence varies with the relative strength of ambient noise to adversarial voices,
i.e., Signal-to-Noise Ratio (SNR) between adversarial voices and ambient noise.
Ambient noise with low volume has limited impact since weak noises are easily overwhelmed by stronger voices~\cite{Metamorph}. 

\smallskip\noindent {\bf Reverberation.}
When played by loudspeakers in indoor environment,
a voice signal may transmit through multiple paths (i.e., direct path and other reflected paths)
with various delays and absorption by many surfaces.
When the direct sound and reflections blend and overlap with each other, reverberation is created.
Reverberation will cause the received voice by microphones to be largely different from the original voice sent out by loudspeakers.
Room Impulse Response (RIR), represented by $r$, can well characterize the acoustic properties
of a room regarding sound transmission and reflection.
The adversarial voice with reverberation is created by convolving $r$ with $x^{adv}$, i.e., $x^{adv}\otimes r$.
RIR varies with the room configuration (e.g., room dimension, reverberation time, and absorption coefficient of reflective materials)
as well as the location of loudspeakers and microphones.
To obtain RIR for a specific room, two different methods are mostly used in practice: simulation approach and real-world measurement.
Simulation approach leverages the well-known Image Source Method~\cite{image-method}
which accepts room configuration and hardware's position as input and returns the simulated RIR.
For real-world measurement, we can transmit a brief input signal (called impulse) by a loudspeaker in a room,
then the response signal recorded by a microphone is the RIR of this room under the current position of loudspeaker and microphone.
However, as the impulse signal also goes through the hardware during transmission,
the RIR obtained by real-world measurement is indeed the composition of impulse responses of the room and hardware~\cite{Metamorph,hang-practical}, i.e., $r\circ h$.

\subsubsection{Robust \attackname}\label{sec:robust-AS2T}




\begin{algorithm}[t]\footnotesize
	\caption{Robust \attackname}
	\label{al:practical-attack}
	\begin{algorithmic}[1]
    \Require Benign voice $x^0$; loss function $\mathcal{L}$; distance metric $d(\cdot)$; hyper-parameter $\lambda$;
    optimization approach $\mathcal{O}$;
    number of iterations \#Iter;
    set of parameterized functions $\mathcal{F}$; 
    sampling size $K$
	\Ensure Adversarial voice $x^{adv}$
	\For{$i=1 \textbf{ to } \text{\#Iter}$}
    \State $L_i=0$
    \For{$j=1 \textbf{ to } K$}
    \State sampling a parameterized function $\mathcal{F}_p$ from $\mathcal{F}$
    \State $L_i^j \gets \mathcal{L}(\mathcal{F}_p(x^{i-1}))$
    \State $L_i\gets L_i+L_i^j$
    \EndFor
    \State $L_i\gets L_i/K + \lambda \times d(x^{i-1}, x^0)$
    \State $g_i \gets \nabla_{x^{i-1}}L_i$
    \State $x^i\gets \mathcal{O}(x^{i-1}, g_i)$
    \EndFor
    \State $x^{adv}\gets x^{\text{\#Iter}}$
	\State \Return{$x^{adv}$}
	\end{algorithmic}
\end{algorithm}

To enhance the robustness of adversarial voices and enable physical over-the-air attack,
we  incorporate the aforementioned distortions into the generation process of adversarial voices.
Our solution is described in Algorithm~\ref{al:practical-attack}, based on a set $\mathcal{F}$ of parameterized functions  modeling the distortions induced by the over-the-air transmission.
In the $i$-th iteration, we randomly sample $K$ functions from $\mathcal{F}$ (Lines 3-4). Each sampled function $\mathcal{F}_p$
is utilized to transform the intermediate voice $x^{i-1}$ (Line 5),
resulting in $K$ transformed voices.
Then we compute the average loss of those $K$ voices based on which the gradient $g_i$ is computed (Lines 7-8).
Finally, the new voice $x^{i}$ is created from $x^{i-1}$ and $g_i$ by invoking the optimization approach $\mathcal{O}$ (Line 9).
In this work, we utilize the following parameterized functions $\mathcal{F}$.

\smallskip\noindent {\bf Addition with random noise.}
We use random noise $n$ (e.g., uniform noise and white Gaussian noise)
to model ambient noise in the physical attack. Formally, we define the functions
$\mathcal{F}=\{\mathcal{F}_{\text{SNR}, n}|\mathcal{F}_{\text{SNR}, n}(x)=x+\Phi(\text{SNR}, x, n),
n\sim \mathcal{Z}, \text{SNR}_l\leq \text{SNR}\leq \text{SNR}_u\}$,
where $\mathcal{Z}$ denotes the distribution of the random noise,
SNR is the Signal-to-Noise Ratio between the adversarial voice $x$
and the random noise $n$ (i.e., $\text{SNR}=10\times\log_{10}\frac{P_{x}}{P_n}$, $P_x$ and $P_n$ are powers of $x$ and $n$ respectively),
SNR$_l$ and SNR$_u$ are the lower and upper bounds of SNR,
and $\Phi(\text{SNR}, x, n)$ scales the magnitude of the random noise $n$ such that the SNR requirement is satisfied.
We denote by \attacknamenospace+RN our attack \attackname with random noise.

\smallskip\noindent {\bf Convolution with room impulse response.}
To improve the robustness of adversarial voices against reverberation in the physical attack,
we define the functions
$\mathcal{F}=\{\mathcal{F}_r|\mathcal{F}_r(x)=r\otimes x, r\sim \mathcal{R}\}$,
where $\mathcal{R}$ denotes the distribution of the RIR.
We denote by \attacknamenospace+RIR our attack \attackname with RIR. 
Furthermore, we denote by \attacknamenospace+RN+RIR our attack with both random noise and RIR, in which
$\mathcal{F}=\{\mathcal{F}_{r,\text{SNR},n}|\mathcal{F}_{r,\text{SNR},n}(x)=r\otimes x+\Phi(\text{SNR},x,n)\}$.

Remark that we do not explicitly model the equipment distortion in this work.
Our experimental results show that the effect of such distortion is not substantial
and can be coped with by \attacknamenospace+RN, \attacknamenospace+RIR, and \attacknamenospace+RN+RIR.
We highlight that in Algorithm~\ref{al:practical-attack}
we randomly sample $K$ parameterized functions from $\mathcal{F}$ instead of using a single function across all the iterations.
In this way, it is expected that the obtained adversarial voices can work in various scenarios
and attack numerous victim devices simultaneously.

\begin{table}[t]
    \centering  \setlength\tabcolsep{6pt}
    \caption{Details of the 14 SR models, where
    Arch and Trans represent architecture and Transformer, respectively}\vspace{-2mm}
    \resizebox{0.42\textwidth}{!}{%
      \begin{tabular}{|c|c|c|c|c|c|}
      \toprule
       {\bf Arch} & {\bf Name} & {\bf \#Params} & \makecell{{\bf Input} {\bf type}} & \makecell{{\bf Training}\\ {\bf dataset}}  & \makecell{{\bf Scoring} \\ {\bf Backend}} \\
      \midrule
      {\bf GMM} & \textbf{Ivector}~\cite{kaldi-ivector-plda} & 80.37M & MFCC &  VoxCeleb1\&2& PLDA \\
      \midrule
      \multirow{3}[7]{*}{\bf TDNN} & \textbf{ECAPA}~\cite{ECAPA-TDNN} & 20.77M & fBank & VoxCeleb1 & COSS  \\
      \cmidrule{2-6}
       & \textbf{Xvector-P}~\cite{kaldi-xvector-plda} & 5.79M & MFCC & VoxCeleb1\&2  & PLDA  \\
       \cmidrule{2-6}
      & \textbf{Xvector-C}~\cite{speech-brain} & 4.21M & fBank & VoxCeleb1  & COSS  \\
      \midrule
      \multirow{8}[12]{*}{\bf CNN} & \textbf{AudioNet}~\cite{becker2018interpreting} & 0.21M & fBank & LibriSpeech  & COSS  \\
      \cmidrule{2-6}
       & \textbf{SincNet}~\cite{SincNet} & 21.84M & wavform & TIMIT & COSS \\
       \cmidrule{2-6}
       & \textbf{Res18-I}~\cite{resnet18} & 11.17M & spectrogram & VoxCeleb1 & COSS \\
       \cmidrule{2-6}
      & \textbf{Res18-V}~\cite{resnet18} & 11.17M & spectrogram & VoxCeleb1 & COSS \\
      \cmidrule{2-6}
      & \textbf{Res34-I}~\cite{resnet34} & 21.28M & spectrogram & VoxCeleb1 & COSS \\
      \cmidrule{2-6}
      & \textbf{Res34-V}~\cite{resnet34} & 21.28M & spectrogram & VoxCeleb1 & COSS \\
      \cmidrule{2-6}
      & \textbf{Auto-I}~\cite{auto-speech} & 15.11M & spectrogram  & VoxCeleb1 & COSS \\
      \cmidrule{2-6}
      & \textbf{Auto-V}~\cite{auto-speech} & 15.11M & spectrogram & VoxCeleb1 & COSS \\
      \midrule
      {\bf LSTM} & \textbf{GE2E}~\cite{GE2E} & 12.13M & fBank & TIMIT  & COSS \\
      \midrule
      {\bf Trans} & \textbf{Hubert}~\cite{HuBert} & 316.61M & wavform & LibriLight & COSS  \\
      \bottomrule
      \end{tabular}%
    }
    \label{tab:system-info}\vspace{-2mm}
  \end{table}%

\section{Evaluation}
In this section, we first present the common evaluation setup
and metrics of our evaluation,
then evaluate effectiveness of \attackname with arbitrary source/target speakers over API and over-the-air,
and finally conduct a thorough transferability study of \attackname among 14 SR models (cf.  \tablename~\ref{tab:system-info}).

\subsection{Common Evaluation Setup}

\noindent{\bf Datasets}. Our evaluation is based on three datasets
derived from Librispeech~\cite{panayotov2015librispeech} and released in \cite{SEC4SR}.
namely, Spk$_{10}$-enroll, Spk$_{10}$-test, and Spk$_{10}$-imposter.
Spk$_{10}$-enroll and Spk$_{10}$-test consist of 10 and 100 distinct voices per speaker from the same ten speakers (five male and five female),
while Spk$_{10}$-imposter consists of 100 voices per speaker from another ten speakers (five male and five female).
\chengk{We also build another three datasets from LibriSpeech with more speakers, 
where Spk$_{100}$-enroll and Spk$_{100}$-test respectively consist of 10 and 100 distinct voices per speaker from the same 100 speakers (52 female and 48 male)
and Spk$_{100}$-imposter consists of 100 voices per speaker from another 100 speakers (49 female and 51 female).}

\smallskip
\noindent{\bf Models}.
In this work, we select 14 SR models for our experiments.
These models cover five architectures (GMM, TDNN, CNN, LSTM, and Transformer),
four input types (waveform, spectrogram, fBank, and MFCC),
and two scoring methods (PLDA and COSS).
More details of these models are shown in \tablename~\ref{tab:system-info}, such as their training dataset, the number of parameters, and etc.

We measure the performance of these models on three tasks using the above datasets
and the results are shown in \tablename~\ref{tab:performance}, where the best cases are highlighted in {blue} color.
Column (Acc) shows accuracy. Column (EER) shows the equal error rate, i.e., when False Acceptance Rate (FAR) equals False Rejection Rate (FRR),
where FAR is the proportion of voices that are uttered by unenrolled speakers but accepted by the model, and
FRR is the proportion of voices that are uttered by
enrolled speakers but rejected (i.e., classified as ${\tt imposter}$).
Column (IER) shows Identification Error Rate, i.e., the rate of voices uttered by enrolled speakers
which are not rejected but incorrectly classified~\cite{chen2019real}.
We tune the threshold $\theta$ for the SV and OSI tasks based on EER.

\smallskip
\noindent{\bf Attacks}.
We implement four optimization approaches for \attackname, namely,
FGSM, PGD, CW$_2$, and FAKEBOB (cf. Section~\ref{sec:optimization}).
We limit the
perturbation budget
$\epsilon$ to 0.002 for $L_\infty$ attacks, the same as \cite{chen2019real,jati2021adversarial}, unless explicitly stated.
Note that the CW$_2$ attack minimizes adversarial perturbations in the loss function, and hence does not have any limitations.


We conduct  experiments on a machine with Ubuntu 18.04, an Intel Xeon E5-2697 v2 2.70GHz CPU, 376GiB memory, and a GeForce RTX 2080Ti GPU.

\begin{table}[t]
    \centering\setlength\tabcolsep{7pt}
    \caption{Performance of the SR models on three tasks}\vspace{-2mm}
    \resizebox{0.4\textwidth}{!}{%
      \begin{tabular}{|c|c|c|c|c|c|c|}
      \toprule
      \multirow{2}{*}{\textbf{model}} & \multicolumn{1}{c|}{\textbf{CSI}} & \multicolumn{2}{c|}{\textbf{SV}} & \multicolumn{3}{c|}{\textbf{OSI}} \\
  \cmidrule{2-7}    \multicolumn{1}{|c|}{} & \multicolumn{1}{c|}{\textbf{Acc (\%)}} & \multicolumn{1}{c|}{\textbf{EER (\%)}} & \multicolumn{1}{c|}{$\mathbf{\theta}$} & \multicolumn{1}{c|}{\textbf{EER (\%)}} & \multicolumn{1}{c|}{\textbf{IER (\%)}} & \multicolumn{1}{c|}{$\mathbf{\theta}$}\\
      \midrule
      \textbf{Ivector} & \textcolor{blue}{\bf 100} & 1.05  & 9.74  & 4.8  & 0 & 12.77 \\
      \midrule
      \midrule
      \textbf{ECAPA} & 99.9  & 0.97  & 0.42  & \textcolor{blue}{\bf 2.2} & 0 & 0.49  \\
      \midrule
      \textbf{Xvector-P} & \textcolor{blue}{\bf 100} & \textcolor{blue}{\bf 0.8} & 13.78 & 6 & 0 & 18.72 \\
      \midrule
      \textbf{Xvector-C} & 96.4  & 4.22  & 0.63  & 9.85 & 0 & 0.7   \\
      \midrule
      \midrule
      \textbf{AudioNet} & 99.9  & 4.2   & 0.82  & 14.01 & 0 & 0.87  \\
      \midrule
      \textbf{SincNet} & 96.3  & 5     & 0.59  & 16.9  & 0 & 0.74  \\
      \midrule
      \textbf{Res18-I} & \textcolor{blue}{\bf 100} & 0.84  & 0.51  & 5.5  & 0 & 0.61  \\
      \midrule
      \textbf{Res18-V} & 99.9  & 1.3   & 0.49  & 6.61  & 0 & 0.59  \\
      \midrule
      \textbf{Res34-I} & \textcolor{blue}{\bf 100} & 0.9   & 0.52  & 6.7   & 0 & 0.61  \\
      \midrule
      \textbf{Res34-V} & 99.8  & 1.2   & 0.51  & 4.3   & 0 & 0.58  \\
      \midrule
      \textbf{Auto-I} & 58.8  & 19.95 & 0.23  & 17.01  & 0 & 0.58 \\
      \midrule
      \textbf{Auto-V} & 99.3  & 1.6   & 0.29  & 3.6   & 0 & 0.39  \\
      \midrule
      \midrule
      \textbf{GE2E} & 78.7  & 10.65 & 0.67  & 23.2  & 0 & 0.87  \\
      \midrule
      \midrule
      \textbf{Hubert} & 95.2  & 9.13  & 0.57  & 17.23 & 0 & 0.62  \\
      \bottomrule
      \end{tabular}%
    }
    \label{tab:performance}\vspace{-3mm}
  \end{table}%

\begin{table*}
    \centering\setlength\tabcolsep{4pt}
    \caption{The \chengk{attack success rate (\%)} of \attackname on the OSI task with FGSM and PGD as the optimization approaches}
    \vspace{-2mm}
    \resizebox{0.93\textwidth}{!}{%
      \begin{tabular}{|c|c|c|cccc|cccc|cccc|cccc|ccc|ccc|cccccc|}
      \toprule
      \multicolumn{3}{|c|}{\multirow{4}[8]{*}{}} & \multicolumn{28}{c|}{\textbf{OSI}} \\
  \cmidrule{4-31}    \multicolumn{3}{|c|}{} & \multicolumn{8}{c|}{\boldmath{}\textbf{Targeted ($t\in G$) Random}\unboldmath{}} & \multicolumn{8}{c|}{\boldmath{}\textbf{Targeted ($t\in G$) Least Likely}\unboldmath{}} & \multicolumn{6}{c|}{\boldmath{}\textbf{Targeted ($t={\tt imposter}$)}\unboldmath{}} & \multicolumn{6}{c|}{\textbf{Untargeted}} \\
  \cmidrule{4-31}    \multicolumn{3}{|c|}{} & \multicolumn{4}{c|}{\boldmath{}\textbf{ASR$_t$ (\%)}\unboldmath{}} & \multicolumn{4}{c|}{\boldmath{}\textbf{ASR$_u$ (\%)}\unboldmath{}} & \multicolumn{4}{c|}{\boldmath{}\textbf{ASR$_t$ (\%)}\unboldmath{}} & \multicolumn{4}{c|}{\boldmath{}\textbf{ASR$_u$ (\%)}\unboldmath{}} & \multicolumn{3}{c|}{\boldmath{}\textbf{ASR$_t$ (\%)}\unboldmath{}} & \multicolumn{3}{c|}{\boldmath{}\textbf{ASR$_u$ (\%)}\unboldmath{}} & \multicolumn{6}{c|}{\boldmath{}\textbf{ASR$_u$ (\%)}\unboldmath{}} \\
  \cmidrule{4-31}    \multicolumn{3}{|c|}{} & \boldmath{}\textbf{$\mathcal{L}_\text{CE}$}\unboldmath{} & \boldmath{}\textbf{$\mathcal{L}_\text{M}$}\unboldmath{} & \boldmath{}\textbf{$\mathcal{L}_{1}$}\unboldmath{} & \boldmath{}\textbf{$\mathcal{L}_{2}$}\unboldmath{} & \boldmath{}\textbf{$\mathcal{L}_\text{CE}$}\unboldmath{} & \boldmath{}\textbf{$\mathcal{L}_\text{M}$}\unboldmath{} & \boldmath{}\textbf{$\mathcal{L}_{1}$}\unboldmath{} & \boldmath{}\textbf{$\mathcal{L}_{2}$}\unboldmath{} & \boldmath{}\textbf{$\mathcal{L}_\text{CE}$}\unboldmath{} & \boldmath{}\textbf{$\mathcal{L}_\text{M}$}\unboldmath{} & \boldmath{}\textbf{$\mathcal{L}_{1}$}\unboldmath{} & \boldmath{}\textbf{$\mathcal{L}_{2}$}\unboldmath{} & \boldmath{}\textbf{$\mathcal{L}_\text{CE}$}\unboldmath{} & \boldmath{}\textbf{$\mathcal{L}_\text{M}$}\unboldmath{} & \boldmath{}\textbf{$\mathcal{L}_{1}$}\unboldmath{} & \boldmath{}\textbf{$\mathcal{L}_{2}$}\unboldmath{} & \boldmath{}\textbf{$\mathcal{L}_\text{CE}^{s}$}\unboldmath{} & \boldmath{}\textbf{$\mathcal{L}_3$}\unboldmath{} & \boldmath{}\textbf{$\mathcal{L}_{1}^{s}$}\unboldmath{} & \boldmath{}\textbf{$\mathcal{L}_\text{CE}^{s}$}\unboldmath{} & \boldmath{}\textbf{$\mathcal{L}_3$}\unboldmath{} & \boldmath{}\textbf{$\mathcal{L}_{1}^{s}$}\unboldmath{} & \boldmath{}\textbf{$\mathcal{L}_\text{CE}^{s}$}\unboldmath{} & \boldmath{}\textbf{$\mathcal{L}_\text{M}^{s}$}\unboldmath{} & \boldmath{}\textbf{$\mathcal{L}_{4}^{s}$}\unboldmath{} & \boldmath{}\textbf{$\mathcal{L}_{2}^{s}$}\unboldmath{} & \boldmath{}\textbf{$\mathcal{L}_{1}^{s}$}\unboldmath{} & \boldmath{}\textbf{$\mathcal{L}^{-}_3$}\unboldmath{} \\
      \midrule
      \multirow{5}[10]{*}{\boldmath{}\textbf{$s\in G$}\unboldmath{}} & \multicolumn{2}{c|}{\textbf{FGSM}} & \textcolor{blue}{\textbf{9.7}} & 9.2   & 9.4   & 9.4   & 30.6  & \textcolor{blue}{\textbf{62.8}} & 22.3  & \textcolor{blue}{\textbf{62.8}} & 0.1   & 0     & \textcolor{blue}{\textbf{0.3}} & 0     & 27.4  & \textcolor{blue}{\textbf{61.8}} & 14.8  & \textcolor{blue}{\textbf{61.8}} & 80.1  & \textcolor{blue}{\textbf{83.9}} & \textcolor{blue}{\textbf{83.9}} & 80.1  & \textcolor{blue}{\textbf{83.9}} & \textcolor{blue}{\textbf{83.9}} & 0     & 41.2  & 37.6  & \textcolor{blue}{\textbf{41.6}} & 0     & \multirow{5}[10]{*}{N/A} \\
  \cmidrule{2-30}          & \multirow{4}[8]{*}{\textbf{PGD}} & \textbf{2} & 65.7  & 49.6  & \textcolor{blue}{\textbf{67.9}} & 62.1  & 77    & \textcolor{blue}{\textbf{92.6}} & 71    & 91.3  & 46.6  & 21.5  & \textcolor{blue}{\textbf{54.2}} & 39.8  & 74.2  & \textcolor{blue}{\textbf{92.7}} & 64    & 91.5  & 96.8  & \textcolor{blue}{\textbf{98.2}} & \textcolor{blue}{\textbf{98.2}} & 97    & \textcolor{blue}{\textbf{98.2}} & \textcolor{blue}{\textbf{98.2}} & 0.2   & 82.4  & 81.4  & \textcolor{blue}{\textbf{84.6}} & 0     &  \\
  \cmidrule{3-30}          &       & \textbf{3} & 92.1  & 75.5  & \textcolor{blue}{\textbf{92.7}} & 91.8  & 95.7  & \textcolor{blue}{\textbf{99.4}} & 93.4  & 98.5  & 85.8  & 53.8  & \textcolor{blue}{\textbf{88.3}} & 85.1  & 93    & \textcolor{blue}{\textbf{99.5}} & 89.4  & 98.1  & 100   & 100   & 100   & 100   & 100   & 100   & 0     & 95.5  & 97.2  & \textcolor{blue}{\textbf{98.7}} & 0     &  \\
  \cmidrule{3-30}          &       & \textbf{4} & \textcolor{blue}{\textbf{99.1}} & 90.7  & 98.9  & 98.8  & 99.8  & \textcolor{blue}{\textbf{100}} & 99    & 99.9  & 97.1  & 82.1  & \textcolor{blue}{\textbf{97.6}} & 97    & 98.8  & \textcolor{blue}{\textbf{100}} & 97.9  & \textcolor{blue}{\textbf{100}} & 100   & 100   & 100   & 100   & 100   & 100   & 0     & 98.6  & 99.8  & \textcolor{blue}{\textbf{100}} & 0     &  \\
  \cmidrule{3-30}          &       & \textbf{5} & 99.8  & 96    & \textcolor{blue}{\textbf{100}} & 99.8  & 100   & 100   & 100   & 100   & 99.4  & 92    & \textcolor{blue}{\textbf{99.8}} & 99.7  & \textcolor{blue}{\textbf{100}} & \textcolor{blue}{\textbf{100}} & 99.8  & \textcolor{blue}{\textbf{100}} & 99.9  & \textcolor{blue}{\textbf{100}} & \textcolor{blue}{\textbf{100}} & 100   & 100   & 100   & 0.1   & 98.8  & \textcolor{blue}{\textbf{100}} & \textcolor{blue}{\textbf{100}} & 0     &  \\
      \midrule
      \multirow{5}[10]{*}{\boldmath{}\textbf{$s\not\in G$}\unboldmath{}} & \multicolumn{2}{c|}{\textbf{FGSM}} & 26    & 10.8  & \textcolor{blue}{\textbf{34.3}} & 33.9  & 26.1  & 10.8  & \textcolor{blue}{\textbf{35.4}} & 34.9  & 0.4   & 0.1   & \textcolor{blue}{\textbf{2.2}} & 2.1   & 0.4   & 0.1   & \textcolor{blue}{\textbf{2.9}} & 2.4   & \multicolumn{6}{c|}{\multirow{5}[10]{*}{N/A}} & \multicolumn{5}{c|}{\multirow{5}[10]{*}{N/A}} & 95 \\
  \cmidrule{2-19}\cmidrule{31-31}          & \multirow{4}[8]{*}{\textbf{PGD}} & \textbf{2} & 86.5  & 43.8  & \textcolor{blue}{\textbf{93.4}} & 92.9  & 86.6  & 43.8  & \textcolor{blue}{\textbf{93.8}} & 93.2  & 64.7  & 6.9   & \textcolor{blue}{\textbf{79.2}} & 78.4  & 64.7  & 6.9   & \textcolor{blue}{\textbf{80.8}} & 79.5  & \multicolumn{6}{c|}{}                         & \multicolumn{5}{c|}{}                 & 100 \\
  \cmidrule{3-19}\cmidrule{31-31}          &       & \textbf{3} & 99.1  & 80.9  & \textcolor{blue}{\textbf{99.7}} & 99.6  & 99.1  & 80.9  & \textcolor{blue}{\textbf{99.7}} & 99.6  & 98.2  & 52.3  & \textcolor{blue}{\textbf{99.3}} & 98.9  & 98.2  & 52.3  & \textcolor{blue}{\textbf{99.3}} & 98.9  & \multicolumn{6}{c|}{}                         & \multicolumn{5}{c|}{}                 & 100 \\
  \cmidrule{3-19}\cmidrule{31-31}          &       & \textbf{4} & 99.7  & 96.5  & \textcolor{blue}{\textbf{99.9}} & \textcolor{blue}{\textbf{99.9}} & 99.7  & 96.5  & \textcolor{blue}{\textbf{99.9}} & \textcolor{blue}{\textbf{99.9}} & 99.6  & 87.7  & \textcolor{blue}{\textbf{99.9}} & \textcolor{blue}{\textbf{99.9}} & 99.6  & 87.7  & \textcolor{blue}{\textbf{99.9}} & \textcolor{blue}{\textbf{99.9}} & \multicolumn{6}{c|}{}                         & \multicolumn{5}{c|}{}                 & 100 \\
  \cmidrule{3-19}\cmidrule{31-31}          &       & \textbf{5} & \textcolor{blue}{\textbf{99.9}} & 98.9  & \textcolor{blue}{\textbf{99.9}} & \textcolor{blue}{\textbf{99.9}} & \textcolor{blue}{\textbf{99.9}} & 98.9  & \textcolor{blue}{\textbf{99.9}} & \textcolor{blue}{\textbf{99.9}} & \textcolor{blue}{\textbf{100}} & 97.6  & 99.9  & 99.9  & \textcolor{blue}{\textbf{100}} & 97.6  & 99.9  & 99.9  & \multicolumn{6}{c|}{}                         & \multicolumn{5}{c|}{}                 & 100 \\
      \bottomrule
      \end{tabular}%
      }
      \label{tab:basic-OSI}\vspace{-2mm}
\end{table*}%

\subsection{Evaluation Metrics} 
We mainly use the following evaluation metrics.

\smallskip
\noindent{\bf Effectiveness}.
To evaluate the effectiveness of an attack,
we adopt untargeted (resp. targeted) attack success rate ASR$_u$ (resp. ASR$_t$), which refers to
the proportion of adversarial voices
that are misclassified (resp. classified as the target result).
Formally, ASR$_u$=100\%-Acc for the CSI task. For the SV/OSI task,
\chengk{ASR$_u$=FAR} when the benign voices are uttered by unenrolled speakers 
and \chengk{ASR$_u$=FRR} when the benign voices are uttered by enrolled speakers.

\smallskip
\noindent{\bf Stealthiness}.
To measure the stealthiness of adversarial voices, we use the standard $L_2$ norm, SNR (cf. Section~\ref{sec:robust-AS2T}), and
Perceptual Evaluation of Speech Quality (PESQ)~\cite{rix2001perceptual}.
PESQ is one of the objective perceptual measures, which simulates the human auditory system~\cite{xiang2017digital}.
The calculation of PESQ is involved.
Intuitively, PESQ first applies an auditory transform to obtain the loudness spectra of the original and adversarial voices, and then compares these two loudness spectra to obtain a metric score whose value is in the range of -0.5 to 4.5.
We refer readers to \cite{rix2001perceptual} for more details.
Smaller $L_2$, larger SNR, and higher PESQ indicate better stealthiness.

\begin{table*}\setlength\tabcolsep{7pt} 
    \centering
    \caption{The \chengk{attack success rate (\%)} of \attackname on the CSI and SV tasks with FGSM and PGD as the optimization approaches}
    \vspace{-2mm}
    \resizebox{.93\textwidth}{!}{%
      \begin{tabular}{|c|c|c|ccc|ccc|ccc|ccc|cccc|c|c|}
      \toprule
      \multicolumn{3}{|c|}{\multirow{4}[8]{*}{}} & \multicolumn{16}{c|}{\textbf{CSI}}                                                                                            & \multicolumn{2}{c|}{\textbf{SV}} \\
  \cmidrule{4-21}    \multicolumn{3}{|c|}{} & \multicolumn{6}{c|}{\textbf{Targeted Random}} & \multicolumn{6}{c|}{\textbf{Targeted Least Likely}} & \multicolumn{4}{c|}{\textbf{Untargeted}} & \multicolumn{2}{c|}{\textbf{Targeted}} \\
  \cmidrule{4-21}    \multicolumn{3}{|c|}{} & \multicolumn{3}{c|}{\boldmath{}\textbf{ASR$_t$ (\%)}\unboldmath{}} & \multicolumn{3}{c|}{\boldmath{}\textbf{ASR$_u$ (\%)}\unboldmath{}} & \multicolumn{3}{c|}{\boldmath{}\textbf{ASR$_t$ (\%)}\unboldmath{}} & \multicolumn{3}{c|}{\boldmath{}\textbf{ASR$_u$ (\%)}\unboldmath{}} & \multicolumn{4}{c|}{\boldmath{}\textbf{ASR$_u$ (\%)}\unboldmath{}} & \multicolumn{2}{c|}{\boldmath{}\textbf{ASR$_t$ (\%)}\unboldmath{}} \\
  \cmidrule{4-21}    \multicolumn{3}{|c|}{} & \boldmath{}\textbf{$\mathcal{L}_\text{CE}$}\unboldmath{} & \boldmath{}\textbf{$\mathcal{L}_\text{M}$}\unboldmath{} & \boldmath{}\textbf{$\mathcal{L}_1$}\unboldmath{} & \boldmath{}\textbf{$\mathcal{L}_\text{CE}$}\unboldmath{} & \boldmath{}\textbf{$\mathcal{L}_\text{M}$}\unboldmath{} & \boldmath{}\textbf{$\mathcal{L}_1$}\unboldmath{} & \boldmath{}\textbf{$\mathcal{L}_\text{CE}$}\unboldmath{} & \boldmath{}\textbf{$\mathcal{L}_\text{M}$}\unboldmath{} & \boldmath{}\textbf{$\mathcal{L}_1$}\unboldmath{} & \boldmath{}\textbf{$\mathcal{L}_\text{CE}$}\unboldmath{} & \boldmath{}\textbf{$\mathcal{L}_\text{M}$}\unboldmath{} & \boldmath{}\textbf{$\mathcal{L}_1$}\unboldmath{} & \boldmath{}\textbf{$\mathcal{L}_\text{CE}^{s}$}\unboldmath{} & \boldmath{}\textbf{$\mathcal{L}_\text{M}^{s}$}\unboldmath{} & \boldmath{}\textbf{$\mathcal{L}_4^{s}$}\unboldmath{} & \boldmath{}\textbf{$\mathcal{L}_1^{s}$}\unboldmath{} & \boldmath{}\textbf{$\mathcal{L}_\text{BCE}/\mathcal{L}'_\text{BCE}$}\unboldmath{} & \boldmath{}\textbf{$\mathcal{L}_{3B}/$$\mathcal{L}^{-}_{3B}$}\unboldmath{} \\
      \midrule
      \multirow{5}[10]{*}{\boldmath{}\textbf{$s\in G$}\unboldmath{}} & \multicolumn{2}{c|}{\textbf{FGSM}} & 13.6  & \textcolor{blue}{\textbf{23.3}} & 11.7  & 13.6  & \textcolor{blue}{\textbf{25.7}} & 11.8  & 1.3   & \textcolor{blue}{\textbf{4.9}} & 0.7   & 1.6   & \textcolor{blue}{\textbf{9.9}} & 1.1   & 28.2  & 28.2  & \textcolor{blue}{\textbf{39.1}} & 16.7  & 68.7  & 68.7 \\
  \cmidrule{2-21}          & \multirow{4}[8]{*}{\textbf{PGD}} & \textbf{2} & 73.6  & \textcolor{blue}{\textbf{81.9}} & 70    & 73.6  & \textcolor{blue}{\textbf{82.5}} & 70    & 64.4  & \textcolor{blue}{\textbf{75.2}} & 60.1  & 64.4  & \textcolor{blue}{\textbf{77.4}} & 60.1  & 82.5  & \textcolor{blue}{\textbf{90.8}} & 81.9  & 73.2  & 95.2  & 95.2 \\
  \cmidrule{3-21}          &       & \textbf{3} & 95    & \textcolor{blue}{\textbf{97.8}} & 93.2  & 95    & \textcolor{blue}{\textbf{97.8}} & 93.2  & 90.4  & \textcolor{blue}{\textbf{94.8}} & 88.9  & 90.4  & \textcolor{blue}{\textbf{95.1}} & 88.9  & 96.2  & \textcolor{blue}{\textbf{99.9}} & 97.3  & 93.7  & 100   & 100 \\
  \cmidrule{3-21}          &       & \textbf{4} & 99.7  & \textcolor{blue}{\textbf{99.9}} & 99    & 99.7  & \textcolor{blue}{\textbf{99.9}} & 99    & 98.5  & \textcolor{blue}{\textbf{99.8}} & 97.7  & 98.5  & \textcolor{blue}{\textbf{99.8}} & 97.7  & 99.6  & \textcolor{blue}{\textbf{100}} & 99.8  & 98.3  & 100   & 100 \\
  \cmidrule{3-21}          &       & \textbf{5} & 100   & 100   & 100   & 100   & 100   & 100   & \textcolor{blue}{\textbf{99.9}} & \textcolor{blue}{\textbf{99.9}} & 99.8  & \textcolor{blue}{\textbf{99.9}} & \textcolor{blue}{\textbf{99.9}} & 99.8  & 100   & 100   & 100   & 99.7  & 100   & 100 \\
      \midrule
      \multirow{5}[10]{*}{\boldmath{}\textbf{$s\notin G$}\unboldmath{}} & \multicolumn{2}{c|}{\textbf{FGSM}} & \textcolor{blue}{\textbf{81.6}} & 70.7  & 76    & 100   & 100   & 100   & \textcolor{blue}{\textbf{57}} & 36.6  & 47.7  & 100   & 100   & 100   & \multicolumn{4}{c|}{\multirow{5}[10]{*}{N/A}} & 56.9  & 56.9 \\
  \cmidrule{2-15}\cmidrule{20-21}          & \multirow{4}[8]{*}{\textbf{PGD}} & \textbf{2} & \textcolor{blue}{\textbf{99.3}} & 98.3  & 98.4  & 100   & 100   & 100   & \textcolor{blue}{\textbf{97.5}} & 92.3  & 94.9  & 100   & 100   & 100   & \multicolumn{4}{c|}{}         & 99.1  & 99.1 \\
  \cmidrule{3-15}\cmidrule{20-21}          &       & \textbf{3} & \textcolor{blue}{\textbf{100}} & 99.9  & \textcolor{blue}{\textbf{100}} & 100   & 100   & 100   & \textcolor{blue}{\textbf{100}} & 99.9  & \textcolor{blue}{\textbf{100}} & 100   & 100   & 100   & \multicolumn{4}{c|}{}         & 100   & 100 \\
  \cmidrule{3-15}\cmidrule{20-21}          &       & \textbf{4} & 100   & 100   & 100   & 100   & 100   & 100   & 100   & 100   & 100   & 100   & 100   & 100   & \multicolumn{4}{c|}{}         & 100   & 100 \\
  \cmidrule{3-15}\cmidrule{20-21}          &       & \textbf{5} & 100   & 100   & 100   & 100   & 100   & 100   & 100   & 100   & 100   & 100   & 100   & 100   & \multicolumn{4}{c|}{}         & 100   & 100 \\
      \bottomrule
      \end{tabular}%
      }
      \label{tab:basic-CSI-SV}\vspace{-2mm}
\end{table*}%

\subsection{Evaluation of \attackname over API}\label{sec:evaluation}

In this section, we evaluate the effectiveness of \attackname with arbitrary source/target speakers over API, i.e.,
directly feeding adversarial voices in the form of audio file to models.
%

\subsubsection{Evaluation Setup}
We target the ECAPA model in \tablename~\ref{tab:system-info}
since this model achieves the best overall performance over the three tasks.
We enroll ten speakers using the dataset Spk$_{10}$-enroll,
forming a speaker group $G$ for the CSI and OSI tasks
and ten speaker models for the SV task.
The  thresholds $\theta$ for the SV and OSI tasks are set to the ones shown in \tablename~\ref{tab:performance}.

We mount attacks using two categories of benign voices, namely,
voices uttered by enrolled speakers in the dataset Spk$_{10}$-test and
voices uttered by unenrolled speakers in the dataset Spk$_{10}$-imposter.
For target speakers,
since the decision space of the OSI task is $\mathcal{D}={G}\cup\{\tt imposter\}$,
we consider two categories of target speakers,
i.e., the enrolled speakers in $G$ and ${\tt imposter}$.
In contrast, since the decision space of the CSI task is $G$,
the target speakers are limited to the enrolled speakers.
We adopt two different ways to set a target speaker, i.e.,
randomly choosing among the enrolled speakers except the source speaker (Targeted Random)
and choosing the enrolled speaker whose score is the least one (Targeted Least Likely).
Note that we use the same target speakers for the CSI and OSI tasks.
Since the SV task makes binary decision, its target speaker is ${\tt imposter}$ (resp. the enrolled speaker)
when the source speaker is the enrolled speaker (resp. unenrolled speaker).

We implemented the loss functions defined in Section~\ref{sec:loss} and the approaches FGSM, PGD, CW$_2$, and FAKEBOB.
The number of iterations for PGD is 2, 3, 4, and 5 with step size $\alpha=\frac{\varepsilon}{5}=0.0004$.
The step size of FGSM is $\varepsilon=0.002$.
For CW$_2$, we set the initial trade-off constant $\lambda=0.1$,
use 9 binary search steps to minimize adversarial perturbations,
run 900-9000 iterations to converge, and set the parameter $\kappa=0$.
For FAKEBOB, the maximum number of iterations is set to 1000
with samples\_per\_draw of NES $m=50,~100$.

\subsubsection{Results}
The results of \attackname with FGSM and PGD as the optimization approaches on the OSI task and the CSI/SV tasks
are showed in \tablename~\ref{tab:basic-OSI} and \tablename~\ref{tab:basic-CSI-SV}, respectively.

\smallskip
\noindent{\bf Loss function comparison.}
In general, with the increase of the number of iterations,
the attack success rate of all the loss functions approaches 100\%.
However, some loss functions are more effective and efficient,
i.e., achieve higher attack success rate within the same number of iterations
and obtain 100\% attack success rate with fewer iterations.

On the OSI task, for targeted attack ($t\in G$), the loss function $\mathcal{L}_1$ often achieves better ASR$_t$
than $\mathcal{L}_\text{CE}$, $\mathcal{L}_\text{M}$ and $\mathcal{L}_2$
(Note that $\mathcal{L}_2$ was adopted by FAKEBOB in \cite{chen2019real}).
In contrast, the most effective loss functions for the untargeted attack
and targeted attack with $t={\tt imposter}$ are $\mathcal{L}_2^s$ and $\mathcal{L}_3$/$\mathcal{L}_1^{s}$, respectively.
Interestingly, $\mathcal{L}_\text{CE}^s$ and $\mathcal{L}_1^{s}$
are rather effective for launching targeted attack ($t={\tt imposter}$),
but achieves extremely limited attack success rate for untargeted attack.

On the CSI task, when the source speaker is an enrolled speaker ($s\in G$),
the loss function $\mathcal{L}_M$/$\mathcal{L}_M^s$ performs better than the others in terms of both ASR$_t$ and ASR$_u$
for targeted/untargeted attacks.
A similar result has been reported in the image domain~\cite{carlini2017towards}.
In contrast, when the source speaker is an unenrolled speaker ($s\not\in G$),
the loss function $\mathcal{L}_\text{CE}^{s}$ outperforms the others.
%
On the SV task, we find that the loss functions have the same performance.
This is because FGSM and PGD optimize them on the signs of the gradients instead of the actual gradients, and
the signs are the same.

\begin{tcolorbox}[size=title,breakable]
    \textbf{Remark 1.}
    The effectiveness of loss functions varies with setting (e.g., task, source/target speakers).
    On the OSI task, the best loss functions for untargeted, targeted ($t\in G$),
    and targeted ($t={\tt imposter}$) are $\mathcal{L}_2^s$, $\mathcal{L}_1$, and $\mathcal{L}_3$, respectively.
    On the CSI task, the best loss functions for untargeted/targeted attack
    are $\mathcal{L}_\text{M}/\mathcal{L}_\text{M}^s$ for $s\in G$  and $\mathcal{L}_\text{CE}^{s}$ for $s\notin G$.
\end{tcolorbox}

\begin{table*}
    \centering\setlength\tabcolsep{5pt}
    \caption{\chengk{The attack success rate (\%)} and stealthiness of \attackname on the CSI task with CW$_2$ and FAKEBOB as optimization approaches}\vspace{-2mm}
    \resizebox{.93\textwidth}{!}{%
      \begin{tabular}{|c|c|c|c|c|c|c|c|c|c|c|c|c|c|c|c|c|}
      \toprule
      \multicolumn{3}{|c|}{\multirow{5}[10]{*}{}} & \multicolumn{14}{c|}{\textbf{CSI}} \\
  \cmidrule{4-17}    \multicolumn{3}{|c|}{} & \multicolumn{5}{c|}{\textbf{Targeted Random}} & \multicolumn{5}{c|}{\textbf{Targeted Least Likely}} & \multicolumn{4}{c|}{\textbf{Untargeted}} \\
  \cmidrule{4-17}    \multicolumn{3}{|c|}{} & \multicolumn{2}{c|}{\textbf{\chengk{Success rate}}} & \multicolumn{3}{c|}{\textbf{Stealthiness}} & \multicolumn{2}{c|}{\textbf{\chengk{Success rate}}} & \multicolumn{3}{c|}{\textbf{Stealthiness}} & \textbf{\chengk{Success rate}} & \multicolumn{3}{c|}{\textbf{Stealthiness}} \\
  \cmidrule{4-17}    \multicolumn{3}{|c|}{} & \boldmath{}\textbf{ASR$_t$ (\%)}\unboldmath{} & \boldmath{}\textbf{ASR$_u$(\%)}\unboldmath{} & \multirow{2}[4]{*}{\textbf{L2}} & \multirow{2}[4]{*}{\begin{tabular}[c]{@{}c@{}}\textbf{SNR} \\ {\bf (dB)} \end{tabular}} & \multirow{2}[4]{*}{\textbf{PESQ}} & \boldmath{}\textbf{ASR$_t$ (\%)}\unboldmath{} & \boldmath{}\textbf{ASR$_u$(\%)}\unboldmath{} & \multirow{2}[4]{*}{\textbf{L2}} & \multirow{2}[4]{*}{\begin{tabular}[c]{@{}c@{}}\textbf{SNR} \\ {\bf (dB)} \end{tabular}} & \multirow{2}[4]{*}{\textbf{PESQ}} & \boldmath{}\textbf{ASR$_u$ (\%)}\unboldmath{} & \multirow{2}[4]{*}{\textbf{L2}} & \multirow{2}[4]{*}{\begin{tabular}[c]{@{}c@{}}\textbf{SNR} \\ {\bf (dB)} \end{tabular}} & \multirow{2}[4]{*}{\textbf{PESQ}} \\
  \cmidrule{4-5}\cmidrule{9-10}\cmidrule{14-14}    \multicolumn{3}{|c|}{} & \multicolumn{2}{c|}{\boldmath{}\textbf{$\mathcal{L}_\text{M}$}\unboldmath{}} &       &       &       & \multicolumn{2}{c|}{\boldmath{}\textbf{$\mathcal{L}_\text{M}$}\unboldmath{}} &       &       &       & \boldmath{}\textbf{$\mathcal{L}_\text{M}^{s}$}\unboldmath{} &       &       &  \\
      \midrule
      \multirow{3}[6]{*}{\boldmath{}\textbf{$s\in G$}\unboldmath{}} & \multicolumn{2}{c|}{\textbf{CW2}} & 100   & 100   & 0.081 & 47.01 & 4.00  & 100   & 100   & 0.102 & 44.54 & 3.85  & 100   & 0.057 & 50.47 & 4.17 \\
  \cmidrule{2-17}          & \multirow{2}[4]{*}{\textbf{FAKEBOB}} & \textbf{m=50} & 86.5  & 86.5  & 0.357 & 31.14 & 2.74  & 78.3  & 79.5  & 0.340  & 31.21 & 2.76  & 93.5  & 0.382 & 30.92 & 2.70  \\
  \cmidrule{3-17}          &       & \textbf{m=100} & 99.3  & 99.3  & 0.390  & 31.17 & 2.69  & 98.4  & 98.4  & 0.376 & 31.38 & 2.72  & 100   & 0.412 & 30.76 & 2.64 \\
      \bottomrule
      \end{tabular}%
  }
  \label{tab:baisc-CSI-CW2-FAKEBOB}\vspace{-2mm}
  \end{table*}%

\noindent{\bf Different source/target speakers.}
We observe that the targeted attack with $t={\tt imposter}$ (i.e., C3) often outperforms
the ones with $t\in G$ (i.e., C1 and C2) on the OSI task.
This is because the former only concentrates on
the relative magnitude of the scores between the enrolled speakers and the threshold $\theta$,
while the latter additionally has to consider the relative magnitude of the scores between the source and target speakers.
Another observation is that the attack with $s\not\in G$
tends to be easier than the one with $s\in G$,
e.g., C2 vs. C1, C5 vs. C4, and C7 vs. C6.
An in-depth analysis reveals that the score of an enrolled speaker is much higher
than the others if the benign voice is uttered by himself (i.e., $s\in G$),
while scores of all the enrolled speakers are similar if the benign voice is uttered by an unenrolled speaker (i.e., $s\not\in G$).
Therefore, it is much easier to increase the score of the target speaker if the source speaker
is an unenrolled one.

\begin{tcolorbox}[size=title,breakable]
    \textbf{Remark 2.}
    Targeted attack with $t={\tt imposter}$  is often easier than the targeted attack with $t\in G$
    and attacks using unenrolled speaker as the source speaker (i.e., $s\not\in G$) is often easier than the attacks with $s\in G$.
\end{tcolorbox}
\begin{table*}[htbp]
    \centering
    \caption{\chengk{The attack success rate (\%) of over-the-air \attackname. RN and RIR denote random noise and reverberation impulse response, respectively.}}
    \resizebox{.98\textwidth}{!}{%
      \begin{tabular}{|c|c|c|c|c|c|c|c|c|c|c|c|c|c|c|c|c|c|c|c|c|c|c|c|}
      \toprule
      \multirow{3}[6]{*}{\boldmath{}\textbf{$\varepsilon$}\unboldmath{}} & \multirow{3}[6]{*}{\diagbox{\bf \chengk{Training}}{\textbf{\chengk{Evaluation}}}} & \multicolumn{2}{c|}{\textbf{Reverberation}} & \multicolumn{20}{c|}{\textbf{Ambient Noise (SNR=?dB)}} \\
  \cmidrule{3-24}          &       & \multirow{2}[4]{*}{\textbf{Sim.}} & \multirow{2}[4]{*}{\textbf{Real}} & \multicolumn{5}{c|}{\textbf{White Noise}} & \multicolumn{5}{c|}{\textbf{Point Source Noise}} & \multicolumn{5}{c|}{\textbf{Music Noise}} & \multicolumn{5}{c|}{\textbf{Speech Babble Noise}} \\
  \cmidrule{5-24}          &       &       &       & \textbf{20} & \textbf{15} & \textbf{10} & \textbf{5} & \textbf{0} & \textbf{20} & \textbf{15} & \textbf{10} & \textbf{5} & \textbf{0} & \textbf{20} & \textbf{15} & \textbf{10} & \textbf{5} & \textbf{0} & \textbf{20} & \textbf{15} & \textbf{10} & \textbf{5} & \textbf{0} \\
      \midrule
      \multirow{4}[8]{*}{\textbf{0.008}} & \textbf{\attackname} & 85.3  & 80.0  & 69.7  & 31.0  & 7.5   & 1.4   & 1.3   & 95.7  & 87.8  & 75.7  & 60.7  & 47.5  & 98.8  & 94.5  & 85.4  & 71.7  & 59.6  & 99.8  & 97.3  & 90.9  & 83.5  & 73.4  \\
  \cmidrule{2-24}          & \boldmath{}\textbf{\attacknamenospace+RN }\unboldmath{} & 100.0  & 99.8  & 100.0  & 100.0  & 99.9  & 96.0  & 89.6  & 100.0  & 100.0  & 99.9  & 99.6  & 97.9  & 100.0  & 100.0  & 100.0  & 100.0  & 99.4  & 100.0  & 100.0  & 100.0  & 100.0  & 99.7  \\
  \cmidrule{2-24}          & \textbf{\attacknamenospace+RIR} & 100.0  & 99.9  & 100.0  & 95.1  & 73.4  & 38.6  & 16.7  & 100.0  & 99.7  & 98.0  & 93.7  & 86.3  & 100.0  & 100.0  & 99.5  & 98.5  & 95.6  & 100.0  & 100.0  & 100.0  & 99.9  & 99.0  \\
  \cmidrule{2-24}          & \textbf{\attacknamenospace+RN+RIR} & 100.0  & 100.0  & 100.0  & 100.0  & 99.8  & 95.5  & 89.4  & 100.0  & 100.0  & 99.8  & 99.2  & 98.0  & 100.0  & 100.0  & 100.0  & 100.0  & 99.7  & 100.0  & 100.0  & 100.0  & 100.0  & 99.9  \\
      \midrule
      \multirow{4}[8]{*}{\textbf{0.01}} & \textbf{\attackname} & 88.5  & 83.1  & 78.1  & 40.8  & 12.3  & 1.8   & 1.9   & 97.2  & 90.7  & 79.8  & 66.6  & 51.5  & 99.6  & 96.1  & 89.1  & 77.8  & 64.1  & 99.9  & 98.3  & 93.1  & 87.2  & 76.8  \\
  \cmidrule{2-24}          & \boldmath{}\textbf{\attacknamenospace+RN }\unboldmath{} & 100.0  & 99.9  & 100.0  & 100.0  & 100.0  & 98.7  & 92.6  & 100.0  & 100.0  & 100.0  & 99.8  & 98.9  & 100.0  & 100.0  & 100.0  & 100.0  & 99.6  & 100.0  & 100.0  & 100.0  & 100.0  & 99.9  \\
  \cmidrule{2-24}          & \textbf{\attacknamenospace+RIR} & 100.0  & 100.0  & 100.0  & 98.5  & 84.9  & 47.0  & 23.3  & 99.9  & 99.7  & 98.6  & 95.8  & 87.9  & 100.0  & 100.0  & 99.9  & 98.7  & 97.0  & 100.0  & 100.0  & 100.0  & 99.9  & 99.3  \\
  \cmidrule{2-24}          & \textbf{\attacknamenospace+RN+RIR} & 100.0  & 100.0  & 100.0  & 100.0  & 100.0  & 98.1  & 92.3  & 100.0  & 100.0  & 100.0  & 99.9  & 99.2  & 100.0  & 100.0  & 100.0  & 100.0  & 100.0  & 100.0  & 100.0  & 100.0  & 100.0  & 99.9  \\
      \midrule
      \multirow{4}[8]{*}{\textbf{0.02}} & \textbf{\attackname} & 96.1  & 89.9  & 94.4  & 78.6  & 39.0  & 11.3  & 2.3   & 99.3  & 96.8  & 90.2  & 80.0  & 67.4  & 99.8  & 99.4  & 96.7  & 89.2  & 81.4  & 100.0  & 100.0  & 98.7  & 94.9  & 89.7  \\
  \cmidrule{2-24}          & \boldmath{}\textbf{\attacknamenospace+RN }\unboldmath{} & 100.0  & 100.0  & 100.0  & 100.0  & 100.0  & 100.0  & 99.6  & 100.0  & 100.0  & 100.0  & 100.0  & 100.0  & 100.0  & 100.0  & 100.0  & 100.0  & 100.0  & 100.0  & 100.0  & 100.0  & 100.0  & 100.0  \\
  \cmidrule{2-24}          & \textbf{\attacknamenospace+RIR} & 100.0  & 100.0  & 100.0  & 100.0  & 99.3  & 83.3  & 42.9  & 100.0  & 100.0  & 99.8  & 98.6  & 94.8  & 100.0  & 100.0  & 100.0  & 99.9  & 99.5  & 100.0  & 100.0  & 100.0  & 100.0  & 99.8  \\
  \cmidrule{2-24}          & \textbf{\attacknamenospace+RN+RIR} & 100.0  & 100.0  & 100.0  & 100.0  & 100.0  & 100.0  & 99.8  & 100.0  & 100.0  & 100.0  & 99.9  & 99.8  & 100.0  & 100.0  & 100.0  & 100.0  & 100.0  & 100.0  & 100.0  & 100.0  & 100.0  & 100.0  \\
      \midrule
      \multirow{4}[8]{*}{\textbf{0.03}} & \textbf{\attackname} & 97.4  & 95.1  & 99.2  & 91.1  & 63.5  & 27.3  & 6.8   & 99.8  & 98.4  & 94.1  & 87.7  & 77.4  & 100.0  & 99.8  & 98.2  & 95.2  & 88.9  & 100.0  & 100.0  & 99.8  & 96.8  & 92.8  \\
  \cmidrule{2-24}          & \boldmath{}\textbf{\attacknamenospace+RN }\unboldmath{} & 100.0  & 100.0  & 100.0  & 100.0  & 100.0  & 100.0  & 100.0  & 100.0  & 100.0  & 100.0  & 100.0  & 100.0  & 100.0  & 100.0  & 100.0  & 100.0  & 100.0  & 100.0  & 100.0  & 100.0  & 100.0  & 100.0  \\
  \cmidrule{2-24}          & \textbf{\attacknamenospace+RIR} & 100.0  & 100.0  & 100.0  & 100.0  & 100.0  & 96.3  & 64.8  & 100.0  & 100.0  & 100.0  & 99.0  & 97.3  & 100.0  & 100.0  & 100.0  & 100.0  & 99.9  & 100.0  & 100.0  & 100.0  & 100.0  & 100.0  \\
  \cmidrule{2-24}          & \textbf{\attacknamenospace+RN+RIR} & 100.0  & 100.0  & 100.0  & 100.0  & 100.0  & 100.0  & 100.0  & 100.0  & 100.0  & 100.0  & 100.0  & 100.0  & 100.0  & 100.0  & 100.0  & 100.0  & 100.0  & 100.0  & 100.0  & 100.0  & 100.0  & 100.0  \\
      \bottomrule
      \end{tabular}%
    }
      \label{tab:over-the-air}\vspace{-2mm}%
  \end{table*}%

\begin{table*}
  \centering
  \caption{The room configuration and hardware setting of over-the-air attack}\vspace{-2mm}
  \resizebox{0.93\textwidth}{!}{%
    \begin{tabular}{|c|c|c|c|c|c|c|c|c|c|c|c|}
    \toprule
    \multicolumn{2}{|c|}{\multirow{2}[4]{*}{}} & \multicolumn{6}{c|}{\textbf{Room configuration}} & \multicolumn{2}{c|}{\textbf{Hardware position}} & \multicolumn{2}{c|}{\textbf{Hardware}} \\
\cmidrule{3-12}    \multicolumn{2}{|c|}{} &  {\bf Room type}  & \begin{tabular}[c]{@{}c@{}}  \textbf{Length}\\ \textbf{(m)} \end{tabular}  & \begin{tabular}[c]{@{}c@{}}  \textbf{Width}\\ \textbf{(m)} \end{tabular}  & \begin{tabular}[c]{@{}c@{}}  \textbf{Height}\\\textbf{(m)}  \end{tabular} & \textbf{\begin{tabular}[c]{@{}c@{}}Absorption \\ coefficient\end{tabular}} & \textbf{ \begin{tabular}[c]{@{}c@{}}Reverberation \\ time (s)\end{tabular} } & \begin{tabular}[c]{@{}c@{}}  \textbf{Distance}\\ \textbf{(m)}  \end{tabular}   & \textbf{Angle ($^{o}$)} & \textbf{Loudspeaker} & \textbf{Microphone} \\
    \midrule
    \multicolumn{2}{|c|}{\multirow{3}[6]{*}{\textbf{Simulated RIR}}} & \textbf{small} & 1-10  & 1-10  & \multirow{3}[6]{*}{2-5} & \multirow{3}[6]{*}{0.2-0.8} & -     &   1.06-11.33    &       & -     & - \\
\cmidrule{3-5}\cmidrule{8-12}    \multicolumn{2}{|c|}{} & \textbf{medium} & 10-30 & 10-30 &       &       & -     &    1.96-35.07   &       & -     & - \\
\cmidrule{3-5}\cmidrule{8-12}    \multicolumn{2}{|c|}{} & \textbf{large} & 30-50 & 30-50 &       &       & -     &   2.63-59.5    &       & -     & - \\
    \midrule
    \multirow{11}[28]{*}{\textbf{Real RIR}} & \multirow{3}[6]{*}{\textbf{RWCP}} & \textbf{\begin{tabular}[c]{@{}c@{}}variable \\ reverberant \\ room\end{tabular}} & 6.66  & 4.18  & -     & -     & 0.3-1.3 & 2     & 10-170 & \multicolumn{1}{c|}{\multirow{3}[6]{*}{\begin{tabular}[c]{@{}c@{}}1) Diatone DS-7 \\ loud speaker \\ 2) B\&K Type 4128 \\ Head-Torso\end{tabular}}} & \multicolumn{1}{c|}{\multirow{3}[6]{*}{\begin{tabular}[c]{@{}c@{}}1) 54ch Spherical array \\ 2) 14ch Linear array\\(2.83cm spacing) \\ 3) 16ch Circle array\end{tabular}}} \\
\cmidrule{3-10}          &       & \textbf{anechoic room} & -     & -     & -     & -     & 0.01-2 & -     & -     &       &  \\
\cmidrule{3-10}          &       & \textbf{office} & -     & -     & -     & -     & 0.01-2 & -     & -     &       &  \\
\cmidrule{2-12}          & \multirow{4}[8]{*}{\textbf{REVERB}} & \textbf{small} & 5.57 & 3.77  & -     & -     & 0.25  & 0.5, 2 & 45, 135 & {BOSE 101MM} & \multicolumn{1}{c|}{\begin{tabular}[c]{@{}c@{}} SONY ECM-77B \end{tabular}} \\
\cmidrule{3-12}          &       & \textbf{medium} & 6.27    & 4.89   & 2.59    & -     & 0.5   & 0.5, 2 &   45, 135  &  Genelec 1029A  & AKG capsules CE20 V17  \\
\cmidrule{3-12}          &       & \textbf{large} & 6.67    & 6.14 & - & -     & 0.7   & 0.5, 2 &   45, 135    &  {BOSE 101MM} & \multicolumn{1}{c|}{\begin{tabular}[c]{@{}c@{}} SONY ECM-77B \end{tabular}}\\
\cmidrule{3-12}          &       & \textbf{\begin{tabular}[c]{@{}c@{}}reverberant \\ meeting room\end{tabular}} & -     & -     & -     & -     & 0.7   & 1, 2.5 &     -  &   \begin{tabular}[c]{@{}c@{}} single stationary\\ speaker  \end{tabular}
   &  \begin{tabular}[c]{@{}c@{}}8-ch circular array \\ with omni-directional\\ microphones\end{tabular}\\
\cmidrule{2-12}          & \multirow{4}[8]{*}{\textbf{AIRD}} & \textbf{studio booth} & 3     & 1.8   & 2.2   & -     & 0.08-0.18 & 0.5, 1.0, 1.5 & -     & \multicolumn{1}{c|}{\multirow{4}[8]{*}{ \begin{tabular}[c]{@{}c@{}}2-way active \\ studio monitor \\ Genelec 8130\end{tabular}}} & \multirow{4}[8]{*}{ \begin{tabular}[c]{@{}c@{}}two Beyerdynamic\\ MM1 omnidirectional \\ condenser measurement\end{tabular}} \\
\cmidrule{3-10}          &       & \textbf{office room} & 5     & 6.4   & 2.9   & -     & 0.37-0.48 & 1.0, 2.0, 3.0 & -     &       &  \\
\cmidrule{3-10}          &       & \textbf{meeting room} & 8     & 5     & 3.1   & -     & 0.21-0.25 & \begin{tabular}[c]{@{}c@{}}1.45, 1.7, 1.9, \\ 2.25, 2.8 \end{tabular} & -     &       &  \\
\cmidrule{3-10}          &       & \textbf{lecture room} & 10.8  & 10.9  & 3.15  & -     & 0.70-0.83 & \begin{tabular}[c]{@{}c@{}}4.0, 5.56, 7.1, \\ 8.68, 10.2 \end{tabular} & -     &       &  \\
    \bottomrule
    \end{tabular}%
  }
  \label{tab:over-the-air-setting}\vspace{-2mm}%
\end{table*}%

\smallskip\noindent{\bf Different tasks.}
We notice that the same loss function often achieves lower targeted attack success rate (ASR$_t$) on the OSI task than on the CSI task.
Recall that we specify the same target speakers for the two tasks.
This indicates that targeted attack on the OSI task is more difficult than that on the CSI task.
It is because the OSI task rejects an input if the scores are less than the threshold $\theta$,
thus a successful attack must simultaneously guarantee that
the score of the target speaker is the maximal one and larger than the threshold $\theta$.

\begin{tcolorbox}[size=title,breakable]
    \textbf{Remark 3.}
   The OSI task is more difficult to attack than the CSI task. 
\end{tcolorbox}

To be exhaustive, {\tablename~\ref{tab:baisc-CSI-CW2-FAKEBOB}} reports the results of \attackname on the CSI task with $s\in G$
using the most effective loss functions $\mathcal{L}_\text{M}$/$\mathcal{L}_\text{M}^{s}$ and FAKEBOB and CW$_2$ approaches.
The results show that FAKEBOB and CW$_2$ are also effective, although
the black-box approach FAKEBOB performs slightly worse than the white-box approach  CW$_2$ in terms of
attack success rate and stealthiness.
{We remark that at the same perturbation budget and comparable attack success rate,
 the single-step approach FGSM produces the least stealthy adversarial voices in terms of L$_2$, SNR, and PESQ,
 while PGD is better than FAKEBOB but worse than CW$_2$.}

\chengk{We also evaluate \attackname on the datasets with more speakers, i.e., Spk$_{100}$-enroll,  Spk$_{100}$-test, and  Spk$_{100}$-imposter,
from which we can draw the same conclusions as Remarks 1-3.
Therefore, the results are omitted here and reported in our technical report~\cite{AS2T}.}

\begin{table*}[htbp]
    \centering
    \caption{\chengk{The attack success rate (\%) of over-the-air \attackname, where RN and RIR denote random noise and reverberation impulse response, respectively}}
    \resizebox{.98\textwidth}{!}{%
      \begin{tabular}{|c|c|c|c|c|c|c|c|c|c|c|c|c|c|c|c|c|c|c|c|c|c|c|c|}
      \toprule
      \multirow{3}[6]{*}{\boldmath{}\textbf{$\varepsilon$}\unboldmath{}} & \multirow{3}[6]{*}{\diagbox{\bf Training}{\textbf{Evaluation}}} & \multicolumn{2}{c|}{\textbf{Reverberation}} & \multicolumn{20}{c|}{\textbf{Ambient Noise (SNR=?dB)}} \\
  \cmidrule{3-24}          &       & \multirow{2}[4]{*}{\textbf{Sim.}} & \multirow{2}[4]{*}{\textbf{Real}} & \multicolumn{5}{c|}{\textbf{White Noise}} & \multicolumn{5}{c|}{\textbf{Point Source Noise}} & \multicolumn{5}{c|}{\textbf{Music Noise}} & \multicolumn{5}{c|}{\textbf{Speech Babble Noise}} \\
  \cmidrule{5-24}          &       &       &       & \textbf{20} & \textbf{15} & \textbf{10} & \textbf{5} & \textbf{0} & \textbf{20} & \textbf{15} & \textbf{10} & \textbf{5} & \textbf{0} & \textbf{20} & \textbf{15} & \textbf{10} & \textbf{5} & \textbf{0} & \textbf{20} & \textbf{15} & \textbf{10} & \textbf{5} & \textbf{0} \\
      \midrule
      \multirow{4}[8]{*}{\textbf{0.008}} & \textbf{\attackname} & 85.3  & 80.0  & 69.7  & 31.0  & 7.5   & 1.4   & 1.3   & 95.7  & 87.8  & 75.7  & 60.7  & 47.5  & 98.8  & 94.5  & 85.4  & 71.7  & 59.6  & 99.8  & 97.3  & 90.9  & 83.5  & 73.4  \\
  \cmidrule{2-24}          & \boldmath{}\textbf{\attacknamenospace+RN }\unboldmath{} & 100.0  & 99.8  & 100.0  & 100.0  & 99.9  & 96.0  & 89.6  & 100.0  & 100.0  & 99.9  & 99.6  & 97.9  & 100.0  & 100.0  & 100.0  & 100.0  & 99.4  & 100.0  & 100.0  & 100.0  & 100.0  & 99.7  \\
  \cmidrule{2-24}          & \textbf{\attacknamenospace+RIR} & 100.0  & 99.9  & 100.0  & 95.1  & 73.4  & 38.6  & 16.7  & 100.0  & 99.7  & 98.0  & 93.7  & 86.3  & 100.0  & 100.0  & 99.5  & 98.5  & 95.6  & 100.0  & 100.0  & 100.0  & 99.9  & 99.0  \\
  \cmidrule{2-24}          & \textbf{\attacknamenospace+RN+RIR} & 100.0  & 100.0  & 100.0  & 100.0  & 99.8  & 95.5  & 89.4  & 100.0  & 100.0  & 99.8  & 99.2  & 98.0  & 100.0  & 100.0  & 100.0  & 100.0  & 99.7  & 100.0  & 100.0  & 100.0  & 100.0  & 99.9  \\
      \midrule
      \multirow{4}[8]{*}{\textbf{0.01}} & \textbf{\attackname} & 88.5  & 83.1  & 78.1  & 40.8  & 12.3  & 1.8   & 1.9   & 97.2  & 90.7  & 79.8  & 66.6  & 51.5  & 99.6  & 96.1  & 89.1  & 77.8  & 64.1  & 99.9  & 98.3  & 93.1  & 87.2  & 76.8  \\
  \cmidrule{2-24}          & \boldmath{}\textbf{\attacknamenospace+RN }\unboldmath{} & 100.0  & 99.9  & 100.0  & 100.0  & 100.0  & 98.7  & 92.6  & 100.0  & 100.0  & 100.0  & 99.8  & 98.9  & 100.0  & 100.0  & 100.0  & 100.0  & 99.6  & 100.0  & 100.0  & 100.0  & 100.0  & 99.9  \\
  \cmidrule{2-24}          & \textbf{\attacknamenospace+RIR} & 100.0  & 100.0  & 100.0  & 98.5  & 84.9  & 47.0  & 23.3  & 99.9  & 99.7  & 98.6  & 95.8  & 87.9  & 100.0  & 100.0  & 99.9  & 98.7  & 97.0  & 100.0  & 100.0  & 100.0  & 99.9  & 99.3  \\
  \cmidrule{2-24}          & \textbf{\attacknamenospace+RN+RIR} & 100.0  & 100.0  & 100.0  & 100.0  & 100.0  & 98.1  & 92.3  & 100.0  & 100.0  & 100.0  & 99.9  & 99.2  & 100.0  & 100.0  & 100.0  & 100.0  & 100.0  & 100.0  & 100.0  & 100.0  & 100.0  & 99.9  \\
      \midrule
      \multirow{4}[8]{*}{\textbf{0.02}} & \textbf{\attackname} & 96.1  & 89.9  & 94.4  & 78.6  & 39.0  & 11.3  & 2.3   & 99.3  & 96.8  & 90.2  & 80.0  & 67.4  & 99.8  & 99.4  & 96.7  & 89.2  & 81.4  & 100.0  & 100.0  & 98.7  & 94.9  & 89.7  \\
  \cmidrule{2-24}          & \boldmath{}\textbf{\attacknamenospace+RN }\unboldmath{} & 100.0  & 100.0  & 100.0  & 100.0  & 100.0  & 100.0  & 99.6  & 100.0  & 100.0  & 100.0  & 100.0  & 100.0  & 100.0  & 100.0  & 100.0  & 100.0  & 100.0  & 100.0  & 100.0  & 100.0  & 100.0  & 100.0  \\
  \cmidrule{2-24}          & \textbf{\attacknamenospace+RIR} & 100.0  & 100.0  & 100.0  & 100.0  & 99.3  & 83.3  & 42.9  & 100.0  & 100.0  & 99.8  & 98.6  & 94.8  & 100.0  & 100.0  & 100.0  & 99.9  & 99.5  & 100.0  & 100.0  & 100.0  & 100.0  & 99.8  \\
  \cmidrule{2-24}          & \textbf{\attacknamenospace+RN+RIR} & 100.0  & 100.0  & 100.0  & 100.0  & 100.0  & 100.0  & 99.8  & 100.0  & 100.0  & 100.0  & 99.9  & 99.8  & 100.0  & 100.0  & 100.0  & 100.0  & 100.0  & 100.0  & 100.0  & 100.0  & 100.0  & 100.0  \\
      \midrule
      \multirow{4}[8]{*}{\textbf{0.03}} & \textbf{\attackname} & 97.4  & 95.1  & 99.2  & 91.1  & 63.5  & 27.3  & 6.8   & 99.8  & 98.4  & 94.1  & 87.7  & 77.4  & 100.0  & 99.8  & 98.2  & 95.2  & 88.9  & 100.0  & 100.0  & 99.8  & 96.8  & 92.8  \\
  \cmidrule{2-24}          & \boldmath{}\textbf{\attacknamenospace+RN }\unboldmath{} & 100.0  & 100.0  & 100.0  & 100.0  & 100.0  & 100.0  & 100.0  & 100.0  & 100.0  & 100.0  & 100.0  & 100.0  & 100.0  & 100.0  & 100.0  & 100.0  & 100.0  & 100.0  & 100.0  & 100.0  & 100.0  & 100.0  \\
  \cmidrule{2-24}          & \textbf{\attacknamenospace+RIR} & 100.0  & 100.0  & 100.0  & 100.0  & 100.0  & 96.3  & 64.8  & 100.0  & 100.0  & 100.0  & 99.0  & 97.3  & 100.0  & 100.0  & 100.0  & 100.0  & 99.9  & 100.0  & 100.0  & 100.0  & 100.0  & 100.0  \\
  \cmidrule{2-24}          & \textbf{\attacknamenospace+RN+RIR} & 100.0  & 100.0  & 100.0  & 100.0  & 100.0  & 100.0  & 100.0  & 100.0  & 100.0  & 100.0  & 100.0  & 100.0  & 100.0  & 100.0  & 100.0  & 100.0  & 100.0  & 100.0  & 100.0  & 100.0  & 100.0  & 100.0  \\
      \bottomrule
      \end{tabular}%
    }
      \label{tab:over-the-air}\vspace{-2mm}%
  \end{table*}%

\subsection{Evaluation of \attackname Over-the-Air}\label{sec:evaluate-practical}

\subsubsection{Evaluation Setup}
It is non-trivial to conduct a large-scale
and thorough evaluation of adversarial attacks over-the-air in the physical world.
Thus, we simulate over-the-air attacks, as done in the speech community~\cite{Rever-dataset,Robust-Model-1,Robust-Model-2,Robust-Model-3},
using the untargeted attack of \attackname on the CSI task,
where the source speakers are enrolled speakers (i.e., $s\in G$) and the loss function is $\mathcal{L}_{\text{M}}^s$ as suggested in Remark 1.


To simulate reverberation, we use 2,000 randomly simulated RIR
and all the 325 publicly available real-world RIR in \cite{Rever-dataset},
and convolve them with the adversarial voices.
The simulated RIR is generated by Image Source Method~\cite{image-method}, covering various room dimension and the position of devices.
The real-world RIR is collected in different physical rooms with various loudspeakers and microphones,
thus also reflects various equipment distortions.
\chengk{The room configurations, positions of loudspeakers and microphones (e.g., distance and angle), and brands of hardware devices
are given in \tablename~\ref{tab:over-the-air-setting}.}
To simulate the ambient noise, we use both the widely-spread white noise
and three types of representative noise occurred in real-world scenarios
provided by the MUSAN dataset~\cite{MUSAN}: point-source, musical, and speech babble noise. 
Point-source noise includes technical sounds (e.g., cellphone noises and dialtones)
and non-technical sounds (e.g., hunder, car horns, and animal sounds).
Musical noise consists of several music genres, e.g., Country, Hip-Hop, and Jazz.
Speech babble noise is introduced when multiple speakers utter at the same time
and part or even the whole speech is mixed with others.
\chengk{Different types of ambient noise simulate different environments where the hardware is placed.}
For ambient noise, we set the SNR between adversarial voices and the noise to 0, 5, 10, 15, and 20 to
imitate different volume of loudspeakers and the environments with different levels of noise.

For \attacknamenospace+RN, we use additive white Gaussian noise, i.e., $\mathcal{Z}=\mathcal{N}(0,1)$
where $\mathcal{N}(0,1)$ is standard normal distribution, and set SNR$_l=0$ and SNR$_h=20$.
For \attacknamenospace+RIR, we approximate the distribution of RIR $\mathcal{R}$ using 1,000 simulated RIR,
and the other disjoint 1,000 simulated RIR and 325 real-world RIR are used for testing.
The sampling size $K$ in Algorithm~\ref{al:practical-attack} is set to 10.
We exploit PGD as the optimization approach
except that the standard gradient descent \chengk{(SGD)} is replaced with Adam~\cite{kingma2014adam},
\chengk{because Adam is more efficient for crafting robust adversarial voices against over-the-air distortions.}
We set the perturbation budget $\varepsilon=0.008, 0.01, 0.02, 0.03$,
the number of iterations \#Iter=400 and step\_size $\alpha=\frac{5\times \varepsilon}{\text{\#Iter}}$.
\chengk{Note that for each crafted adversarial voice, we only simulate the over-the-air transmission {\it one} time.}
\begin{table*}[htbp]
  \centering\setlength\tabcolsep{3.5pt}
  \caption{Transferability of \attackname with the FGSM and PGD optimization approaches.
  For each pair of source and target systems, 
  the result indicates the accuracy drop of the target system. CSI task, untargeted attack, $s\in G$.}\vspace{-2mm}
  \resizebox{.9\textwidth}{!}{%
    \begin{tabular}{|c|c|c||c|c|c||c|c|c|c|c|c|c|c||c||c|}
    \toprule
    \multicolumn{2}{|c|}{\diagbox{\bf Source}{\bf $\Downarrow$ Accuracy (\%)}{\bf Target}} & \textbf{Ivector} & \textbf{ECAPA} & \textbf{Xvector-P} & \textbf{Xvector-C} & \textbf{AudioNet} & \textbf{SincNet} & \textbf{Res18-I} & \textbf{Res18-V} & \textbf{Res34-I} & \textbf{Res34-V} & \textbf{Auto-I} & \textbf{Auto-V} & \textbf{GE2E} & \textbf{Hubert} \\
    \midrule
    {\bf GMM} & \textbf{Ivector} & \cellcolor{gray!40} 100.0 & 7.3   & \textcolor{blue}{\bf 30.5} & \textcolor{blue}{\bf 26.7} & 21.2  & 10.3  & \textcolor{red}{\bf 0.5} & \textcolor{red}{\bf 0.6} & 3.2   & \textcolor{red}{\bf 2.9} & 11    & 6.8   & 23    & \textcolor{blue}{\bf 33.8} \\
    \midrule
    \midrule
    \multirow{3}[3]{*}{\bf TDNN} & \textbf{ECAPA} & 8.6   & \cellcolor{gray!40} 99.9 & \textcolor{blue}{\bf 14.3} & 11.4  & 10.9  & 8.6   & \textcolor{red}{\bf 1.1} & \textcolor{red}{\bf 1.0} & 7.1   & \textcolor{red}{\bf 1.4} & 12.5  & 12.2  & \textcolor{blue}{\bf 14.5} & \textcolor{blue}{\bf 28.8} \\
    \cmidrule{2-16}
    & \textbf{Xvector-P} & \textcolor{blue}{\bf 44.4} & 22.5  & \cellcolor{gray!40} 100.0 & \textcolor{blue}{\bf 34.4} & 18.3  & 9.4   & \textcolor{red}{\bf 1.3} & \textcolor{red}{\bf 1.2} & 4.2   & \textcolor{red}{\bf 2.5} & 12.8  & 8.2   & 21.1  & \textcolor{blue}{\bf 34.9} \\
    \cmidrule{2-16}
    & \textbf{Xvector-C} & \textcolor{blue}{\bf 51.5} & \textcolor{blue}{\bf 42.7} & \textcolor{blue}{\bf 58.6} & \cellcolor{gray!40} 96.4 & 13.5  & 7.7   & \textcolor{red}{\bf 1.9} & \textcolor{red}{\bf 2.1} & \textcolor{red}{\bf 7.0} & 7.5   & 15.8  & 14.2  & 18.9  & 25.6 \\
    \midrule
    \midrule
    \multirow{8}[9]{*}{\bf CNN} & \textbf{AudioNet} & 4.3   & \textcolor{red}{\bf 0.1} & \textcolor{red}{\bf 0.5} & \textcolor{red}{\bf -1.9} & \cellcolor{gray!40} 99.9 & 8.1   & 0.9   & 0.7   & 4     & 2.8   & \textcolor{blue}{\bf 8.2} & 4     & \textcolor{blue}{\bf 18.5} & \textcolor{blue}{\bf 26.9} \\
    \cmidrule{2-16}
    & \textbf{SincNet} & 4.5   & \textcolor{red}{\bf 0.5} & \textcolor{red}{\bf 0.7} & 4.2   & \textcolor{blue}{\bf 20.8} & \cellcolor{gray!40} 67.6 & \textcolor{red}{\bf 2.2} & 3.6   & 6.3   & 5.9   & \textcolor{blue}{\bf 14.0} & 8.7   & \textcolor{blue}{\bf 36.6} & 10.2 \\
    \cmidrule{2-16}
    & \textbf{Res18-I} & 1.4   & \textcolor{red}{\bf 0.1} & \textcolor{red}{\bf 0.4} & \textcolor{red}{\bf -2.9} & 15.2  & 8.6   & \cellcolor{gray!40} 100.0 & \textcolor{blue}{\bf 34.2} & \textcolor{blue}{\bf 41.8} & \textcolor{blue}{\bf 41.4} & 23.9  & 27.3  & 18.3  & 23.5 \\
    \cmidrule{2-16}
    & \textbf{Res18-V} & 2     & \textcolor{red}{\bf 0.2} & \textcolor{red}{\bf 0.3} & \textcolor{red}{\bf -2.9} & 14.8  & 8.2   & \textcolor{blue}{\bf 36.6} & \cellcolor{gray!40} 99.9 & \textcolor{blue}{\bf 47.2} & \textcolor{blue}{\bf 48.7} & 28.5  & 34.8  & 18.4  & 23.3 \\
    \cmidrule{2-16}
    & \textbf{Res34-I} & 1.2   & \textcolor{red}{\bf 0.1} & \textcolor{red}{\bf 0.3} & \textcolor{red}{\bf -3.1} & 15.3  & 9.5   & 15.2  & 20    & \cellcolor{gray!40} 100.0 & \textcolor{blue}{\bf 27.1} & 18.2  & \textcolor{blue}{\bf 22.6} & 17.6  & \textcolor{blue}{\bf 23.9} \\
    \cmidrule{2-16}
    & \textbf{Res34-V} & 1.6   & \textcolor{red}{\bf 0.2} & \textcolor{red}{\bf 0.2} & \textcolor{red}{\bf -3.1} & 15.4  & 8     & 20    & \textcolor{blue}{\bf 24.9} & \textcolor{blue}{\bf 38.9} & \cellcolor{gray!40} 99.8 & 23.4  & \textcolor{blue}{\bf 28.4} & 18.7  & 22.6 \\
    \cmidrule{2-16}
    & \textbf{Auto-I} & 2.6   & \textcolor{red}{\bf 0.1} & \textcolor{red}{\bf 0.4} & \textcolor{red}{\bf -3.1} & 12.2  & 7.3   & 9.4   & 10.2  & 18    & \textcolor{blue}{\bf 23.5} & \cellcolor{gray!40} 58.8 & \textcolor{blue}{\bf 47.3} & 16.5  & \textcolor{blue}{\bf 22.4} \\
    \cmidrule{2-16}
    & \textbf{Auto-V} & 2.6   & \textcolor{red}{\bf 0.6} & \textcolor{red}{\bf 1.1} & \textcolor{red}{\bf -2.2} & 13.9  & 8.9   & 12.1  & 17.8  & \textcolor{blue}{\bf 25.2} & \textcolor{blue}{\bf 31.8} & \textcolor{blue}{\bf 44.2} & \cellcolor{gray!40} 99.3 & 18.5  & 20.2 \\
    \midrule
    \midrule
    {\bf LSTM} & \textbf{GE2E} & 3.5   & \textcolor{red}{\bf 0.1} & \textcolor{red}{\bf 0.3} & \textcolor{red}{\bf -2.7} & \textcolor{blue}{\bf 25.2} & 4.3   & 1.1   & 3.2   & 2.7   & 5.2   & \textcolor{blue}{\bf 12.1} & 9.8   & \cellcolor{gray!40} 68.7 & \textcolor{blue}{\bf 28.2} \\
    \midrule
    \midrule
    {\bf Trans} & \textbf{Hubert} & 1.7   & 0.3   & 0.2   & \textcolor{red}{\bf -3.2} & 3.9   & 4.3   & \textcolor{red}{\bf 0.1} & \textcolor{red}{\bf -0.1} & 0.7   & 0.7   & \textcolor{blue}{\bf 9.7} & \textcolor{blue}{\bf 8.0} & \textcolor{blue}{\bf 7.6} & \cellcolor{gray!40} 95.2 \\
    \bottomrule
    \end{tabular}%
    }
  \label{tab:transfer-CSI-Untargeted}\vspace{-2mm}%
\end{table*}%
\begin{table*}
  \centering
  \caption{The input gradient size of different SRS models. CSI task, untargeted attack, $s\in G$.}\vspace*{-2mm}
  \resizebox{0.95\textwidth}{!}{%
    \begin{tabular}{|c|c|c|c|c|c|c|c|c|c|c|c|c|c|c|}
    \toprule
      {\bf Model}    & \textbf{IV} & \textbf{ECAPA} & \textbf{XV-P} & \textbf{XV-C} & \textbf{AudioNet} & \textbf{SincNet} & \textbf{Res18-I} & \textbf{Res18-V} & \textbf{Res34-I} & \textbf{Res34-V} & \textbf{Auto-I} & \textbf{Auto-V} & \textbf{GE2E} & \textbf{Hubert} \\
    \midrule
    $\mathbf{\|\nabla_x \mathcal{L}_t^x\|_1}$     & 1.75e+09 & 0.173  & 8.54e+08 & 0.110  & 0.156  & 0.007  & 0.187  & 0.151  & 0.285  & 0.231  & 0.131  & 0.165  & 0.026  & 0.248  \\
    \bottomrule
    \end{tabular}%
  }
  \label{tab:input-gradient-size}%
\end{table*}%

\subsubsection{Results}
The results are shown in \tablename~\ref{tab:over-the-air}.

\smallskip\noindent{\bf Different sources of distortions.}
We observe that the success rate of \attackname increases with the budget $\varepsilon$,
indicating a trade-off between the robust and stealthiness of adversarial voices.
The success rate of \attackname is positively correlated with the SNR between adversarial voices and ambient noise.
This is not surprising, as the adversarial voices with larger magnitude can overwhelm the weaker ambient noise.


We notice that the attack under the real-world RIR is slightly less effective than that under simulated RIR.
This is because the real-world RIR contains additional distortion from the loudspeakers and microphones.
Nevertheless, the difference is minor, indicating that the equipment distortion is not substantial compared to the other distortions.

\smallskip\noindent{\bf \attackname vs. \attacknamenospace+X.}
With the budget $\varepsilon=0.008$,
\attacknamenospace+RIR achieves 19\% higher success rate than \attackname under the real-world reverberation.
This indicates that incorporating the simulated RIR into \attackname
improves the robustness of adversarial voices against real-world reverberation.
Similarly, modeling white-noise in \attackname
enhances the robustness of adversarial voices  against white, point-source, musical, and speech babble noises,
with at least 26\% attack success rate improvement when $\varepsilon=0.008$ and SNR=0 dB.
Unsurprisingly, the combination of \attacknamenospace+RIR and \attacknamenospace+RN, i.e., \attacknamenospace+RN+RIR,
improves the practicability of \attackname under both real-world reverberation and various ambient noises.

\begin{tcolorbox}[size=title,breakable]
  \textbf{Remark 4.}
  The impact of equipment distortion is minor compared to that of reverberation and ambient noise.
  By incorporating the simulated reverberation and white noise into the generation of adversarial voices,
  \attackname can be improved towards robust over-ther-air attack
  against real-world reverberation and different types of ambient noises.
\end{tcolorbox}

\begin{table*}
    \centering\setlength\tabcolsep{4.5pt}
    \caption{Transferability of \attackname using FGSM and PGD. 
    For each pair of source and target systems, 
    the result indicates the increase of ASR$_t$. 
    (Note that some target models misclassify some benign voices).
    CSI task, targeted attack, $s\in G$.}\vspace{-2mm}
    \resizebox{.9\textwidth}{!}{%
      \begin{tabular}{|c|c|c||c|c|c||c|c|c|c|c|c|c|c||c||c|}
      \toprule
      \multicolumn{2}{|c|}{\diagbox{\bf Source}{\bf $\Uparrow$ ASR$_t$ (\%)}{\bf Target}} & \textbf{Ivector} & \textbf{ECAPA} & \textbf{Xvector-P} & \textbf{Xvector-C} & \textbf{AudioNet} & \textbf{SincNet} & \textbf{Res18-I} & \textbf{Res18-V} & \textbf{Res34-I} & \textbf{Res34-V} & \textbf{Auto-I} & \textbf{Auto-V} & \textbf{GE2E} & \textbf{Hubert} \\
      \midrule
      {\bf GMM} & \textbf{Ivector} & \cellcolor{gray!40} 100.0 & 2.5   & \textcolor{blue}{\bf 6.4} & \textcolor{blue}{\bf 6.4} & 2.2   & 1.5   & \textcolor{red}{\bf 0.1} & \textcolor{red}{\bf 0.1} & 0.3   & 0.3   & 1.4   & 0.9   & \textcolor{blue}{\bf 3.8} & \textcolor{red}{\bf -0.3} \\
      \midrule
      \midrule
      \multirow{3}[4]{*}{\bf TDNN} & \textbf{ECAPA} & \textcolor{blue}{\bf 2.0} & \cellcolor{gray!40} 100.0 & \textcolor{blue}{\bf 4.5} & \textcolor{blue}{\bf 3.3} & 1.5   & 1     & \textcolor{red}{\bf 0.1} & 0.3   & 0.9   & \textcolor{red}{\bf 0.3} & 1.9   & 1.9   & 1.8   & \textcolor{red}{\bf -0.2} \\
      \cmidrule{2-16}
      & \textbf{Xvector-P} & \textcolor{blue}{\bf 13.3} & \textcolor{blue}{\bf 7.1} & \cellcolor{gray!40} 100.0 & \textcolor{blue}{\bf 8.9} & 1.8   & 1.5   & \textcolor{red}{\bf 0.0} & \textcolor{red}{\bf 0.1} & 0.4   & 0.2   & 2.3   & 1.2   & 2.7   & \textcolor{red}{\bf 0.0} \\
      \cmidrule{2-16}
      & \textbf{Xvector-C}  & \textcolor{blue}{\bf 16.0} & \textcolor{blue}{\bf 15.5} & \textcolor{blue}{\bf 17.8} & \cellcolor{gray!40} 99.7 & 2     & 1.2   & \textcolor{red}{\bf 0.3} & \textcolor{red}{\bf 0.2} & 0.6   & 0.5   & 3     & 2     & 5.7   & \textcolor{red}{\bf 0.1} \\
      \midrule
      \midrule
      \multirow{8}[12]{*}{\bf CNN} & \textbf{AudioNet}  & 0.4   & 0.1   & 0.1   & \textcolor{red}{\bf -0.2} & \cellcolor{gray!40} 99.9 & \textcolor{blue}{\bf 1.4} & \textcolor{red}{\bf 0.1} & 0.2   & \textcolor{red}{\bf 0.1} & 0.3   & \textcolor{blue}{\bf 0.6} & 0.5   & \textcolor{blue}{\bf 3.3} & 0.3 \\
      \cmidrule{2-16}
      & \textbf{SincNet}  & 0.8   & \textcolor{red}{\bf 0.2} & \textcolor{red}{\bf 0.2} & 0.9   & \textcolor{blue}{\bf 5.6} & \cellcolor{gray!40} 34.4 & \textcolor{red}{\bf 0.3} & 0.5   & 1.2   & 0.9   & 1.4   & \textcolor{blue}{\bf 1.9} & \textcolor{blue}{\bf 13.6} & 0.5 \\
      \cmidrule{2-16}
      & \textbf{Res18-I}  & 0.2   & \textcolor{red}{\bf 0.0} & \textcolor{red}{\bf 0.0} & \textcolor{red}{\bf -0.3} & 1.7   & 1.3   & \cellcolor{gray!40} 100.0 & 6.4   & \textcolor{blue}{\bf 9.0} & \textcolor{blue}{\bf 10.1} & \textcolor{blue}{\bf 7.7} & 6     & 1.9   & 0.2 \\
      \cmidrule{2-16}
      & \textbf{Res18-V}  & 0.3   & \textcolor{red}{\bf 0.1} & \textcolor{red}{\bf 0.1} & \textcolor{red}{\bf -0.3} & 2.2   & 1.1   & \textcolor{blue}{\bf 11.2} & \cellcolor{gray!40} 99.9 & \textcolor{blue}{\bf 14.3} & \textcolor{blue}{\bf 13.7} & 9.2   & 9     & 1.9   & 0.2 \\
      \cmidrule{2-16}
      & \textbf{Res34-I}  & 0.3   & \textcolor{red}{\bf 0.0} & \textcolor{red}{\bf 0.1} & \textcolor{red}{\bf -0.3} & 1.9   & 1     & 3     & 3     & \cellcolor{gray!40} 100.0 & \textcolor{blue}{\bf 7.3} & \textcolor{blue}{\bf 5.7} & \textcolor{blue}{\bf 4.7} & 1.7   & 0.2 \\
      \cmidrule{2-16}
      & \textbf{Res34-V}  & 0.3   & \textcolor{red}{\bf 0.0} & 0.1   & \textcolor{red}{\bf -0.3} & 2.1   & 1.1   & 4.6   & 4.7   & \textcolor{blue}{\bf 8.6} & \cellcolor{gray!40} 100.0 & \textcolor{blue}{\bf 7.6} & \textcolor{blue}{\bf 6.1} & 1.7   & \textcolor{red}{\bf 0.1} \\
      \cmidrule{2-16}
      & \textbf{Auto-I}  & 0.2   & \textcolor{red}{\bf 0.1} & 0.2   & \textcolor{red}{\bf -0.3} & 1.5   & 1.2   & 1.9   & 1.7   & \textcolor{blue}{\bf 3.8} & \textcolor{blue}{\bf 4.5} & \cellcolor{gray!40} 73.2 & \textcolor{blue}{\bf 11.6} & 1.3   & \textcolor{red}{\bf 0.1} \\
      \cmidrule{2-16}
      & \textbf{Auto-V}  & 0.4   & \textcolor{red}{\bf 0.1} & 0.4   & \textcolor{red}{\bf -0.2} & 2.2   & 1.1   & 3.9   & 4.5   & \textcolor{blue}{\bf 7.2} & \textcolor{blue}{\bf 8.4} & \textcolor{blue}{\bf 15.6} & \cellcolor{gray!40} 99.9 & 2.5   & \textcolor{red}{\bf 0.2} \\
      \midrule
      \midrule
      {\bf LSTM} & \textbf{GE2E} & 0.5   & 0.1   & \textcolor{red}{\bf 0.1} & \textcolor{red}{\bf -0.1} & \textcolor{blue}{\bf 5.3} & \textcolor{blue}{\bf 1.2} & \textcolor{red}{\bf 0.0} & 0.2   & 0.4   & 0.7   & 0.6   & \textcolor{blue}{\bf 1.2} & \cellcolor{gray!40} 72.6 & 0.4 \\
      \midrule
      \midrule
      {\bf Trans} & \textbf{Hubert} & 0.2   & \textcolor{red}{\bf 0.0} & 0.1   & \textcolor{red}{\bf -0.3} & \textcolor{blue}{\bf 0.7} & 0.5   & \textcolor{red}{\bf 0.0} & 0.1   & 0.1   & 0.1   & 0.5   & \textcolor{blue}{\bf 0.8} & \textcolor{blue}{\bf 1.2} & \cellcolor{gray!40} 88.1 \\
      \bottomrule
      \end{tabular}%
      }
      \label{tab:transfer-CSI-targeted}\vspace{-3mm}%
  \end{table*}%

\subsection{Transferability Analysis}\label{sec:transfer-analysis}

In this section, we conduct a thorough transferability study of \attackname among 14 SR models (cf.  \tablename~\ref{tab:system-info}) along two major axes:
model-specific factors (e.g., model architecture, training dataset, and input types)
and attack-specific factors (e.g., number of iterations, step\_size, and perturbation budget).

%

  \begin{figure*}
    \centering
    \begin{subfigure}[t]{0.24\textwidth}
        \centering
        \includegraphics[width=1\textwidth]{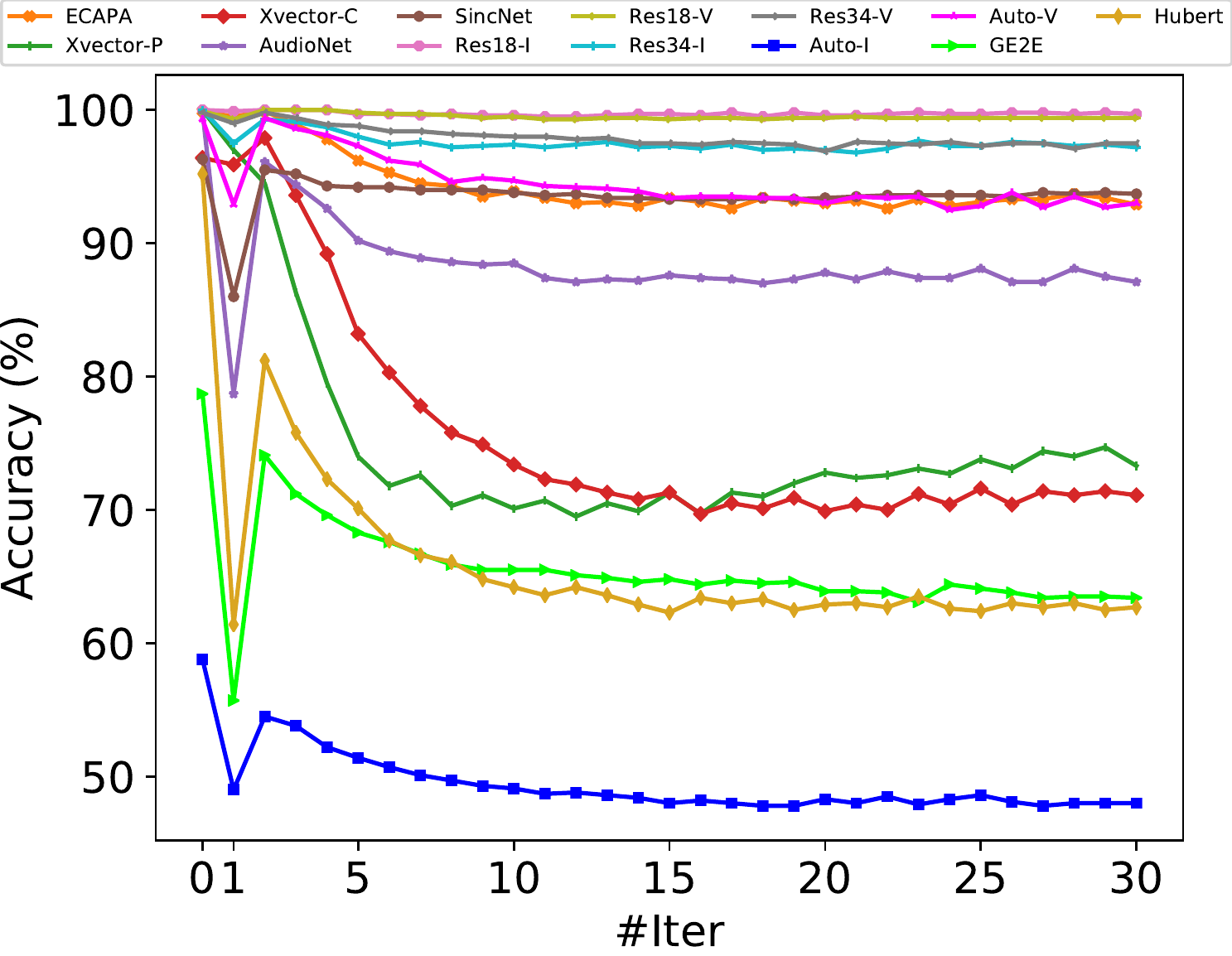}\vspace{-2mm}
        \caption{Ivector}
    \end{subfigure}
    \begin{subfigure}[t]{0.24\textwidth}
        \centering
        \includegraphics[width=1\textwidth]{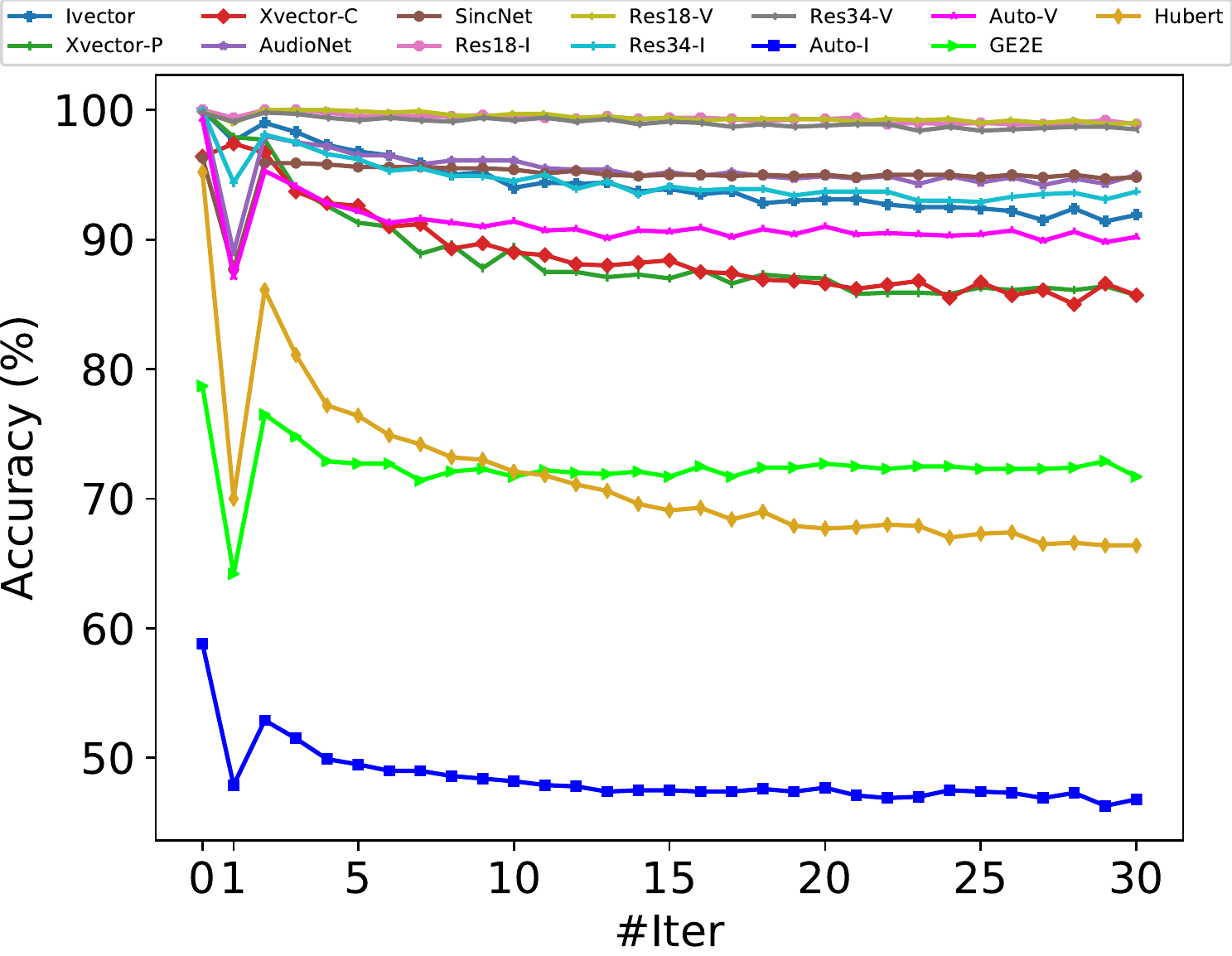}\vspace{-2mm}
        \caption{ECAPA}
    \end{subfigure}
    \begin{subfigure}[t]{0.24\textwidth}
        \centering
        \includegraphics[width=1\textwidth]{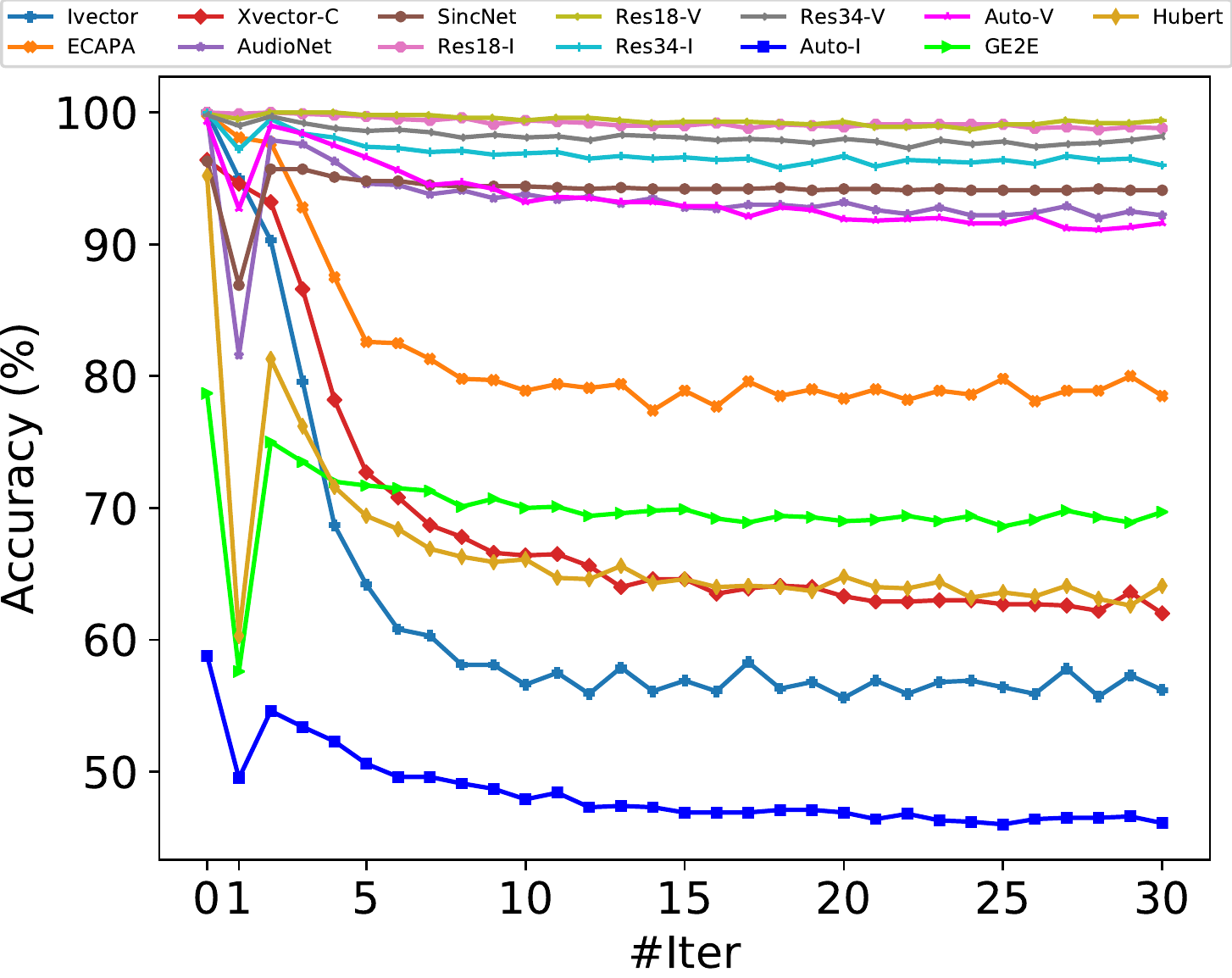}\vspace{-2mm}
        \caption{Xvector-P}
    \end{subfigure}
    \begin{subfigure}[t]{0.24\textwidth}
        \centering
        \includegraphics[width=1\textwidth]{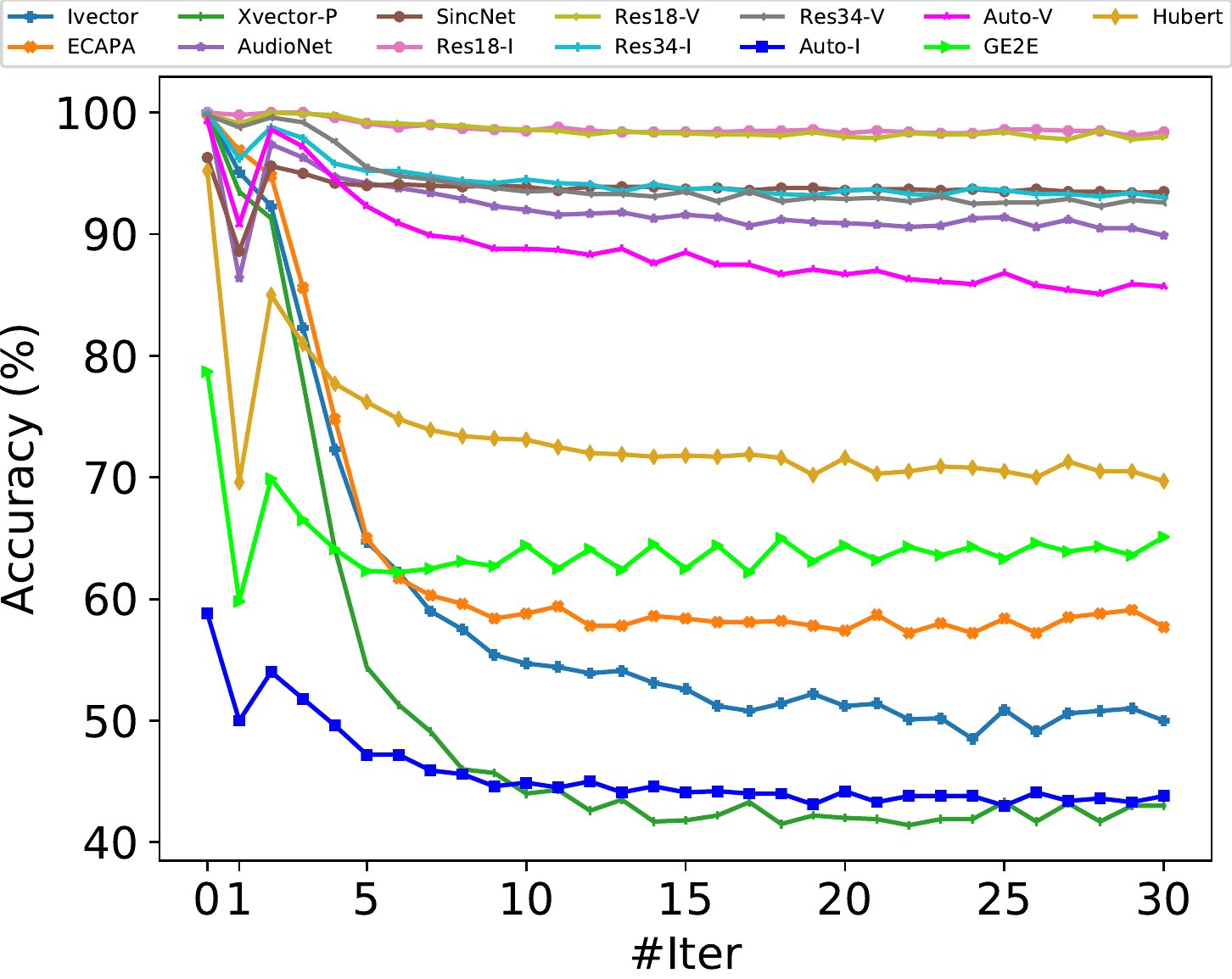}\vspace{-2mm}
        \caption{Xvector-C}
    \end{subfigure}
    \begin{subfigure}[t]{0.186\textwidth}
        \centering
        \includegraphics[width=1\textwidth]{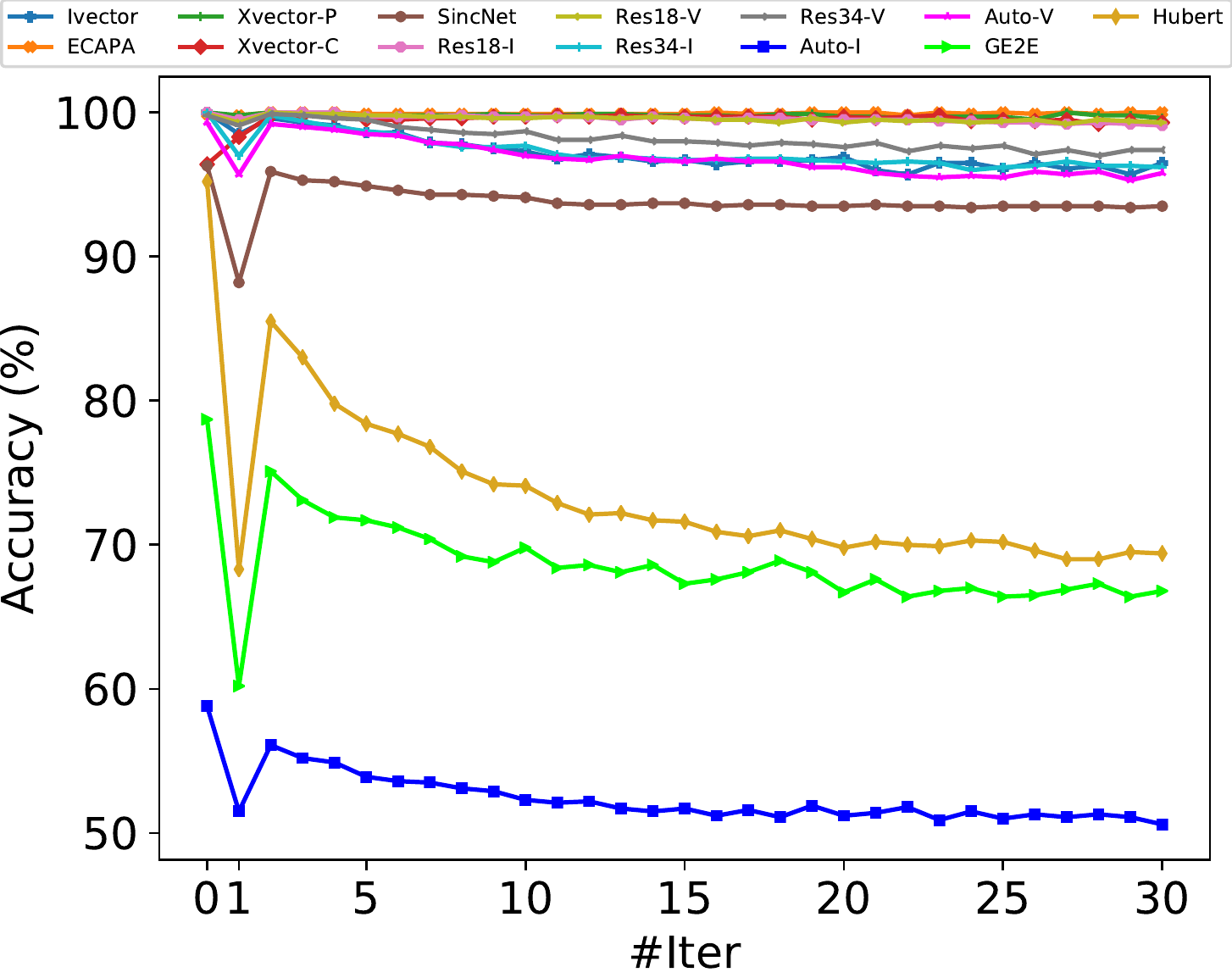}\vspace{-2mm}
        \caption{AudioNet}
    \end{subfigure}
    \begin{subfigure}[t]{0.187\textwidth}
        \centering
        \includegraphics[width=1\textwidth]{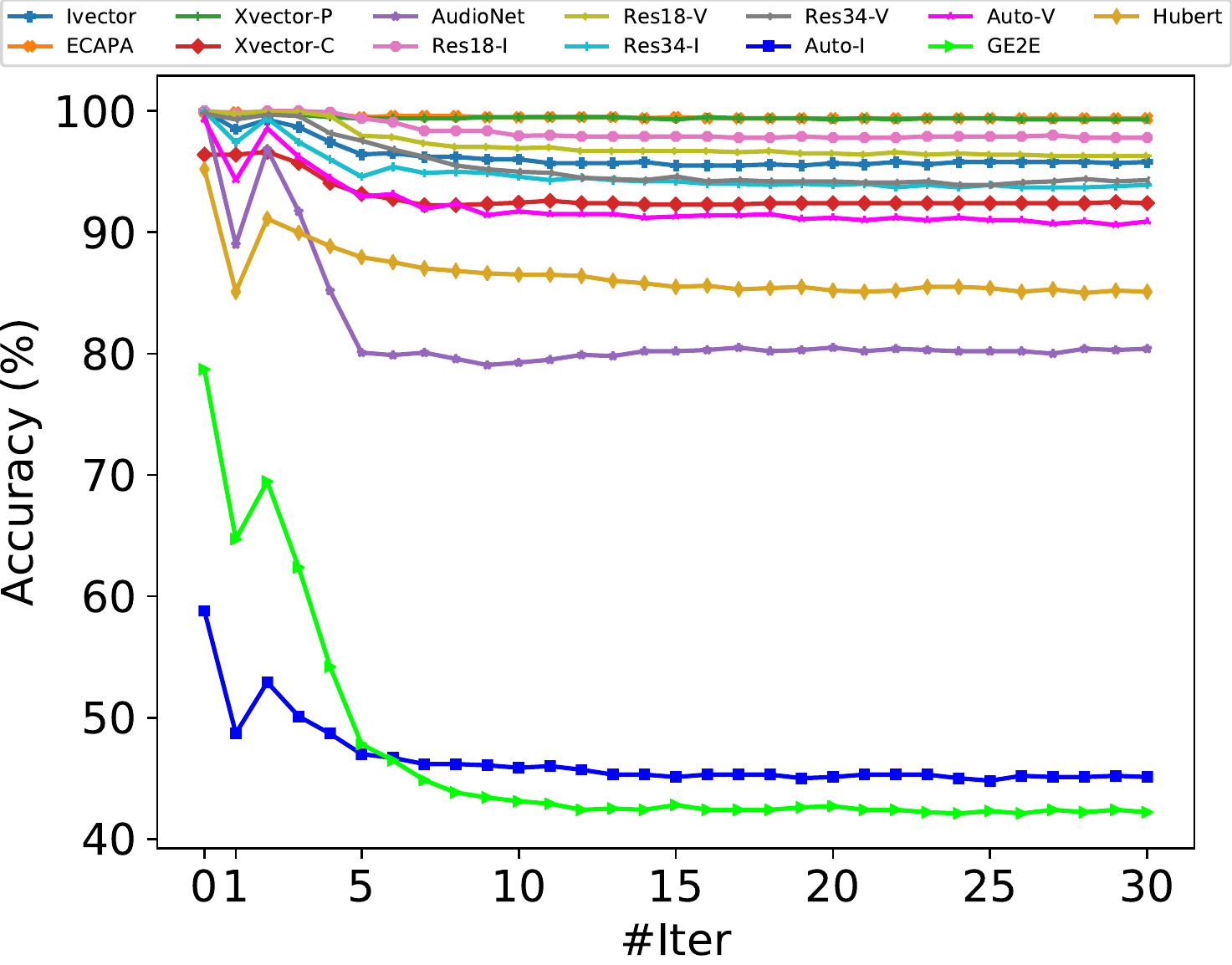}\vspace{-2mm}
        \caption{SincNet}
    \end{subfigure}
    \begin{subfigure}[t]{0.187\textwidth}
        \centering
        \includegraphics[width=1\textwidth]{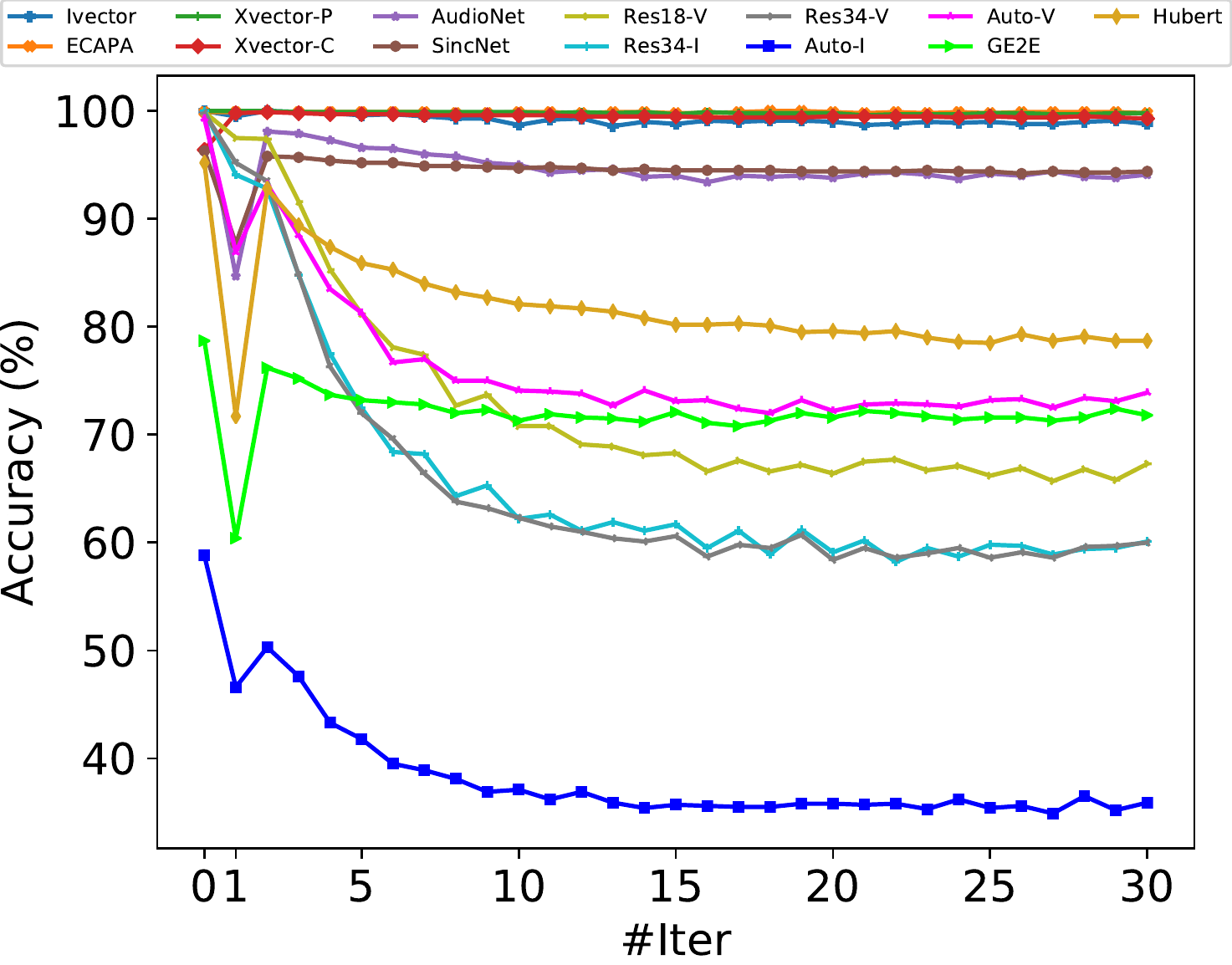}\vspace{-2mm}
        \caption{Res18-I}
    \end{subfigure}
    \begin{subfigure}[t]{0.187\textwidth}
        \centering
        \includegraphics[width=1\textwidth]{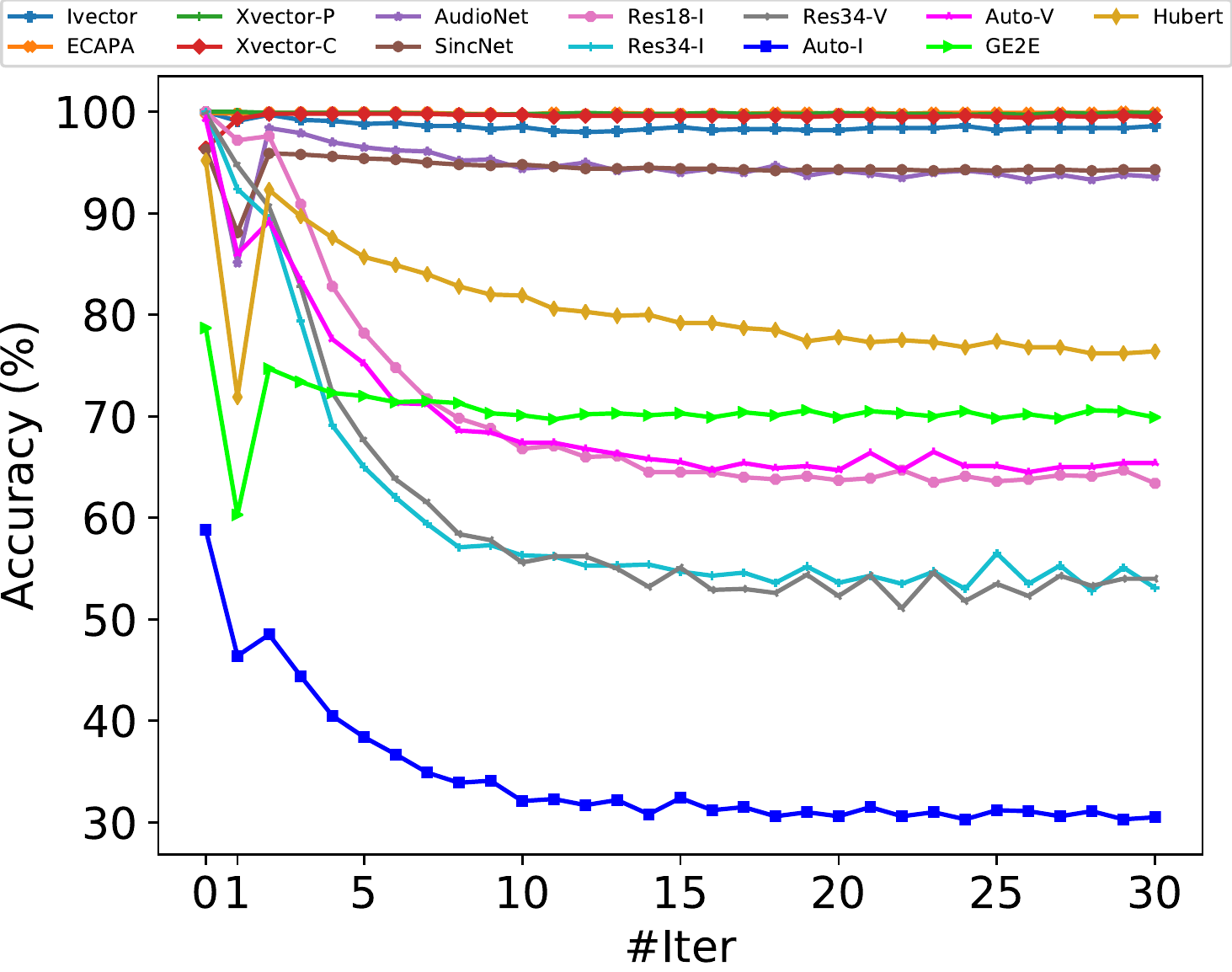}\vspace{-2mm}
        \caption{Res18-V}
    \end{subfigure}
    \begin{subfigure}[t]{0.187\textwidth}
        \centering
        \includegraphics[width=1\textwidth]{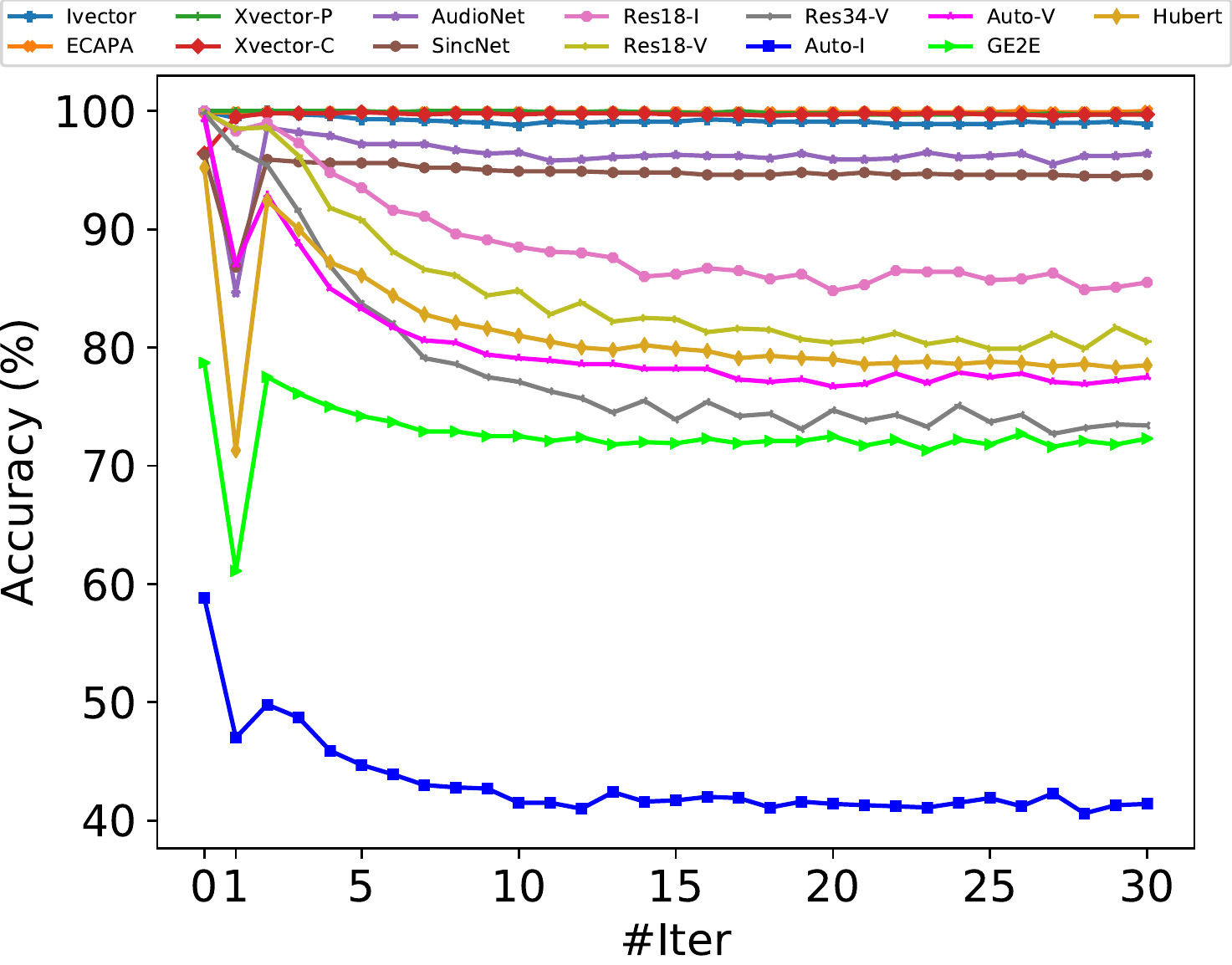}\vspace{-2mm}
        \caption{Res34-I}
    \end{subfigure}
    \begin{subfigure}[t]{0.187\textwidth}
        \centering
        \includegraphics[width=1\textwidth]{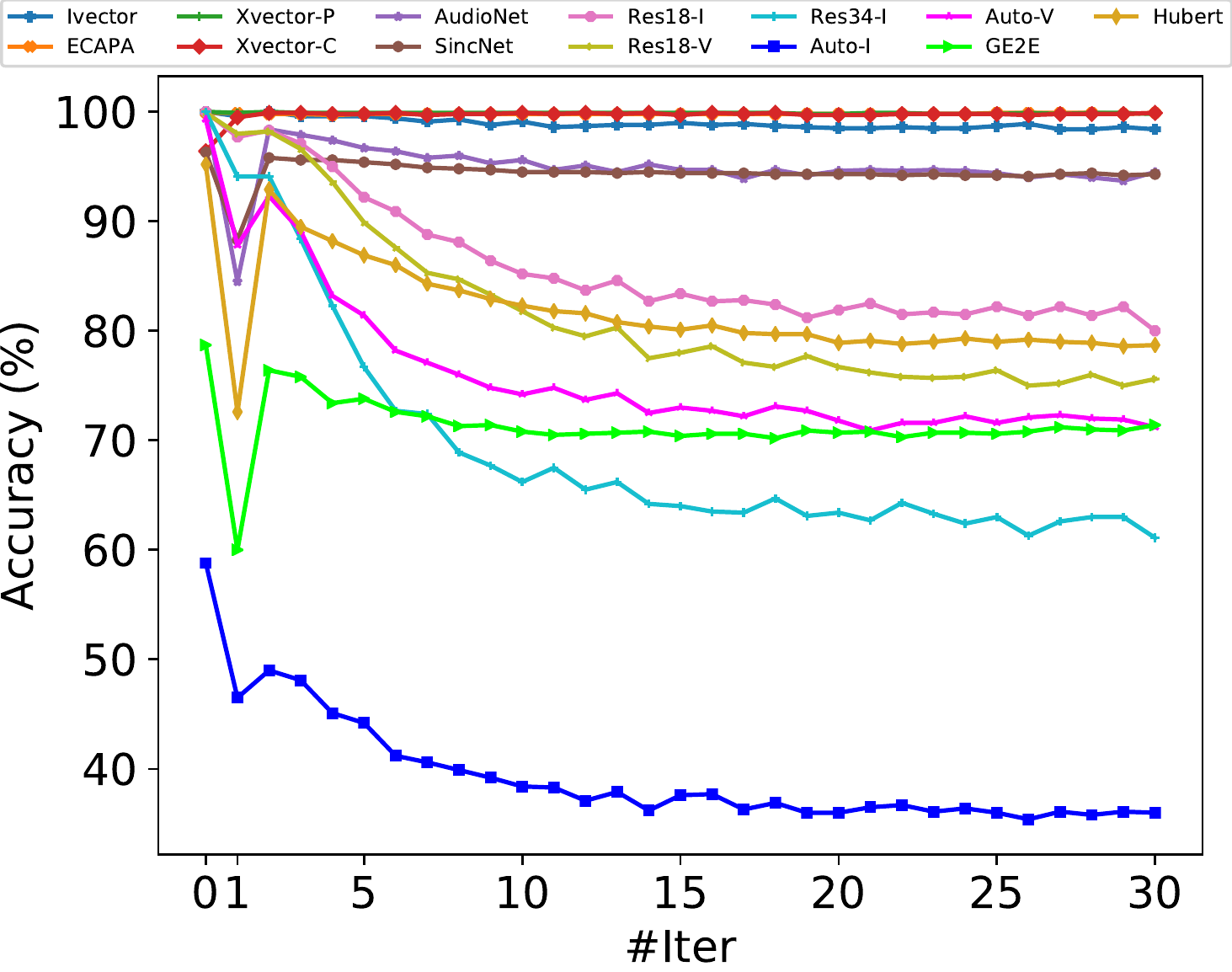}\vspace{-2mm}
        \caption{Res34-V}
    \end{subfigure}
    \begin{subfigure}[t]{0.187\textwidth}
        \centering
        \includegraphics[width=1\textwidth]{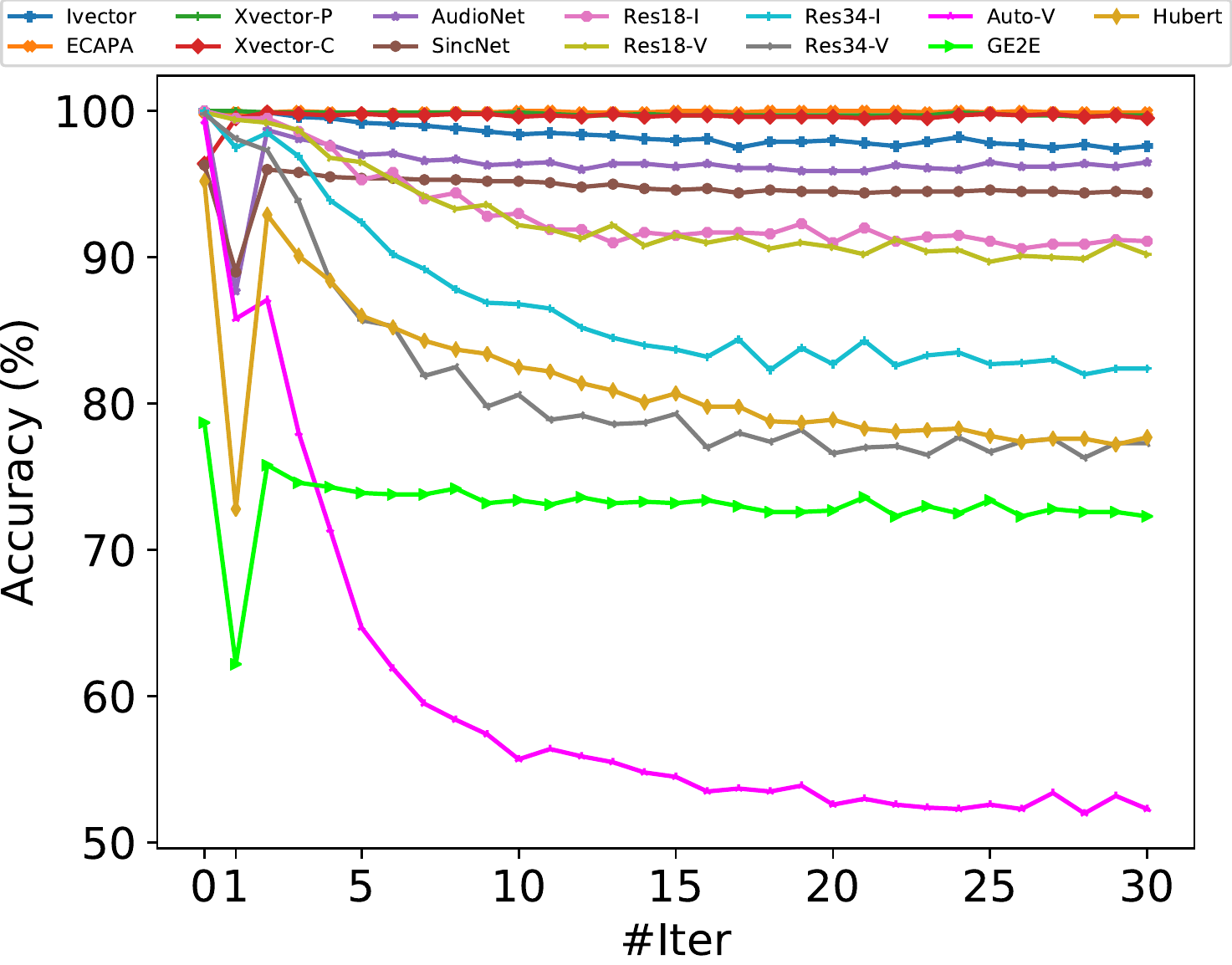}\vspace{-2mm}
        \caption{Auto-I}
    \end{subfigure}
    \begin{subfigure}[t]{0.187\textwidth}
        \centering
        \includegraphics[width=1\textwidth]{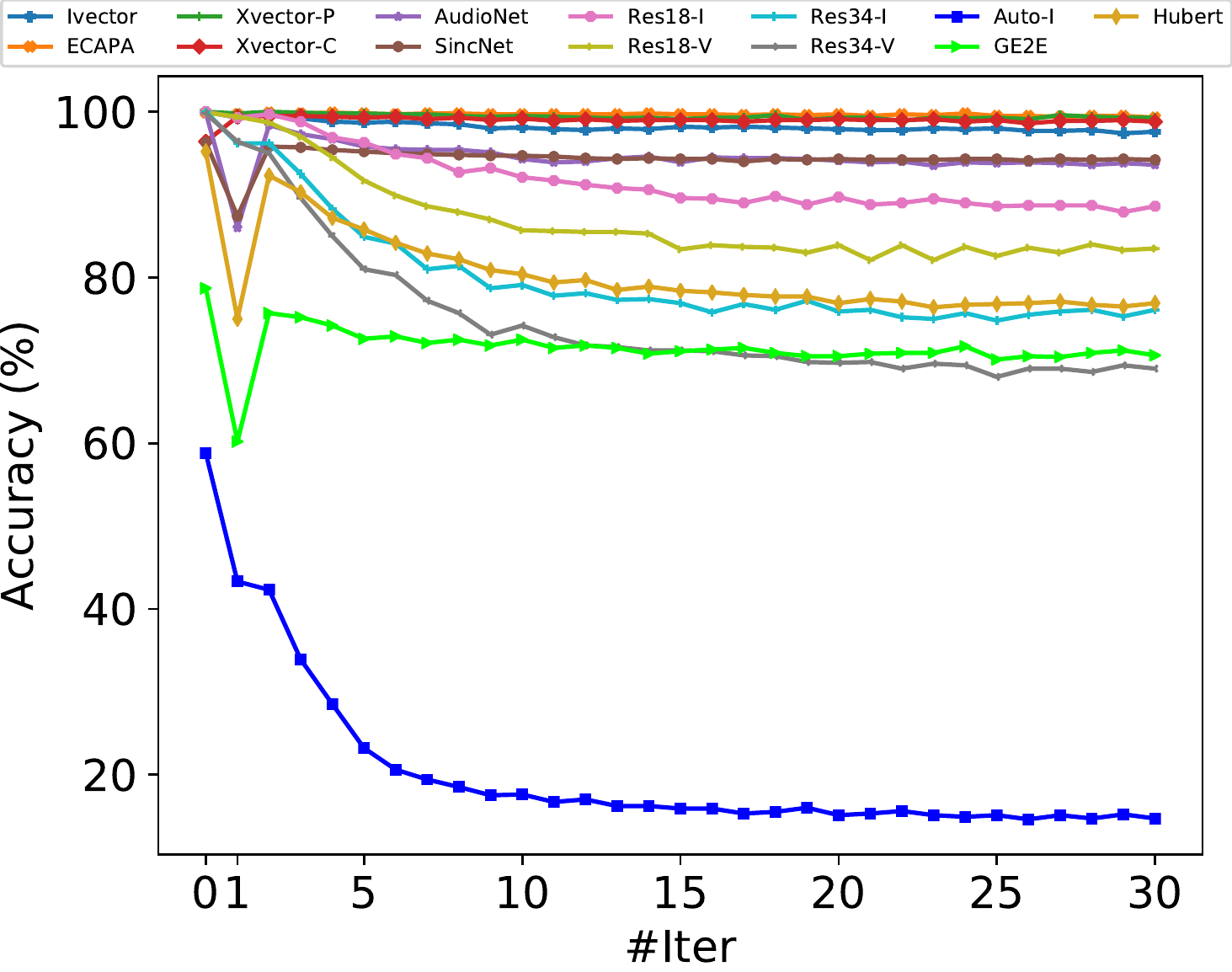}\vspace{-2mm}
        \caption{Auto-V}
    \end{subfigure}
    \begin{subfigure}[t]{0.187\textwidth}
        \centering
        \includegraphics[width=1\textwidth]{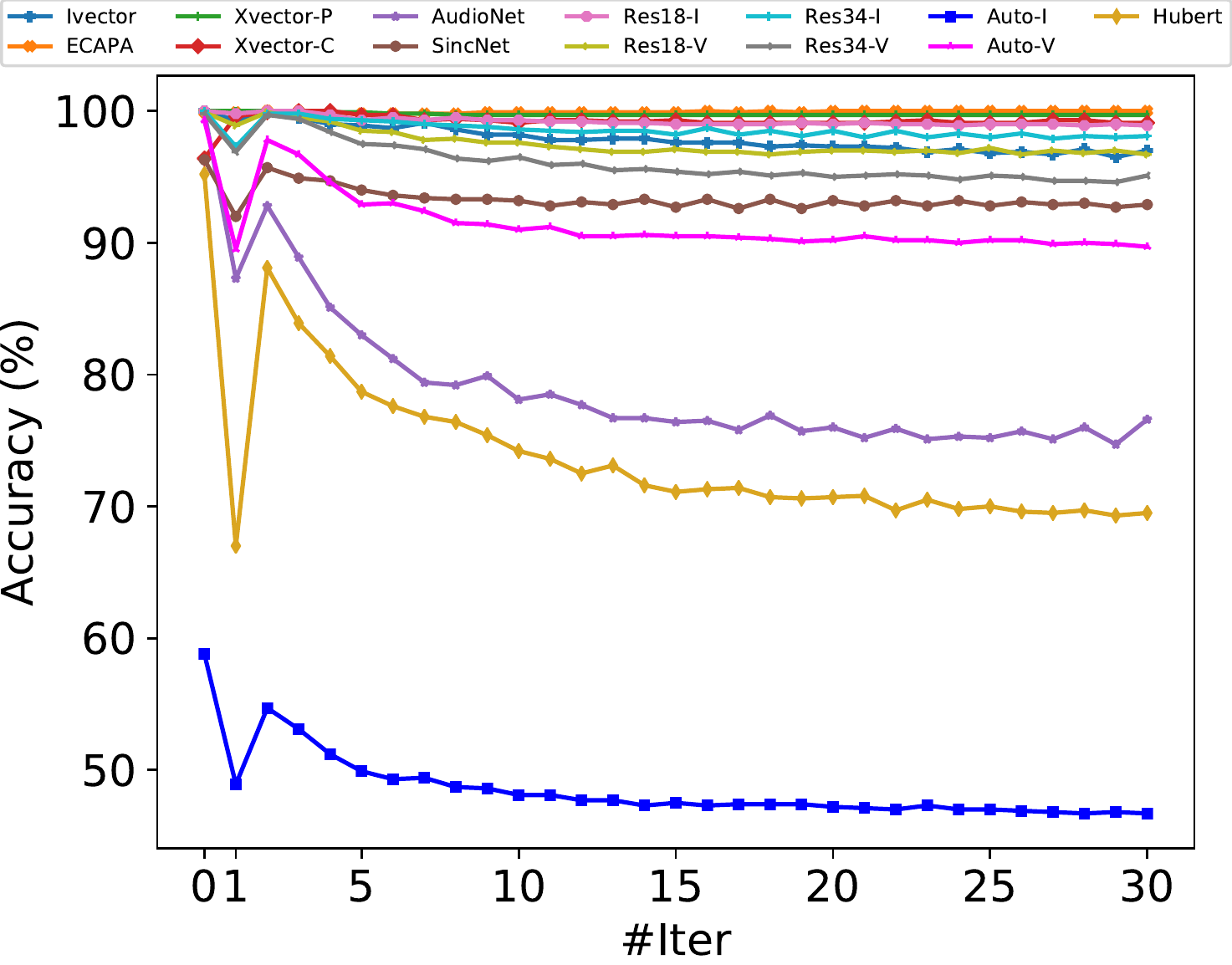}\vspace{-2mm}
        \caption{GE2E}
    \end{subfigure}
    \begin{subfigure}[t]{0.187\textwidth}
        \centering
        \includegraphics[width=1\textwidth]{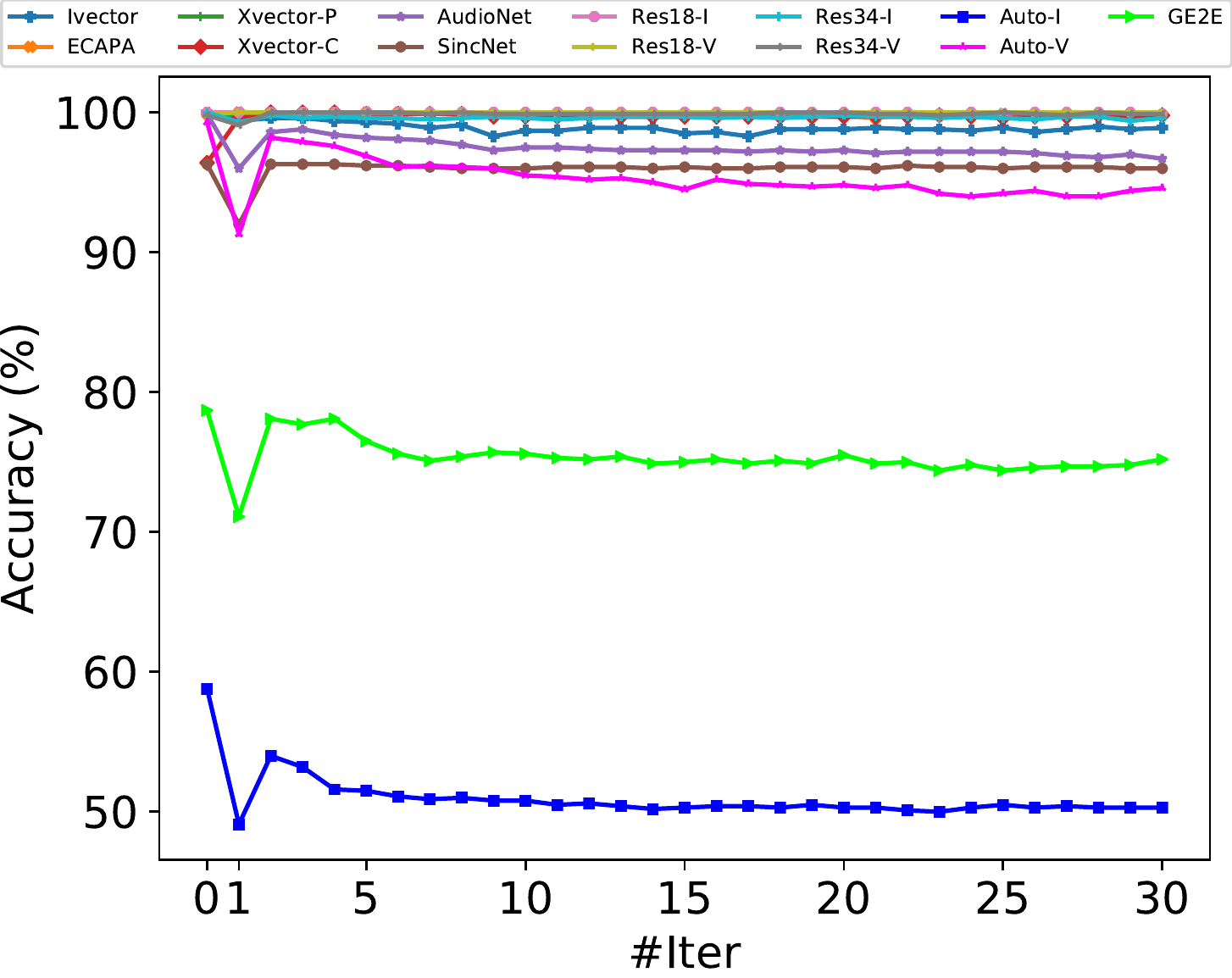}\vspace{-2mm}
        \caption{Hubert}
    \end{subfigure}\vspace{-2mm}
    \caption{Transferability of \attackname using FGSM and PGD 
    w.r.t. \#Iter.
    The subfigures' captions and the legends indicate the source and target models, respectively.
    \#Iter=0 and \#Iter=1 are no attack and \attackname attack with FGSM, respectively.}
    \label{fig:transfer-vs-iteration-PGD}\vspace{-3mm}
\end{figure*}
\subsubsection{Model-specific Factors}\label{sec:transfer-model}
{\bf Evaluation setup}.
We mainly evaluate the transferability of \attackname with FGSM and PGD optimization approaches,
where the perturbation budget $\varepsilon$ is $0.002$,
step\_size $\alpha$ of FGSM (resp. PGD) is $\varepsilon$ (resp. $\frac{\varepsilon}{5}=0.0004$).
To avoid the bias introduced by attack-specific factors,
we vary the number of iterations (i.e., \#Iter) of PGD from 2 to 30 with step 1
and report the best one among FGSM and PGD of all the \#Iter.

\smallskip
\noindent
{\bf Results}. \tablename~\ref{tab:transfer-CSI-Untargeted}
reports the results of \attackname with untargeted attack  on the CSI task, where the source speakers are enrolled speakers, i.e., C8 in \tablename~\ref{tab:s-t-setting}.

We find that Ivector and all the TDNN models transfer to each other quite well,
especially for the transfer attacks {\it Xvector-C} $\to$ {\it Ivector} and {\it Xvector-C} $\to$ {\it Xvector-P},
with 51.5\% and 58.6\% accuracy drop, respectively.

Among CNN-based models, 
ResNet (i.e., Res18/34-I/V) and AutoSpeech (i.e., AutoI/V) transfer to each other quite well with at least 9.4\% accuracy drop,
but they are much less transferable to AudioNet and SincNet, and the vice versa
(except for {\it AudioNet} $\to$ {\it Auto-I} and {\it SincNet} $\to$ {\it Auto-I}).
This gap is possibly due to:
1) difference between training datasets: ResNet and AutoSpeech are trained on VoxCeleb1, while AudioNet and SincNet are on LibriSpeech and TIMIT, respectively;
2) difference between input types: the input type of ResNet and AutoSpeech is spectrogram while
the input types of AudioNet and SincNet are fBank and raw waveform, respectively.
Indeed, the most transferable model to AudioNet is GE2E which has the same input type.

We find that the transferability from CNN models to TDNN models is rather limited,
indicating the adversarial examples often do not transfer well to the target models
with different architectures from the source model.

Remarkably, the transferability between two models are not necessarily symmetric,
even they have the same architecture.
For instance, 
the accuracy drop of {\it SincNet} $\to$ {\it AudioNet} (same architecture) is over 20\%,
while the accuracy drop of {\it AudioNet} $\to$ {\it SincNet} is merely 8.1\%;
the accuracy drop of {\it SincNet} $\to$ {\it GE2E} (different architecture) is 36.6\%,
while the accuracy drop of {\it GE2E} $\to$ {\it SincNet} is less than 5\%.

\chengk{To understand asymmetry of transferability,
consider the source SRS model with parameter $\theta_s$, the target SRS model with parameter $\theta_t$,
the original voice $x$, and the adversarial perturbation $\delta$ crafted against the source model,
the loss attained by the target model on the adversarial voice $x'=x+\delta$, denoted by $\mathcal{L}_{t}^{x'}$,
can be  simplified through a linear approximation as:
$\mathcal{L}_{t}^{x'} = \mathcal{L}_{t}^{x} + \delta^{T}\nabla_x\mathcal{L}_{t}^{x}.$}

\chengk{For $L_\infty$ attack, $\delta=\varepsilon \frac{\nabla_x \mathcal{L}_s^x}{\|\nabla_x \mathcal{L}_s^x\|_\infty}$, where $\mathcal{L}_s^x$ is the loss attained by the source model on the original voice.
Then, according to Cauchy-Schwartz inequality, we can derive the upper bound of the change of loss as follows:
\begin{align*}
  \delta^{T}\nabla_x\mathcal{L}_t^x = \varepsilon \frac{(\nabla_x\mathcal{L}_s^x)^T}{\|\nabla_x \mathcal{L}_s^x\|_\infty}\nabla_x\mathcal{L}_t^x \leq \varepsilon \|\nabla_x \mathcal{L}_t^x\|_1
\end{align*}
Obviously, the upper bound of the change of the loss is positively co-related with $\|\nabla_x \mathcal{L}_t^x\|_1$, called the input gradient size in \cite{why-transfer}.
\tablename~\ref{tab:input-gradient-size} shows the input gradient size of all the models.
We can observe that the input gradient size of AudioNet and GE2E is much larger than that of SincNet,
which results in asymmetry of transferability.}


%


We also report the results of \attackname with randomly chosen target speakers on the CSI task in \tablename~\ref{tab:transfer-CSI-targeted}, where the source speakers are enrolled speakers, i.e., C6 in \tablename~\ref{tab:s-t-setting}.
We can draw a similar conclusion.
However, the best attack {\it Xvector-C} $\to$ {\it Xvector-P} only achieves 17.8\% success rate for targeted attack,
compared to 58.6\% for untargeted attack.
This indicates that targeted transfer attack is much difficult than untargeted transfer attack.

\begin{tcolorbox}[size=title,breakable]
    \textbf{Remark 5.}
    Transfer attack is  effective in general when models have the same architecture,
   but the same architecture is neither sufficient nor necessary, 
    as the transferability is also affected by the training dataset and input type.
\end{tcolorbox}

We also evaluated the transferability of \attackname on the OSI and SV tasks, namely, C5 and C10, from which
 a similar conclusion can be drawn. Therefore, the results are omitted here and reported in our technical report~\cite{AS2T}.

%

\subsubsection{Attack-specific Factors}\label{sec:transfer-attack}

According to the above results,
we study the impact of the number of iterations, step\_size, and perturbation budget on the transferability of \attackname
with the untargeted attack on CSI task, where the source speakers are enrolled speakers, i.e., C8 in \tablename~\ref{tab:s-t-setting}.

\smallskip\noindent{\bf The number of iterations (\#Iter).}
We craft adversarial examples using \attackname with the PGD and FGSM optimization approaches,
where $\varepsilon$=0.002, $\alpha$=$\varepsilon/5$ for PGD, and $\alpha$=0.002 for FGSM.
We vary \#Iter from 2 to 30 for PGD with step 1.

The results are plotted in \figurename~\ref{fig:transfer-vs-iteration-PGD}.
In general, the transferability rate increases with \#Iter.
This is possibly because large \#Iter makes adversarial examples
far from the decision boundary, hence can fool the other models with high probability.
However, for some pairs of source and target models,
especially when they have different architectures,
increasing \#Iter fails to improve the transferability.

We also find that the transferability of PGD with small \#Iter is lower than that of FGSM.
With the increase of \#Iter, the transferability of PGD is higher than that of FGSM
on some pairs of source and target models (e.g., {\it Ivector} $\to$ {\it Xvector-C}),
but not on some pairs (e.g., {\it Xvector-P} $\to$ {\it Hubert}).
The latter is reasonable since FGSM tends to craft adversarial examples with larger perturbation.
The former is interesting since it differs from the conclusion in the image domain
that multiple iterations attack is less transferable than single iteration attack~\cite{kurakin2016adversarial}.

\smallskip\noindent{\bf The step\_size ($\alpha$).}
To explore the effect of step\_size of PGD on the transferability,
we choose two pairs of source and target models
according to the results in \tablename~\ref{tab:transfer-CSI-Untargeted}:
{\it Xvector-C} $\to$ {\it Xvector-P} and {\it Hubert} $\to$ {\it Res34-V},
where the former (resp. latter) achieves the best (worst) transferability
among all the pairs in \tablename~\ref{tab:transfer-CSI-Untargeted}.
We set $\varepsilon=0.002$ and vary $\alpha$ from $\frac{\varepsilon}{\text{\#Iter}_b}$ to $\varepsilon$
where $\text{\#Iter}_b$ is the number of iterations
yielding the best transferability in \figurename~\ref{fig:transfer-vs-iteration-PGD}.
$\text{\#Iter}_b=22$ for {\it Xvector-C} $\to$ {\it Xvector-P} and $\text{\#Iter}_b=23$ for {\it Hubert} $\to$ {\it Res34-V}.

\begin{figure}[t]\centering
      \begin{subfigure}[t]{0.24\textwidth}
          \includegraphics[width=.92\textwidth]{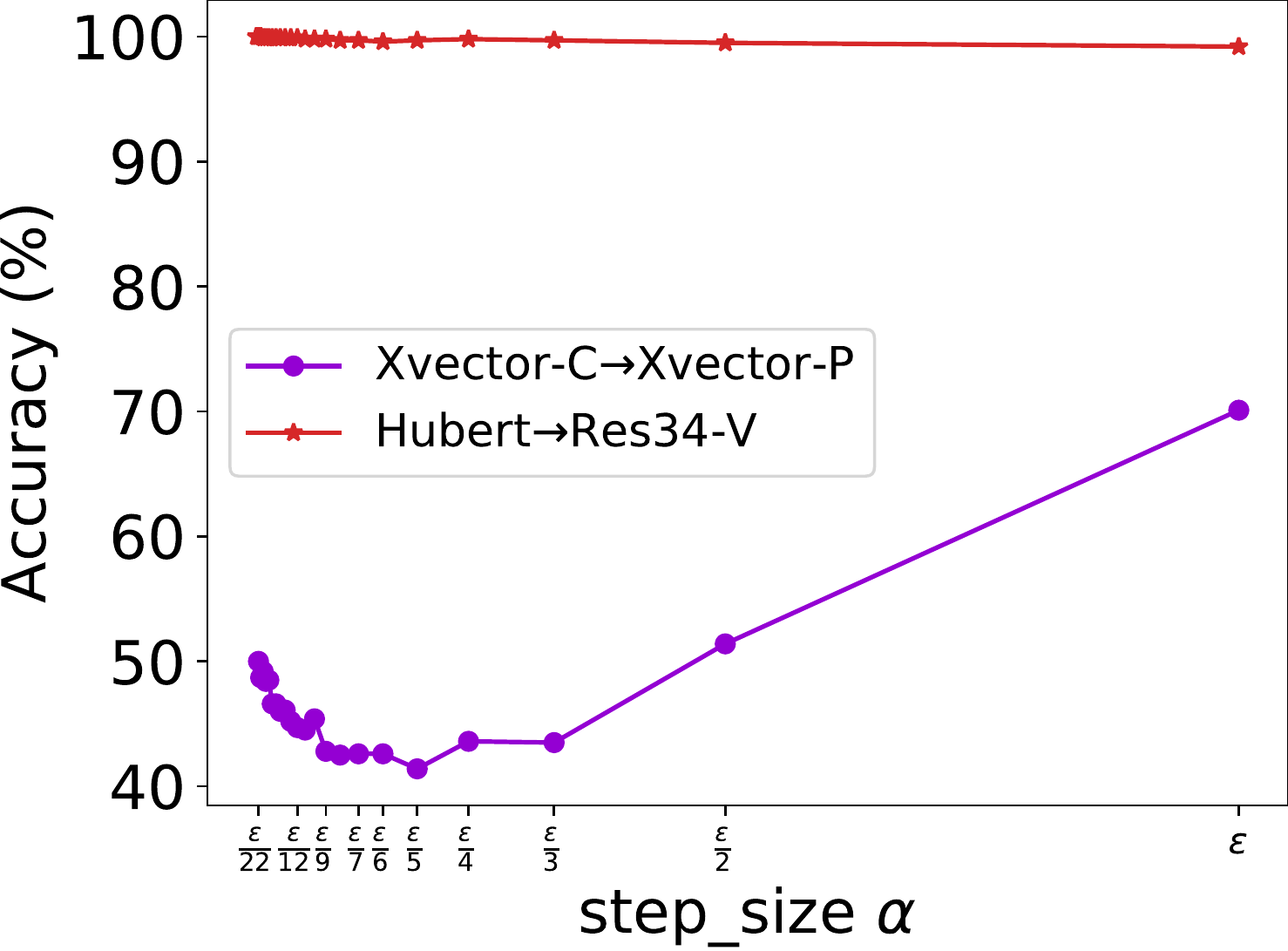}
          \caption{Accuracy w.r.t. $\alpha$} 
          \label{fig:transfer-vs-step-size-PGD}
      \end{subfigure}\hspace{.5mm}
      \begin{subfigure}[t]{0.24\textwidth}
        \includegraphics[width=1\textwidth]{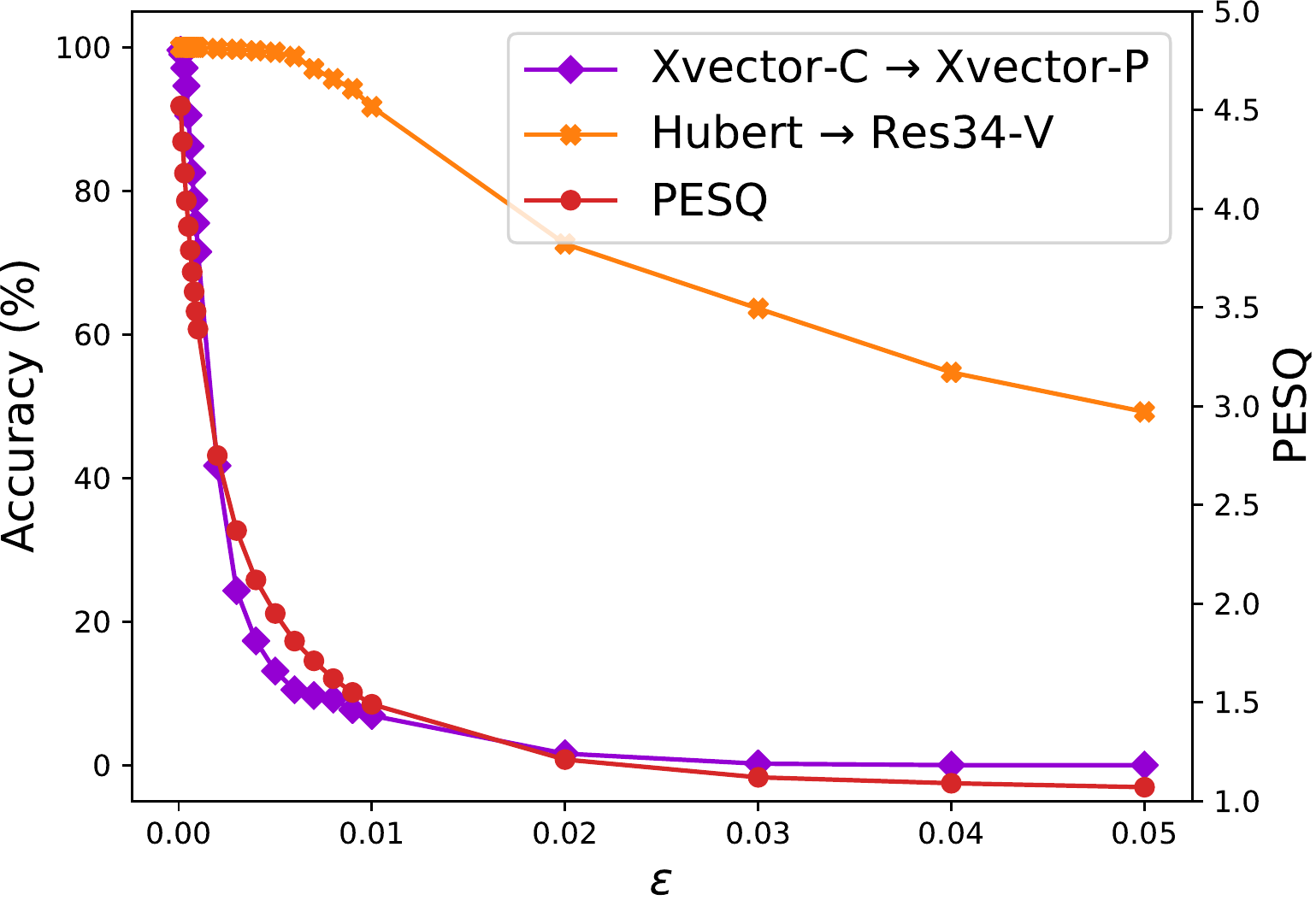}
        \caption{Accuracy and stealthiness (PESQ) w.r.t. $\varepsilon$} 
        \label{fig:transfer-vs-epsilon-PGD}
    \end{subfigure}\vspace{-2mm}
    \caption{Transferability of \attackname w.r.t. step\_size $\alpha$ and  perturbation budge $\epsilon$}\vspace{-5mm}
  \end{figure}

The results are plotted in \figurename~\ref{fig:transfer-vs-step-size-PGD}.
We find that the step\_size is a crucial parameter for {\it Xvector-C} $\to$ {\it Xvector-P},
where too small or too large $\alpha$ will harm the transferability of {\it Xvector-C} $\to$ {\it Xvector-P}.
However, it seems that $\alpha$ has negligible effect on the transferability of {\it Hubert} $\to$ {\it Res34-V}.

\smallskip\noindent{\bf The perturbation budget ($\varepsilon$).}
We study the impact of the budget $\varepsilon$ on the transfer attacks {\it Xvector-C} $\to$ {\it Xvector-P}
and {\it Hubert} $\to$ {\it Res34-V},
%
with $\text{\#Iter}=22$ 
and $\text{\#Iter}=23$ respectively, 
$\alpha=\frac{\varepsilon}{5}$, and $\varepsilon$ ranging from 1e-4 to 0.05,
according to the results in \figurename~\ref{fig:transfer-vs-step-size-PGD}.

The results are plotted in \figurename~\ref{fig:transfer-vs-epsilon-PGD}.
The accuracy 
decreases with the increase of the perturbation budget,
though the first attack drops more quickly.
However, the PESQ also decreases with the increase of $\varepsilon$.
\chengk{To validate whether PESQ is consistent with human perception,
we conduct a human study (cf. section~\ref{sec:human-study}) and the results confirmed this.
Therefore, we can conclude that transferability rate can be improved by allowing large perturbation budget,
but at the cost of sacrificing the stealthiness.}

\subsection{\chengk{Human Study}}\label{sec:human-study}
\chengk{To demonstrate that the perceptual objective metric PESQ is consistent with human perception in quantifying the stealthiness of adversarial voices, we conduct a human study on MTurk~\cite{MTurk}
under the approval of the Institutional Review Board (IRB) of our institutes.}

\smallskip\noindent
\chengk{{\bf Setup of human study.}
We recruit participants from MTurk and ask them to tell whether the voices in a pair are uttered by the same speaker (The three options are {\it same}, {\it different}, and {\it not sure}).
Specifically, we randomly select 4 speakers (2 male and 2 female), and randomly choose 1 normal voice per speaker (called reference voice).
Then for each speaker, we randomly select 1 normal voice with different text from reference voice,
1 distinct adversarial voice per perturbation budget (we set $\varepsilon=0.05,0.01,0.001$) that are crafted from other normal voices of the same speaker,
and 1 normal voice from other speakers with the same gender.
Together, we build 24 pairs of voices: 4 pairs are {\it normal pairs} (one reference voice and one normal voice from the same speaker),
4 pairs are {\it other pairs} (one reference voice and one normal voice from another speaker)
and 16 pairs are {\it adversarial pairs} (one reference voice and one adversarial voice from the same speaker; 4 pairs per perturbation budget).}

\chengk{To guarantee the quality of our questionnaire and validity of the results, we filter out the questionnaires that are randomly chosen by participants.
In particular, we insert three pairs of voices, where each pair contains one male voice and one female voice as a concentration test.
Only when all of them are correctly answered (i.e, the answer {\it different} is selected), we regard it as a valid questionnaire, otherwise, we exclude it.}

\begin{figure}[t]
\centering
  \includegraphics[width=0.36\textwidth]{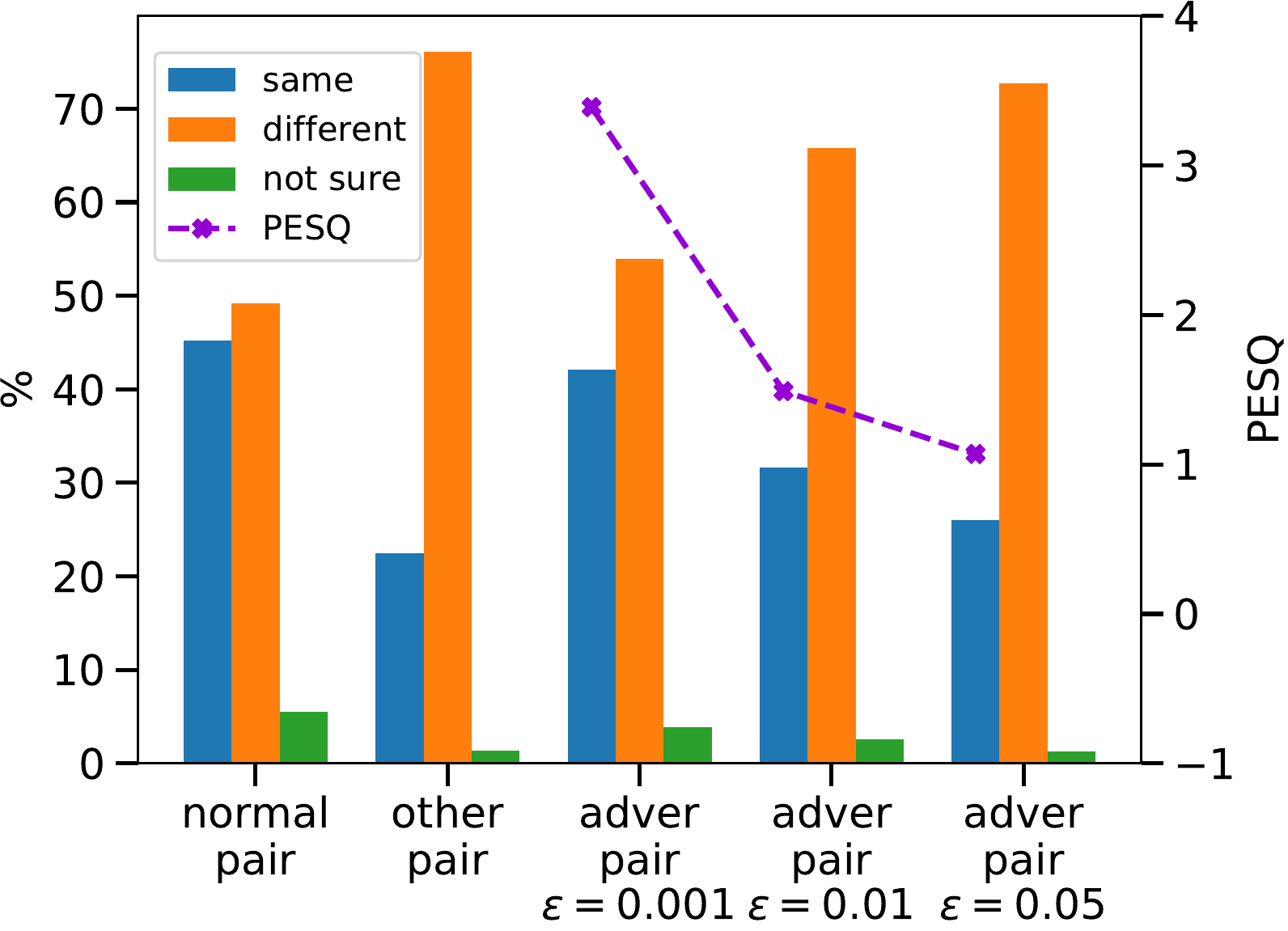}\vspace*{-2mm}
  \caption{Result of human study where adver is short for adversarial.}\vspace*{-2mm}
  \label{fig:human-study}
\end{figure}

\smallskip
\noindent
\chengk{{\bf Results of human study.} We finally received 100 questionnaires where 13 questionnaires are filtered out as they failed to pass our concentration tests. Therefore, there are 77 valid questionnaires.
The results of the human study are shown in \figurename~\ref{fig:human-study}.
76.1\% of participants believe that voices in each other pair are uttered by different speakers, indicating the quality of collected questionnaires.
For the adversarial pairs with $\varepsilon=0.001$, 42.1\% of participants believe that voices in each pair are uttered by the same speaker, very close to the baseline 45.2\% of normal pairs.
However, with the increase of $\varepsilon$, more participants think that voices in each pair are uttered by different speakers.
We note that the PESQ also decreases with the increase of $\varepsilon$, indicating that PESQ is consistent with human perception to some extent in quantifying the stealthiness.}

\begin{tcolorbox}[size=title,breakable]
    \textbf{Remark 6.}
    Increasing the number of iterations and perturbation budget improves transferability 
    at the cost of sacrificing stealthiness.
    However, for some pairs of source and target models, strengthening the attack-specific factors is ineffective in improving transferability,
    indicating that model-specific factors are dominant factors over attack-specific ones.
\end{tcolorbox}

\section{Related Work}
Adversarial attacks recently have attracted intensive attention in various domains, e.g., \cite{LiuSZW20,GuoWZZSW21,ZhangZCSC21,bu2019taking,zhao2021attack,SongLCFL21}.
In this section, we mainly discuss related works in the speaker recognition (SR) domain.

\smallskip
\noindent
{\bf Adversarial attacks}.
There exist both white-box 
and black-box attacks 
in the SR domain.
Existing white-box attacks  utilize FGSM and PGD from the image domain to create adversarial voices. 
The works in \cite{jati2021adversarial,shamsabadi2021foolhd,DBLP:journals/corr/abs-1711-03280}
demonstrated their attacks for C6/C8 (cf. \tablename~\ref{tab:s-t-setting}),
while other white-box attacks~\cite{abs-1801-03339, li2020adversarial, zhang2021attack, LiZJXZWM020, xie2020enabling, WangGX20}
focused on the SV task with C10 only.

FAKEBOB~\cite{chen2019real} and SirenAttack~\cite{du2020sirenattack} are two black-box attacks targeting SRSs.
SirenAttack~\cite{du2020sirenattack} only considers C6/C8 
and is less effective than FAKEBOB~\cite{chen2019real},
thus we do not integrate it into \attackname.
FAKEBOB~\cite{chen2019real} covers more settings than the above works 
including C2, C6, and C10.

Compared over those works, we highlight the following three contributions.
(i) To our knowledge, \attackname is the first attack
that covers all the combinations of the source
and target speakers on the three tasks (i.e., C1-C10 in \tablename~\ref{tab:s-t-setting}) and is capable to launch arbitrary source-to-target speaker adversarial attack
by utilizing various loss functions. In contrast, prior works consider only a few settings.
We remark that each setting (i.e., C1-C10) is meaningful
since it can be used to achieve at least one goal by the adversary.
(ii)  \attackname is applicable to different levels of knowledge about the victim model
by integrating our loss functions into any optimization approaches, e.g.,
white-box FGSM, PGD, and CW, and black-box FAKEBOB,
while prior works are limited to either white-box or black-box attacks.
(iii) We explore different choices of loss functions and conduct a thorough evaluation by which
we find the optimal loss functions leading to the most effective and efficient attacks.
We remark that the loss function for C2 (resp. C8) adopted by FAKEBOB
(resp. \cite{jati2021adversarial} and \cite{DBLP:journals/corr/abs-1711-03280})
achieved inferior performance than our optimal loss functions.

Some previous works also demonstrate the practicability of transfer attacks,
where the source and target models differ in at least one of the following factors:
architecture, input type, training dataset, and scoring method.
All the existing works consider only two architectures, namely,
GMM and TDNN in \cite{chen2019real,li2020adversarial}, CNN and TDNN in \cite{jati2021adversarial},
and LSTM and TDNN in \cite{xie2020enabling}.
In this work, we consider all these four architectures and an additional Transformer.
The works in \cite{chen2019real, abs-1801-03339} train the SR models using two datasets,
while our work involve five training datasets.
The works in \cite{chen2019real,abs-1801-03339,li2020adversarial} cover two input types,
e.g., MFCC and PLP in \cite{chen2019real}, MFCC and fBank in \cite{abs-1801-03339},
and MFCC and spectrogram in \cite{li2020adversarial},
while this work covers more input types, including waveform, spectrogram, fBank, and MFCC.
In addition, our work covers both PLDA and COSS as the scoring method,
while previous works merely consider one.
In summary, we performed thus far the largest-scale transferability analysis in the speaker recognition domain among 14 SR models.
We also evaluated the impact of attack-specific factors on the transferability, which has not been considered in the previous works.

\smallskip
\noindent
{\bf Robust over-the-air adversarial attacks}.
To enhance the robustness of adversarial voices and enable physical over-the-air attack,
prior works either simply craft adversarial voices with high-confidence~\cite{chen2019real}
or integrate the distortions occurred in the over-the-air transmission into the
generation process of adversarial voices~\cite{hang-practical,xie2021real,xie-universal,LiW00020}.

Chen et al.~\cite{chen2019real} argue that the high sensitivity of adversarial voices to the distortions
incurred by over-the-air transmission is attributed to their closeness from the decision boundary.
Motivated by this, they 
craft high-confidence adversarial voices using FAKEBOB
and show that these adversarial voices
are robust against over-the-air distortion,
and are device-/environment-independent to some extent.

To improve the robustness of adversarial voices against reverberation in physical environment,
\cite{hang-practical,xie2021real,xie-universal,LiW00020}
handle reverberation  by pre-coding RIR into the loss function,
where \cite{xie2021real,xie-universal} use simulated RIR,
while \cite{LiW00020,hang-practical} use real-world RIR.
Additionally, \cite{LiW00020} uses isotropic noise and band-pass filter
to cope with the ambient noise and equipment distortion
and incorporate them into the loss function.

Our solution towards robust adversarial voices under the over-the-air setting
is similar to \cite{hang-practical,xie2021real,xie-universal,LiW00020} except that:
(i) although we do not model the equipment distortion as in \cite{LiW00020},
we find that this distortion is not substantial
and modeling the reverberation and ambient noise is sufficient.
(ii) we incorporate random noise with different levels of SNR
to simulate different acoustic environments in the real-life scenarios,
which enables our attack to remain effective under various physical environments.

Regrading the evaluation of over-the-air attack,
\cite{chen2019real,hang-practical,LiW00020} adopt real-world evaluation,
i.e. repeatedly playing and recording voices by hand,
while \cite{xie2021real,xie-universal} and our work use a simulated manner.
Compared to real-world evaluation, simulated evaluation makes it possible to perform a large-scale and thorough evaluation,
covering different scenarios such as attacking scenes, acoustic environments with various ambient noises,
hardware devices, and positions of the adversary and the victim.
Compared to \cite{xie2021real,xie-universal} which only consider reverberation in the simulated evaluation,
our work additionally considers equipment distortions and four types of real-life representative ambient noises with various SNR,
making our simulated evaluation more realizable in the real-world evaluation.
We hope that the simulated evaluation manner in this work will serve as a benchmark
for evaluating over-the-air attack for future works.


\section{Conclusion and Future Work}

We proposed \attackname, the first attack in speaker recognition domain
that covers all the combinations of source and target speakers on all the three tasks.
It features novel source-/target-oriented loss functions
and enables the adversary to create adversarial voices using arbitrary source and target speakers
to achieve various goals in diverse attack scenarios.
The loss functions can be freely composed with both white-box and black-box optimization approaches
to adapt to the adversary's knowledge about the target model.

We improved the robustness of \attackname
for launching over-the-air attacks in the physical world 
by utilizing various parameterized transformation functions to model diverse distortions and incorporate them into the generation of adversarial voices.
The effectiveness of our approach is confirmed by a thorough evaluation.

We leveraged \attackname to conduct thus far the larger-scale transferability analysis among 14 SR models,
covering five architectures, five training datasets, four input types, and two scoring backends.
The transferability analysis reveals how model-specific and attack-specific factors
impact the transferability of \attackname and results in many useful findings and insights.

Our work motivates the following future works.
(i) The transferability rate is limited in some cases, probably because the loss functions
tailored for non-transfer attack are not optimal for transfer attack.
One future work is to explore better loss functions for transfer attack.
(ii) The transferability between models may be asymmetric, contradicting the decision boundary similarity based explanation of   transferable
adversarial examples in the image domain, \chengk{because} similarity is a systematical metric.
An interesting future work is to study explanation that is consistent with this asymmetric phenomenon.
(iii) Another future work is to investigate defense solutions against adversarial attacks in the speaker recognition domains under various settings.

\section*{Acknowledgment}
This work was supported by the National Natural Science Foundation of China (NSFC)
under Grant No. 62072309, Ant Group and the National Key Research and Development Program under Grant No. 2020AAA0107800.

\ifCLASSOPTIONcaptionsoff
  \newpage
\fi



\begin{IEEEbiography}[{\includegraphics[width=1in,clip,keepaspectratio]{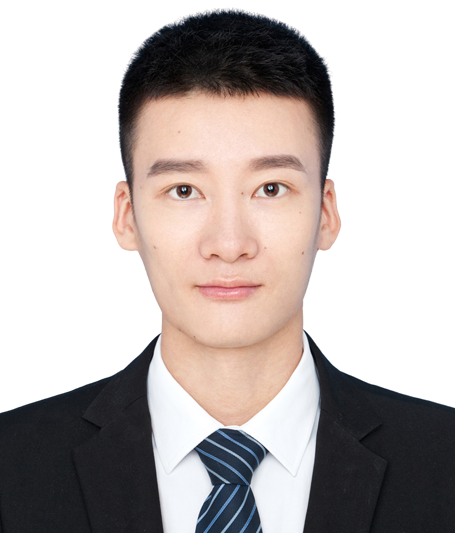}}]{Guangke Chen}
received his BEng degree from South China University of Technology, Guangzhou, China, in 2019.
He is currently pursuing the Ph.D. degree with ShanghaiTech University, advised by Dr. Song. 
His research interest lies in the area of multimedia and machine learning security and privacy.
He is currently doing research on the security issues of speaker and speech recognition systems.
More information is available at \url{http://guangkechen.site/}.
\end{IEEEbiography}

\begin{IEEEbiography}[{\includegraphics[width=1in,clip,keepaspectratio]{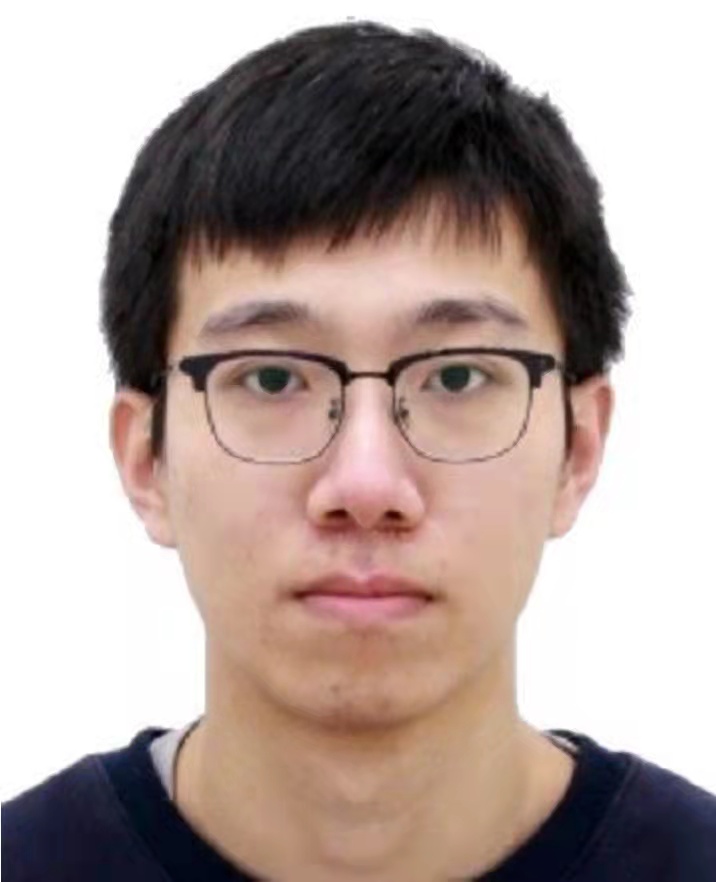}}]{Zhe Zhao}
received his B.S. degree from Ocean University of China, Tsingtao, China, in 2016.
From 2016 to 2018, he was a software engineer at Hewlett-Packard Company.
Now he is a Ph.D. student at School of Information Science and Technology, ShanghaiTech University.
His research interest lies in the area of software engineering and testing.
He is currently doing research in trusted artificial intelligence.
His supervisor is Dr. Song.
\end{IEEEbiography}

\begin{IEEEbiography}[{\includegraphics[width=1in,height=1.25in,clip,keepaspectratio]{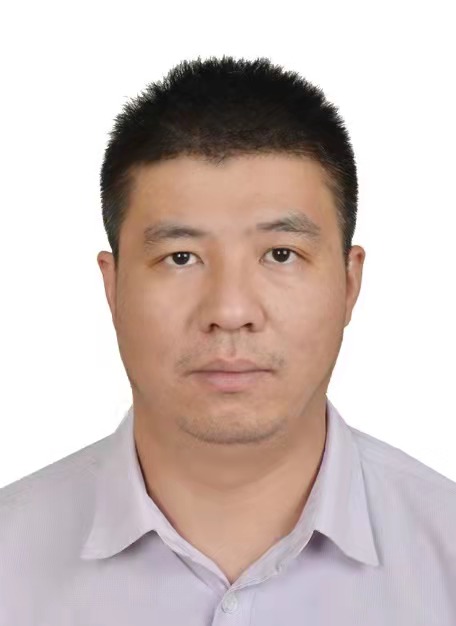}}]{Fu Song}
    received the B.S. degree from Ningbo
    University, Ningbo, China, in 2006, the M.S. degree from East China Normal University, Shanghai,
    China, in 2009, and the Ph.D. degree in computer science from University Paris-Diderot,
    Paris, France, in 2013.
    From 2013 to 2016, he was a Lecturer and Associate Research Professor at East China Normal University.
    From August 2016 to July 2021, he is an Assistant Professor with ShanghaiTech University, Shanghai, China.
    Since July 2021, he is an Associate Professor with ShanghaiTech University.
    His research interests include formal methods and computer/AI security. 
    \end{IEEEbiography}

\begin{IEEEbiography}[{\includegraphics[width=1.12in,clip,keepaspectratio]{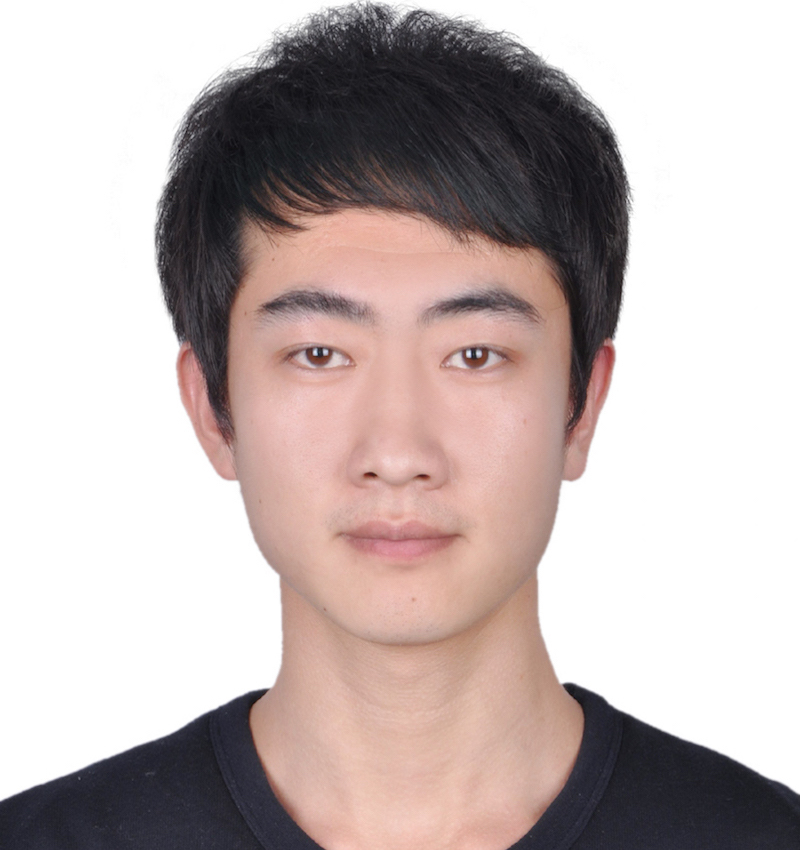}}]{Sen Chen} (Member, IEEE) is an Associate Professor at Tianjin University, China. Before that, he was a Research Assistant Professor at Nanyang Technological University (NTU), Singapore, and a Research Assistant of NTU from 2016 to 2019 and a Research Fellow from 2019-2020. He received his Ph.D. degree in Computer Science  East China Normal University, China, in 2019. His research focuses on Security and Software Engineering. 
More information is available on {\url{https://sen-chen.github.io/}.}
\end{IEEEbiography}

\begin{IEEEbiography}[{\includegraphics[width=1.12in,clip,keepaspectratio]{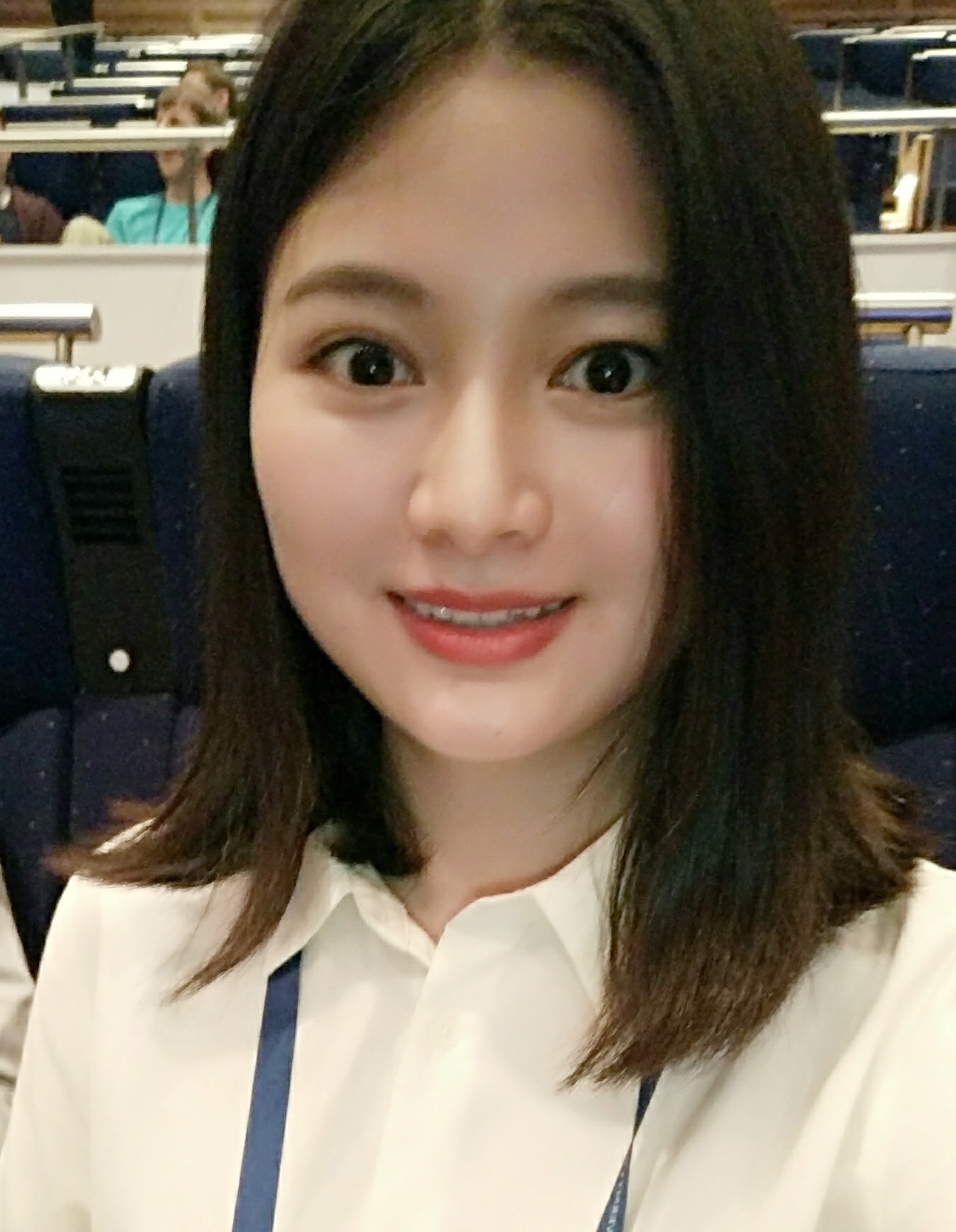}}]{Lingling Fan} is an Associate Professor at Nankai University, China. She received her Ph.D and BEng degrees in computer science from East China Normal University, Shanghai, China in June 2019 and June 2014, respectively. In 2017, she joined Nanyang Technological University (NTU), Singapore as a Research Assistant and then had been as a Research Fellow of NTU since 2019. Her research focuses on program analysis and testing, software security. She got an ACM SIGSOFT Distinguished Paper Award at ICSE 2018.
\end{IEEEbiography}

\begin{IEEEbiography}[{\includegraphics[width=1.12in,clip,keepaspectratio]{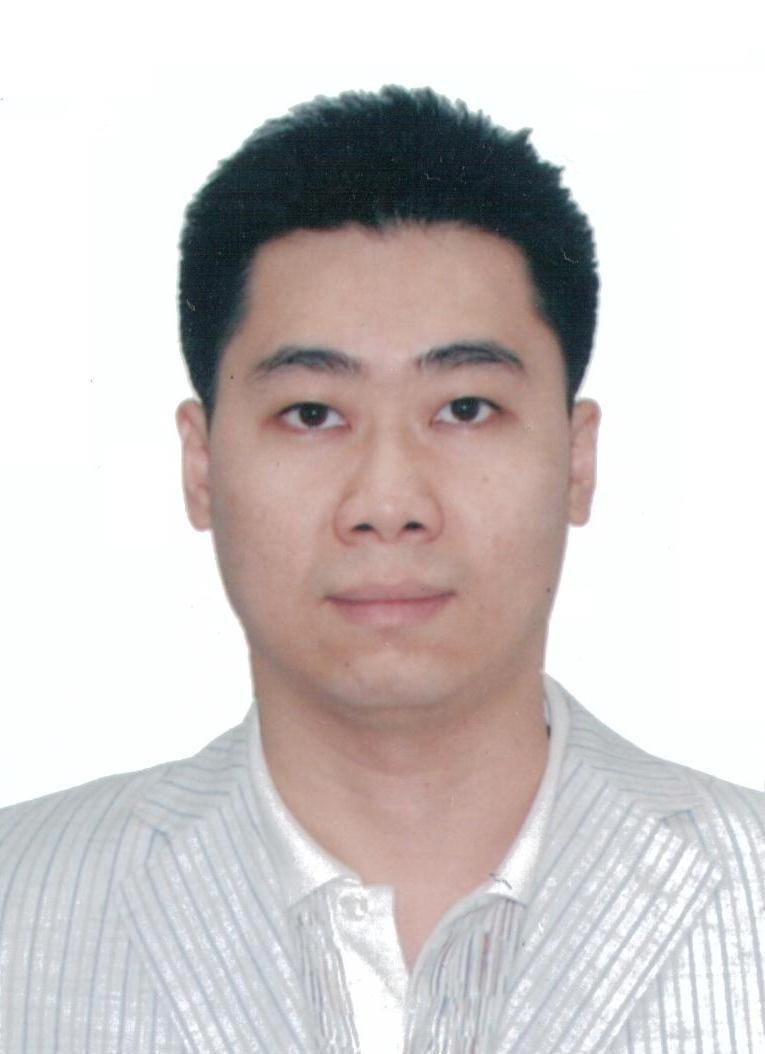}}]{Liu Yang} (Senior Member, IEEE)
graduated in 2005 with a Bachelor of Computing (Honours) in the National University of Singapore (NUS). In 2010, he obtained his PhD and started his post doctoral work in NUS, MIT and SUTD. In 2011, 
In 2012 fall, he joined Nanyang Technological University (NTU) as a Nanyang Assistant Professor. He is currently a full professor and the director of the cybersecurity lab in NTU. He specializes in software verification, security and software engineering. His research has bridged the gap between the theory and practical usage of formal methods and program analysis to evaluate the design and implementation of software for high assurance and security.
He has more than 300 publications and 6 best paper awards in top tier conferences and journals.
\end{IEEEbiography}

\appendix
 
\section{Appendix}
  
\begin{table*}
  \centering\setlength\tabcolsep{3.5pt}
  \caption{Transferability of \attackname with the FGSM and PGD optimization approaches.
  For each pair of source and target systems, 
  the result indicates the increase of FAR of the target system. 
  OSI task, untargeted attack, $s\notin G$.}
  \resizebox{0.95\textwidth}{!}{%
    \begin{tabular}{|c|c|c||c|c|c||c|c|c|c|c|c|c|c||c||c|}
    \toprule
    \multicolumn{2}{|c|}{\diagbox{\bf Source}{\bf $\Uparrow$ FAR (\%)}{\bf Target}} & \textbf{Ivector} & \textbf{ECAPA} & \textbf{Xvector-P} & \textbf{Xvector-C} & \textbf{AudioNet} & \textbf{SincNet} & \textbf{Res18-I} & \textbf{Res18-V} & \textbf{Res34-I} & \textbf{Res34-V} & \textbf{Auto-I} & \textbf{Auto-V} & \textbf{GE2E} & \textbf{Hubert} \\
    \midrule
    {\bf GMM} & \textbf{Ivector} & \cellcolor{gray!40} 95.2 & \textcolor{blue}{\bf 32.6} & \textcolor{blue}{\bf 13.0} & \textcolor{blue}{\bf 8.9} & 3.3   & \textcolor{red}{\bf -2.7} & 0.9   & \textcolor{red}{\bf -1.3} & 0.3   & 0.5   & -0.5  & 4     & \textcolor{red}{\bf -1.4} & 4.3 \\
    \midrule
    \midrule
    \multirow{3}[4]{*}{\bf TDNN} & \textbf{ECAPA} & \textcolor{blue}{\bf 15.8} & \cellcolor{gray!40} 97.8 & \textcolor{blue}{\bf 17.7} & \textcolor{blue}{\bf 10.7} & 1.1   & \textcolor{red}{\bf -3.6} & 1.2   & 0.1   & 3.4   & 4.5   & -0.1  & 5.5   & \textcolor{red}{\bf -1.7} & \textcolor{red}{\bf -0.2} \\
    \cmidrule{2-16}
    & \textbf{Xvector-P} & \textcolor{blue}{\bf 33.0} & \textcolor{blue}{\bf 55.9} & \cellcolor{gray!40} 94.0 & \textcolor{blue}{\bf 21.2} & 2.5   & \textcolor{red}{\bf -3.0} & 0.8   & \textcolor{red}{\bf -1.1} & 0.5   & 1.7   & -0.4  & 3.9   & \textcolor{red}{\bf -1.3} & 2.9 \\
    \cmidrule{2-16}
    & \textbf{Xvector-C} & \textcolor{blue}{\bf 16.3} & \textcolor{blue}{\bf 44.7} & \textcolor{blue}{\bf 21.9} & \cellcolor{gray!40} 90.2 & 0.1   & \textcolor{red}{\bf -3.0} & 1     & -0.5  & 2.2   & 4.9   & -0.4  & 3.4   & \textcolor{red}{\bf -1.9} & \textcolor{red}{\bf -2.8} \\
    \midrule
    \midrule
    \multirow{8}[12]{*}{\bf CNN} & \textbf{AudioNet} & 0.7   & 0.9   & -0.1  & \textcolor{red}{\bf -3.4} & \cellcolor{gray!40} 86.0 & \textcolor{red}{\bf -2.8} & \textcolor{blue}{\bf 1.7} & 0.2   & 1.5   & \textcolor{blue}{\bf 6.1} & 0.1   & \textcolor{blue}{\bf 4.6} & -0.5  & \textcolor{red}{\bf -6.2} \\
    \cmidrule{2-16}
    & \textbf{SincNet} & 3.5   & 4.2   & 0.8   & \textcolor{red}{\bf -3.4} & \textcolor{blue}{\bf 9.0} & \cellcolor{gray!40} 61.5 & 3.6   & 2.5   & \textcolor{blue}{\bf 8.9} & \textcolor{blue}{\bf 8.8} & \textcolor{red}{\bf 0.1} & 2.9   & 5.6   & \textcolor{red}{\bf -4.0} \\
    \cmidrule{2-16}
    & \textbf{Res18-I} & 0.2   & 3.9   & -0.1  & \textcolor{red}{\bf -3.3} & 3.5   & \textcolor{red}{\bf -3.2} & \cellcolor{gray!40} 94.5 & 12.6  & \textcolor{blue}{\bf 16.9} & \textcolor{blue}{\bf 30.9} & 3.1   & \textcolor{blue}{\bf 15.1} & -0.8  & \textcolor{red}{\bf -1.6} \\
    \cmidrule{2-16}
    & \textbf{Res18-V} & 0     & 4.7   & -0.1  & \textcolor{red}{\bf -2.8} & 4.9   & \textcolor{red}{\bf -2.9} & \textcolor{blue}{\bf 22.3} & \cellcolor{gray!40} 93.4 & \textcolor{blue}{\bf 20.5} & \textcolor{blue}{\bf 37.6} & 2.7   & 17.2  & -0.2  & \textcolor{red}{\bf -1.9} \\
    \cmidrule{2-16}
    & \textbf{Res34-I} & 0.1   & 2.6   & -0.2  & \textcolor{red}{\bf -2.6} & 1.4   & \textcolor{red}{\bf -3.2} & \textcolor{blue}{\bf 8.5} & 6.3   & \cellcolor{gray!40} 93.3 & \textcolor{blue}{\bf 25.0} & 1.4   & \textcolor{blue}{\bf 8.7} & -0.4  & \textcolor{red}{\bf -2.1} \\
    \cmidrule{2-16}
    & \textbf{Res34-V} & 0.4   & 3.8   & -0.1  & \textcolor{red}{\bf -3.6} & 2     & \textcolor{red}{\bf -3.2} & 7.1   & \textcolor{blue}{\bf 7.4} & \textcolor{blue}{\bf 13.2} & \cellcolor{gray!40} 95.7 & 1.8   & \textcolor{blue}{\bf 13.3} & -1    & \textcolor{red}{\bf -2.1} \\
    \cmidrule{2-16}
    & \textbf{Auto-I} & 0.5   & 5.1   & -0.2  & \textcolor{red}{\bf -3.6} & 0.8   & \textcolor{red}{\bf -3.5} & 6.4   & 2     & \textcolor{blue}{\bf 7.2} & \textcolor{blue}{\bf 13.3} & \cellcolor{gray!40} 83.0 & \textcolor{blue}{\bf 28.9} & -0.6  & \textcolor{red}{\bf -1.7} \\
    \cmidrule{2-16}
    & \textbf{Auto-V} & 0.9   & 9.8   & 0     & \textcolor{red}{\bf -3.6} & 2.4   & \textcolor{red}{\bf -3.5} & \textcolor{blue}{\bf 13.9} & 7.8   & \textcolor{blue}{\bf 15.5} & \textcolor{blue}{\bf 28.6} & 11.6  & \cellcolor{gray!40} 96.3 & -0.8  & \textcolor{red}{\bf -1.0} \\
    \midrule
    \midrule
    {\bf LSTM} & \textbf{GE2E} & -1    & \textcolor{blue}{\bf 2.1} & -0.3  & \textcolor{red}{\bf -4.4} & \textcolor{blue}{\bf 1.9} & \textcolor{red}{\bf -3.1} & 0.1   & -1.5  & -0.6  & \textcolor{blue}{\bf 0.5} & -0.3  & -1.2  & \cellcolor{gray!40} 69.7 & \textcolor{red}{\bf -1.7} \\
    \midrule
    \midrule
    {\bf Trans} & \textbf{Hubert} & \textcolor{blue}{\bf 0.1} & \textcolor{blue}{\bf 1.9} & \textcolor{red}{\bf -5.9} & \textcolor{red}{\bf -5.6} & \textcolor{blue}{\bf 2.9} & \textcolor{red}{\bf -3.7} & -0.7  & -2.2  & -3.3  & -1.6  & 0     & -0.4  & -0.5  & \cellcolor{gray!40} 81.8 \\
    \bottomrule
    \end{tabular}%
    }
  \label{tab:transfer-OSI}%
\end{table*}%

\subsection{Evaluation of \attackname with More Speakers}
We also evaluate \attackname on the datasets with more speakers, i.e., Spk$_{100}$-enroll,  Spk$_{100}$-test, and  Spk$_{100}$-imposter.
We enroll 100 speakers using the dataset Spk$_{100}$-enroll, forming a speaker group $G$ for the CSI and OSI tasks
and 100 speaker models for the SV task. We mount attacks using two categories of benign voices:
voices uttered by enrolled speakers in
Spk$_{100}$-test ($s\in G$) and voices uttered by unenrolled speakers in
Spk$_{100}$-imposter ($s\notin G$).
To better compare the effectiveness of different loss functions, we consider the PGD optimization approach with \#Iter=2.
The attack success rates of \attackname are reported in \tablename~\ref{tab:basic-OSI-100} and \tablename~\ref{tab:basic-CSI-SV-100}.
The best loss function for each setting is consistence with the results obtained on Spk$_{10}$-enroll, Spk$_{10}$-test, and Spk$_{10}$-imposter.

\begin{table*}
  \centering\setlength\tabcolsep{3.5pt}
  \caption{Transferability of \attackname with the FGSM and PGD optimization approaches. 
  For each pair of source and target systems, 
  the result indicates the increase of FAR of the target system. 
  SV task, $s\notin G$.}
  \resizebox{0.95\textwidth}{!}{%
    \begin{tabular}{|c|c|c||c|c|c||c|c|c|c|c|c|c|c||c||c|}
    \toprule
    \multicolumn{2}{|c|}{\diagbox{\bf Source}{\bf $\Uparrow$ FAR (\%)}{\bf Target}} & \textbf{Ivector} & \textbf{ECAPA} & \textbf{Xvector-P} & \textbf{Xvector-C} & \textbf{AudioNet} & \textbf{SincNet} & \textbf{Res18-I} & \textbf{Res18-V} & \textbf{Res34-I} & \textbf{Res34-V} & \textbf{Auto-I} & \textbf{Auto-V} & \textbf{GE2E} & \textbf{Hubert} \\
    \midrule
    {\bf GMM} & \textbf{Ivector} & \cellcolor{gray!40} 98.3 & \textcolor{blue}{\bf 11.8} & \textcolor{blue}{\bf 12.2} & \textcolor{blue}{\bf 4.9} & 4.8   & \textcolor{red}{\bf 0.1} & \textcolor{red}{\bf 0.1} & 0.5   & 0.8   & 0.8   & 4.8   & \textcolor{red}{\bf 0.4} & 0.8   & 0.5 \\
    \midrule
    \midrule
    \multirow{3}[4]{*}{\bf TDNN} & \textbf{ECAPA} & \textcolor{blue}{\bf 8.0} & \cellcolor{gray!40} 98.0 & \textcolor{blue}{\bf 12.1} & 4.4   & 1.9   & \textcolor{red}{\bf 0.2} & 1.8   & 1.2   & 2.8   & 2.3   & \textcolor{blue}{\bf 6.6} & 3.3   & \textcolor{red}{\bf -0.2} & \textcolor{red}{\bf -0.9} \\
    \cmidrule{2-16}
    & \textbf{Xvector-P} & \textcolor{blue}{\bf 20.6} & \textcolor{blue}{\bf 20.1} & \cellcolor{gray!40} 98.5 & \textcolor{blue}{\bf 9.2} & 2.8   & \textcolor{red}{\bf 0.0} & 0.7   & 0.7   & 1.1   & 1.3   & 5.4   & 0.4   & \textcolor{red}{\bf 0.4} & \textcolor{red}{\bf 0.0} \\
    \cmidrule{2-16}
    & \textbf{Xvector-C} & \textcolor{blue}{\bf 15.4} & \textcolor{blue}{\bf 24.5} & \textcolor{blue}{\bf 17.8} & \cellcolor{gray!40} 94.9 & 3.2   & \textcolor{red}{\bf 0.2} & 1.3   & \textcolor{red}{\bf 1.2} & 2.4   & 3.1   & 7.6   & 2.3   & 2.3   & \textcolor{red}{\bf -1.7} \\
    \midrule
    \midrule
    \multirow{8}[12]{*}{\bf CNN} & \textbf{AudioNet} & 0.5   & 0.2   & 0.2   & \textcolor{red}{\bf -1.9} & \cellcolor{gray!40} 94.1 & \textcolor{red}{\bf -0.1} & \textcolor{blue}{\bf 1.2} & 0.4   & \textcolor{blue}{\bf 1.5} & \textcolor{blue}{\bf 2.1} & 1.1   & 0.3   & 0.6   & \textcolor{red}{\bf -3.0} \\
    \cmidrule{2-16}
    & \textbf{SincNet} & 1.3   & 2     & \textcolor{red}{\bf 1.1} & \textcolor{red}{\bf 0.2} & \textcolor{blue}{\bf 7.9} & \cellcolor{gray!40} 38.0 & 3.3   & 3     & 3.5   & 4.6   & \textcolor{blue}{\bf 6.2} & 2.6   & \textcolor{blue}{\bf 14.6} & \textcolor{red}{\bf -2.6} \\
    \cmidrule{2-16}
    & \textbf{Res18-I} & 0.5   & 1.5   & 0.4   & \textcolor{red}{\bf -2.1} & 2.6   & \textcolor{red}{\bf 0.0} & \cellcolor{gray!40} 98.2 & 7.4   & \textcolor{blue}{\bf 8.1} & \textcolor{blue}{\bf 13.0} & \textcolor{blue}{\bf 13.5} & 6.4   & 0.1   & \textcolor{red}{\bf -1.5} \\
    \cmidrule{2-16}
    & \textbf{Res18-V} & 0.6   & 1.9   & 0.5   & \textcolor{red}{\bf -2.1} & 4.7   & \textcolor{red}{\bf 0.1} & \textcolor{blue}{\bf 12.5} & \cellcolor{gray!40} 97.8 & 11    & \textcolor{blue}{\bf 14.6} & \textcolor{blue}{\bf 15.6} & 8.8   & 0.7   & \textcolor{red}{\bf -1.8} \\
    \cmidrule{2-16}
    & \textbf{Res34-I} & 0.5   & 1.5   & 0.6   & \textcolor{red}{\bf -2.2} & 2.1   & 0     & \textcolor{blue}{\bf 7.3} & 5.6   & \cellcolor{gray!40} 98.6 & \textcolor{blue}{\bf 10.4} & \textcolor{blue}{\bf 12.1} & 6.3   & \textcolor{red}{\bf -0.4} & \textcolor{red}{\bf -1.2} \\
    \cmidrule{2-16}
    & \textbf{Res34-V} & 0.4   & 1.9   & 0.4   & \textcolor{red}{\bf -2.1} & 2.6   & -0.1  & \textcolor{blue}{\bf 7.1} & 5.8   & \textcolor{blue}{\bf 7.4} & \cellcolor{gray!40} 98.0 & \textcolor{blue}{\bf 12.3} & 7     & \textcolor{red}{\bf -0.3} & \textcolor{red}{\bf -1.4} \\
    \cmidrule{2-16}
    & \textbf{Auto-I} & 1.3   & 3.2   & 1.4   & \textcolor{red}{\bf -2.2} & 2.3   & 0     & 6.1   & 4.7   & \textcolor{blue}{\bf 7.1} & \textcolor{blue}{\bf 9.7} & \cellcolor{gray!40} 80.1 & \textcolor{blue}{\bf 21.7} & \textcolor{red}{\bf -0.3} & \textcolor{red}{\bf -1.1} \\
    \cmidrule{2-16}
    & \textbf{Auto-V} & 0.8   & 4.5   & 1.3   & \textcolor{red}{\bf -2.0} & 4.2   & 0.2   & \textcolor{blue}{\bf 9.5} & 6.8   & 8.7   & \textcolor{blue}{\bf 13.2} & \textcolor{blue}{\bf 26.9} & \cellcolor{gray!40} 96.8 & \textcolor{red}{\bf -0.2} & \textcolor{red}{\bf -1.2} \\
    \midrule
    \midrule
    {\bf LSTM} & \textbf{GE2E} & 0.1   & 0.4   & \textcolor{red}{\bf -0.2} & \textcolor{red}{\bf -2.5} & \textcolor{blue}{\bf 6.6} & \textcolor{blue}{\bf 1.1} & 0.6   & 0.3   & 0.9   & 0.5   & \textcolor{blue}{\bf 1.1} & 0.8   & \cellcolor{gray!40} 61.1 & \textcolor{red}{\bf -3.1} \\
    \midrule
    \midrule
    {\bf Trans} & \textbf{Hubert} & \textcolor{blue}{\bf 1.3} & 0.7   & \textcolor{blue}{\bf 0.8} & \textcolor{red}{\bf -2.3} & \textcolor{blue}{\bf 1.0} & 0     & -0.3  & -0.1  & 0.4   & \textcolor{red}{\bf -0.7} & 0.4   & -0.4  & \textcolor{red}{\bf -1.1} & \cellcolor{gray!40} 92.3 \\
    \bottomrule
    \end{tabular}%
  }
  \label{tab:transfer-SV}%
\end{table*}%

\begin{table*}
  \centering\setlength\tabcolsep{4pt}
  \caption{The attack success rate  (\%)  of \attackname on the OSI task with PGD (\#Iter=2) as the optimization approaches}\vspace{-2mm}
  \resizebox{0.97\textwidth}{!}{%
    \begin{tabular}{|c|cccc|cccc|cccc|cccc|ccc|ccc|cccccc|}
    \toprule
    \multicolumn{1}{|c|}{\multirow{4}[8]{*}{}} & \multicolumn{28}{c|}{\textbf{OSI}} \\
\cmidrule{2-29}    \multicolumn{1}{|c|}{} & \multicolumn{8}{c|}{\boldmath{}\textbf{Targeted ($t\in G$) Random}\unboldmath{}} & \multicolumn{8}{c|}{\boldmath{}\textbf{Targeted ($t\in G$) Least Likely}\unboldmath{}} & \multicolumn{6}{c|}{\boldmath{}\textbf{Targeted ($t={\tt imposter}$)}\unboldmath{}} & \multicolumn{6}{c|}{\textbf{Untargeted}} \\
\cmidrule{2-29}    \multicolumn{1}{|c|}{} & \multicolumn{4}{c|}{\boldmath{}\textbf{ASR$_t$ (\%)}\unboldmath{}} & \multicolumn{4}{c|}{\boldmath{}\textbf{ASR$_u$ (\%)}\unboldmath{}} & \multicolumn{4}{c|}{\boldmath{}\textbf{ASR$_t$ (\%)}\unboldmath{}} & \multicolumn{4}{c|}{\boldmath{}\textbf{ASR$_u$ (\%)}\unboldmath{}} & \multicolumn{3}{c|}{\boldmath{}\textbf{ASR$_t$ (\%)}\unboldmath{}} & \multicolumn{3}{c|}{\boldmath{}\textbf{ASR$_u$ (\%)}\unboldmath{}} & \multicolumn{6}{c|}{\boldmath{}\textbf{ASR$_u$ (\%)}\unboldmath{}} \\
\cmidrule{2-29}    \multicolumn{1}{|c|}{} & \boldmath{}\textbf{$\mathcal{L}_\text{CE}$}\unboldmath{} & \boldmath{}\textbf{$\mathcal{L}_\text{M}$}\unboldmath{} & \boldmath{}\textbf{$\mathcal{L}_{1}$}\unboldmath{} & \boldmath{}\textbf{$\mathcal{L}_{2}$}\unboldmath{} & \boldmath{}\textbf{$\mathcal{L}_\text{CE}$}\unboldmath{} & \boldmath{}\textbf{$\mathcal{L}_\text{M}$}\unboldmath{} & \boldmath{}\textbf{$\mathcal{L}_{1}$}\unboldmath{} & \boldmath{}\textbf{$\mathcal{L}_{2}$}\unboldmath{} & \boldmath{}\textbf{$\mathcal{L}_\text{CE}$}\unboldmath{} & \boldmath{}\textbf{$\mathcal{L}_\text{M}$}\unboldmath{} & \boldmath{}\textbf{$\mathcal{L}_{1}$}\unboldmath{} & \boldmath{}\textbf{$\mathcal{L}_{2}$}\unboldmath{} & \boldmath{}\textbf{$\mathcal{L}_\text{CE}$}\unboldmath{} & \boldmath{}\textbf{$\mathcal{L}_\text{M}$}\unboldmath{} & \boldmath{}\textbf{$\mathcal{L}_{1}$}\unboldmath{} & \boldmath{}\textbf{$\mathcal{L}_{2}$}\unboldmath{} & \boldmath{}\textbf{$\mathcal{L}_\text{CE}^{s}$}\unboldmath{} & \boldmath{}\textbf{$\mathcal{L}_3$}\unboldmath{} & \boldmath{}\textbf{$\mathcal{L}_{1}^{s}$}\unboldmath{} & \boldmath{}\textbf{$\mathcal{L}_\text{CE}^{s}$}\unboldmath{} & \boldmath{}\textbf{$\mathcal{L}_3$}\unboldmath{} & \boldmath{}\textbf{$\mathcal{L}_{1}^{s}$}\unboldmath{} & \boldmath{}\textbf{$\mathcal{L}_\text{CE}^{s}$}\unboldmath{} & \boldmath{}\textbf{$\mathcal{L}_\text{M}^{s}$}\unboldmath{} & \boldmath{}\textbf{$\mathcal{L}_{4}^{s}$}\unboldmath{} & \boldmath{}\textbf{$\mathcal{L}_{2}^{s}$}\unboldmath{} & \boldmath{}\textbf{$\mathcal{L}_{1}^{s}$}\unboldmath{} & \boldmath{}\textbf{$\mathcal{L}^{-}_3$}\unboldmath{} \\
    \midrule
    \multirow{1}{*}{\boldmath{}\textbf{$s\in G$}\unboldmath{}} & 77.18  & 53.69  & \textcolor{blue}{\textbf{82.7}} &  68.23 &  86.12   & \textcolor{blue}{\textbf{98.31}} &  86.74  & 98.26 & 34.43 &  11.34 & \textcolor{blue}{\textbf{47.33}} & 25.87 & 83.74  & \textcolor{blue}{\textbf{98.39}} & 81.48  &  98.36 &  99.65 & 99.84 & \textcolor{blue}{\textbf{99.87}} &   99.79  & 99.85 & \textcolor{blue}{\textbf{99.88}} & 0.14   & 98.5  & 98.73  & \textcolor{blue}{\textbf{99.44}} & 0.01 & \multirow{1}{*}{N/A} \\
    \midrule
    \multirow{1}{*}{\boldmath{}\textbf{$s\not\in G$}\unboldmath{}} &  85.25  &  23.12 & \textcolor{blue}{\textbf{91.68}} & 91.24 & 85.55 & 23.12 & \textcolor{blue}{\textbf{91.92}} & 91.24  & 38.66  & 0.59  & \textcolor{blue}{\textbf{54.08}} & 53.72  &  38.89 &  0.60 & \textcolor{blue}{\textbf{54.59}} & 53.97 & \multicolumn{6}{c|}{N/A}   & \multicolumn{5}{c|}{N/A}  & 100  \\
    \bottomrule
    \end{tabular}%
    }
    \label{tab:basic-OSI-100}\vspace{-2mm}
\end{table*}%

\begin{table*}\setlength\tabcolsep{7pt} 
  \centering
  \caption{The attack success rate (\%) of \attackname on the CSI and SV tasks with PGD (\#Iter=2) as the optimization approaches}\vspace{-2mm}
  \resizebox{.97\textwidth}{!}{%
    \begin{tabular}{|c|ccc|ccc|ccc|ccc|cccc|c|c|}
    \toprule
    \multicolumn{1}{|c|}{\multirow{4}[8]{*}{}} & \multicolumn{16}{c|}{\textbf{CSI}}                                                                                            & \multicolumn{2}{c|}{\textbf{SV}} \\
\cmidrule{2-19}    \multicolumn{1}{|c|}{} & \multicolumn{6}{c|}{\textbf{Targeted Random}} & \multicolumn{6}{c|}{\textbf{Targeted Least Likely}} & \multicolumn{4}{c|}{\textbf{Untargeted}} & \multicolumn{2}{c|}{\textbf{Targeted}} \\
\cmidrule{2-19}    \multicolumn{1}{|c|}{} & \multicolumn{3}{c|}{\boldmath{}\textbf{ASR$_t$ (\%)}\unboldmath{}} & \multicolumn{3}{c|}{\boldmath{}\textbf{ASR$_u$ (\%)}\unboldmath{}} & \multicolumn{3}{c|}{\boldmath{}\textbf{ASR$_t$ (\%)}\unboldmath{}} & \multicolumn{3}{c|}{\boldmath{}\textbf{ASR$_u$ (\%)}\unboldmath{}} & \multicolumn{4}{c|}{\boldmath{}\textbf{ASR$_u$ (\%)}\unboldmath{}} & \multicolumn{2}{c|}{\boldmath{}\textbf{ASR$_t$ (\%)}\unboldmath{}} \\
\cmidrule{2-19}    \multicolumn{1}{|c|}{} & \boldmath{}\textbf{$\mathcal{L}_\text{CE}$}\unboldmath{} & \boldmath{}\textbf{$\mathcal{L}_\text{M}$}\unboldmath{} & \boldmath{}\textbf{$\mathcal{L}_1$}\unboldmath{} & \boldmath{}\textbf{$\mathcal{L}_\text{CE}$}\unboldmath{} & \boldmath{}\textbf{$\mathcal{L}_\text{M}$}\unboldmath{} & \boldmath{}\textbf{$\mathcal{L}_1$}\unboldmath{} & \boldmath{}\textbf{$\mathcal{L}_\text{CE}$}\unboldmath{} & \boldmath{}\textbf{$\mathcal{L}_\text{M}$}\unboldmath{} & \boldmath{}\textbf{$\mathcal{L}_1$}\unboldmath{} & \boldmath{}\textbf{$\mathcal{L}_\text{CE}$}\unboldmath{} & \boldmath{}\textbf{$\mathcal{L}_\text{M}$}\unboldmath{} & \boldmath{}\textbf{$\mathcal{L}_1$}\unboldmath{} & \boldmath{}\textbf{$\mathcal{L}_\text{CE}^{s}$}\unboldmath{} & \boldmath{}\textbf{$\mathcal{L}_\text{M}^{s}$}\unboldmath{} & \boldmath{}\textbf{$\mathcal{L}_4^{s}$}\unboldmath{} & \boldmath{}\textbf{$\mathcal{L}_1^{s}$}\unboldmath{} & \boldmath{}\textbf{$\mathcal{L}_\text{BCE}/\mathcal{L}'_\text{BCE}$}\unboldmath{} & \boldmath{}\textbf{$\mathcal{L}_{3B}/$$\mathcal{L}^{-}_{3B}$}\unboldmath{} \\
    \midrule
    \multirow{1}{*}{\boldmath{}\textbf{$s\in G$}\unboldmath{}} &   82.85 & \textcolor{blue}{\textbf{92.09}} &  85.21 & 82.86 & \textcolor{blue}{\textbf{92.99}} & 85.23 & 64.24  & \textcolor{blue}{\textbf{74.78}} & 67.55  & 64.35  & \textcolor{blue}{\textbf{82.38}} & 67.82  & 93.06  & \textcolor{blue}{\textbf{99.6}} & 98.73  & 86.63  &  98.13 & 98.13 \\
    \midrule
    \multirow{1}{*}{\boldmath{}\textbf{$s\notin G$}\unboldmath{}}  & 99.47  &  97.25 & \textcolor{blue}{\textbf{99.52}}  &  100  &  100  & 100 &  \textcolor{blue}{\textbf{94.63}} & 70.68  &  93.66 &  100 & 100  & 100 & \multicolumn{4}{c|}{\multirow{1}{*}{N/A}} & 99.32  & 99.32 \\
    \bottomrule
    \end{tabular}%
    }
    \label{tab:basic-CSI-SV-100}\vspace{-2mm}
\end{table*}%

\subsection{More Transferability Results}\label{sec:more-transfer-result}
For the OSI task, we consider the untargeted attack with with unenrolled speakers as the source speaker, i.e., C5.
For the SV task, we consider the targeted attack with unenrolled speakers as the source speaker and the enrolled speaker as the target speaker, i.e., C10.
The results of C5 and C10 are shown in \tablename~\ref{tab:transfer-OSI} and \tablename~\ref{tab:transfer-SV}, respectively.

\end{document}